\documentclass[reprint,notitlepage,aps,prd,superscriptaddress,showpacsnofootinbibz,nofootinbib,maxnames=4,nobibnotes,longbibliography]{revtex4-2}

\usepackage{amsmath,amssymb, graphicx, setspace}
\usepackage[pdftex,hidelinks]{hyperref}
\usepackage{commath}
\usepackage{upgreek}
\usepackage{cleveref} 
\usepackage{float}
\usepackage[dvipsnames]{xcolor}
\usepackage[normalem]{ulem}
\usepackage[separate-uncertainty = true, mode =math]{siunitx}
\usepackage[acronym,hyperfirst=false,nonumberlist,nowarn]{glossaries-extra}
\usepackage{siunitx}
\usepackage{enumitem}
\usepackage[caption=false]{subfig}
\usepackage{titlesec}
\usepackage{booktabs}

\titlespacing*{\subsection}{0pt}{10pt}{5pt}

\allowdisplaybreaks 

\setabbreviationstyle[acronym]{long-short}
\glssetcategoryattribute{acronym}{nohyperfirst}{true}

\newcommand{\mathsym}[1]{{}}
\newcommand{\unicode}[1]{{}}

\usepackage{calligra} 
\DeclareMathAlphabet{\mathcalligra}{T1}{calligra}{m}{n} 
\DeclareFontShape{T1}{calligra}{m}{n}{<->s*[2.2]callig15}{} 
\DeclareMathOperator{\sign}{sign}

\Crefname{equation}{Eq.}{Eqs.}
\Crefname{figure}{Fig.}{Figs.}
\Crefname{tabular}{Tab.}{Tabs.}
\Crefname{section}{Sec.}{Secs.}

\newcommand\ttl{\mathrm{TTL}}

\newcommand{\delay}[1]{{\mathcal{D}_{#1}}}

\newcolumntype{R}[1]{>{\raggedright}m{#1}}
\newcolumntype{L}[1]{>{\raggedleft}m{#1}}
\newcolumntype{C}[1]{>{\centering}m{#1}}

\newacronym{CoM}{CoM}{center of mass}
\newacronym[longplural={degrees of freedom},\glsshortpluralkey=DoF]{DOF}{DoF}{degree of freedom}
\newacronym{DWS}{DWS}{differential wavefront sensing}
\newacronym{Ifo}{Ifo}{interferometer}
\newacronym{FS}{FS}{free space}
\newacronym{GRS}{GRS}{gravitational reference sensor}
\newacronym{LAI}{LAI}{long arm interferometer}
\newacronym{LO}{LO}{local oscillator}
\newacronym{MF}{MF}{MOSA coordinate frame}
\newacronym[longplural={movable optical subassemblies}]{MOSA}{MOSA}{movable optical subassembly}
\newacronym[longplural={optical benches}]{OB}{OB}{optical bench}
\newacronym{PAAM}{PAAM}{point ahead angle mechanism}
\newacronym{PD}{PD}{photodiode}
\newacronym{Ref-Ifo}{Ref-Ifo}{reference interferometer}
\newacronym{RX-beam}{RX-beam}{received beam}
\newacronym{SC}{SC}{spacecraft}
\newacronym{SF}{SF}{spacecraft coordinate frame}
\newacronym{TDI}{TDI}{time-delay interferometry}
\newacronym[longplural=test masses]{TM}{TM}{test mass}
\newacronym{TMI}{TMI}{test mass interferometer}
\newacronym{TX-beam}{TX-beam}{transmitted beam}
\newacronym{TTL}{TTL}{tilt-to-length}
\newacronym[longplural={amplitude spectral densities}]{ASD}{ASD}{amplitude spectral density}
\newacronym[longplural={power spectral densities}]{PSD}{PSD}{power spectral density}

\begin{document}
%%%%%%%%%%%%%%%%%%%%%%%%%%%%%%%%%%%%%%%%%%%%%%%%%%%%%%%%%%%%%
\title{In-Depth Modeling of Tilt-To-Length Coupling in LISA's Interferometers and TDI Michelson Observables}
%%%%%%%%%%%%%%%%%%%%%%%%%%%%%%%%%%%%%%%%%%%%%%%%%%%%%%%%%%%%%
\def\addressa{Max Planck Institute for Gravitational Physics (Albert Einstein Institute), 30167 Hannover, Germany}
\def\addressb{Institute for Gravitational Physics of the Leibniz Universität Hannover, 30167 Hannover, Germany}
\def\addressc{Hamburger Sternwarte, University of Hamburg, Gojenbergsweg 112, D-21029 Hamburg, Germany}

\author{G.~Wanner}\affiliation{\addressb}\affiliation{\addressa}
\author{S.~Shah}\affiliation{\addressa}\affiliation{\addressb}\affiliation{\addressc}
\author{M.~Staab}\affiliation{\addressa}\affiliation{\addressb}
\author{H.~Wegener}\affiliation{\addressa}\affiliation{\addressb}
\author{S.~Paczkowski}\affiliation{\addressa}\affiliation{\addressb}
\date{\today}
%%%%%%%%%%%%%%%%%%%%%%%%%%%%%%%%%%%%%%%%%%%%%%%%%%%%%%%%%%%%%
\begin{abstract}
We present first-order models for \gls{TTL} coupling in LISA, both for the individual interferometers, as well as in the \gls{TDI} Michelson observables. These models include the noise contributions from angular and lateral jitter coupling of the six test masses, six \glspl{MOSA}, and three spacecraft. We briefly discuss which terms are considered to be dominant and reduce the TTL model for the second-generation \gls{TDI} Michelson~X observable to these primary noise contributions to estimate the resulting noise level. We show that the expected TTL noise will initially violate the entire mission displacement noise budget, resulting in the known necessity to fit and subtract TTL noise in data post-processing. By comparing the noise levels for different assumptions prior to subtraction, we show why noise mitigation by realignment prior to subtraction is favorable.  \\
We then discuss that the TTL coupling in the individual interferometers will have noise contributions that will not be present in the \gls{TDI} observables. Models for TTL coupling noise in \gls{TDI} and in the individual interferometers are therefore different, and commonly made assumptions are valid as such only for \gls{TDI}, but not for the individual interferometers. \\
Finally, we analyze what implications can be drawn from the presented models for the subsequent fit-and-subtraction in post-processing. We show that noise contributions from the test mass and inter-satellite interferometers are indistinguishable, such that only the combined coefficients can be fit and used for subtraction. However, a distinction is considered not necessary. 
Additionally, we show a correlation between coefficients for transmitter and receiver jitter couplings in each individual \gls{TDI} Michelson observable. This full correlation, however, can be resolved by using all three Michelson observables for fitting the TTL coefficients.  
\end{abstract}
%%%%%%%%%%%%%%%%%%%%%%%%%%%%%%%%%%%%%%%%%%%%%%%%%%%%%%%%%%%%%
\maketitle
%
%%%%%%%%%%%%%%%%%%%%%%%%%%%%%%%%%%%%%%%%%%%%%%%%%%%%%%%%%%%%%
%%%%%%%%%%%%%%%%%%%%%%%%%%%%%%%%%%%%%%%%%%%%%%%%%%%%%%%%%%%%%
%%%%%%%%%%%%%%%%%%%%%%%%%%%%%%%%%%%%%%%%%%%%%%%%%%%%%%%%%%%%%
\section{Introduction}
\label{Sec:intro}
%%%%%%%%%%%%%%%%%%%%%%%%%%%%%%%%%%%%%%%%%%%%%%%%%%%%%%%%%%%%%
The space-based gravitational wave detector `Laser Interferometer Space Antenna' (LISA) \cite{2017arXiv170200786A}  is an ESA-led mission with contributions from NASA and the European member states. It is planned for launch in the mid-2030s. Once in its final orbit and switched to science mode, LISA will measure gravitational waves in the \SI{0.1}{mHz} to \SI{1}{Hz} frequency band. It, thereby, complements the very successful ground-based gravitational wave detectors \cite{Aasi2015short,Acernese2023short,Abe2022short,Dooley2016}, 
which address frequencies in the Hz to kHz range, as well as pulsar timing arrays \cite{Agazie2023short,Antoniadis2023,Reardon2023,Xu2023}, which address gravitational waves in the nanohertz regime. 
By measuring gravitational waves originating from binary systems of white dwarfs, neutron stars, and black holes of a wide range of masses, LISA will address a rich and versatile set of scientific questions \cite{2017arXiv170200786A}.

LISA consists of three \gls{SC} in an (almost) equilateral triangular configuration with a mean interspacecraft separation of approximately 2.5\,million\, km. Each SC will follow an individual heliocentric orbit such that the triangular constellation is tilted by \SI{60}{\degree} out of the ecliptic and trails the Earth by \SI{\sim20}{\degree} at a distance of \SIrange{50}{60}{million\,km}. Each LISA SC will house two effectively identical optical benches (OBs) and two freely-falling \glspl{TM}~\cite{Armano2016, Armano2018}. 

Gravitational waves will induce variations of the proper distance between the freely-falling \glspl{TM} aboard two different \gls{SC}. LISA's requirements are set to measure such distance variations down to the picometer level using its heterodyne laser interferometry.  For technical reasons, the measurement of the distance variation between two test masses is split up into three individual measurements. A \gls{TMI} measures the distance variation between a test mass and a reference point on the local satellite. The \acrlong{LAI} \glsunset{LAI}(\acrshort{LAI}, also known as inter-satellite interferometer or science interferometer) measures the distance variation between the very same reference point and a comparable reference point on a remote satellite. Finally, another \gls{TMI} measures the distance variations between the remote reference point and the freely falling test mass on that satellite. The motion of the reference points on the corresponding optical benches cancels when the three measurements are combined appropriately.  
The well-known basic working principle of LISA is, therefore, as follows: Laser beams sent from one spacecraft to another get phase-shifted by the variations in the proper distance caused by the gravitational waves. Consequently, the gravitational wave signals are measured as phase variations in the \glspl{LAI}.

Celestial mechanics will influence the individual SC orbits, resulting in arm length variations by up to \SI{\approx \pm1}{\%}, which is $\approx \pm$\SI{25000}{km} during a year. Since all three arm lengths vary differently, the equilateral triangular form of LISA will be slightly deformed during the course of a year. This variation of the arm lengths is unlike in the ground-based detectors and causes a significant coupling of laser frequency noise into the interferometric readout signals. It makes laser frequency noise coupling a primary noise source in LISA with an equivalent of mm-level displacement noise, which is orders of magnitude larger than the pm-level of proper distance variations caused by gravitational waves, which LISA is designed to measure. Fortunately, the coupling of laser frequency noise into the longitudinal phase measurement as well as its suppression in the final recombined signal with time-delay interferometry (\gls{TDI})~\cite{Armstrong1999, Tinto2004,Otto2015,Muratore2020,Tinto2023} techniques is well-understood. The so-called \gls{TDI} observables are formed by linearly combining the various interferometric readout signals with suitable time delays, which suppresses the laser frequency noise coupling below the LISA requirements.

After the suppression of laser frequency noise, a variety of secondary noise sources remain. These include clock phase noise~\cite{Hartwig2021}, relative intensity noise~\cite{Wissel2023}, and others. Among these secondary noises is the \acrfull{TTL} coupling noise, i.e.\ the cross-coupling of angular or lateral vibrations (also commonly known as ``jitters") into the LISA interferometric phase readout. 
This noise type is a cross-talk since the interferometric phase signal is intended to only sense distance variations, while angular and lateral motions are nominally orthogonal \glspl{DOF} and should not be sensed. Nevertheless, motion in all \gls{DOF} can - and usually does - couple into the interferometric phase. \gls{TTL} coupling noise was already one of the major noise sources in the LISA Pathfinder mission~\cite{Hartig2023-LPF-DA,Wanner2017,Armano2018,Armano2016} and will be even more significant in LISA.

For this reason, a three-fold TTL suppression scheme is planned for LISA. This means TTL coupling noise is suppressed in LISA 
\begin{enumerate}[topsep=0pt,itemsep=-1ex,partopsep=1ex,parsep=1ex]
	\item by design,
	\item by realignment, and
	\item by fit and subtraction in post-processing.
\end{enumerate}

The first suppression step consists of two parts: The first part is the split-interferometry concept where the \glspl{TMI} and \glspl{LAI} both measure the motion of the optical bench, ideally along the same axis. When the signals from these interferometers are then combined in the \gls{TDI}, the commonly sensed longitudinal optical bench motion caused by angular or lateral jitter (a strong contributor to TTL noise) cancels. \\
The second part of noise suppression by design is given by the use of dedicated imaging optics, which are known and proven to suppress TTL coupling noise~\cite{Chwalla2020,Troebs2018,Chwalla2016,Schuster2016}.

A second stage of TTL noise mitigation is by fine-tuning the alignment in the interferometer. This concept is theoretically understood (e.g.~\cite{Hartig2022-G}) and experimentally proven both in on-ground experiments, as well as by the LISA-Pathfinder mission~\cite{Hartig2023-LPF-DA,Chwalla2020}. It is, therefore, also planned for LISA that the alignment can be adapted either prior to launch, or in flight, or both, to reduce the TTL coupling noise in LISA.

It is currently not expected that the required noise levels can be achieved by the first two mitigation strategies alone. Therefore, it is planned to fit a linear TTL model to the obtained \gls{TDI} observables and subtract the noise from the measured data. Such a fit and subtraction was already successfully performed in LISA Pathfinder~\cite{Armano2016}, and was successfully tested for LISA by simulations~\cite{Paczkowski2022, LCST-INST-RP-002, LCST-INST-TN-017}.

A number of articles have already been published on the topic of TTL coupling in LISA. There are publications focused on the validation of TTL noise suppression by imaging optics (e.g.~\cite{Chwalla2020,Troebs2018,Schuster2016,Wang2020}). Others focus on specific noise contributions, such as from the far-field wavefront distortions of light transmitted through the telescope (e.g.~\cite{Weaver2022,Sasso2019,Sasso2018far-field,Wang2020,Lin2023}), or influences of alignment (e.g.~\cite{Zhao2020}). Particular focus is currently given on the noise reduction by fit and subtraction (e.g.~\cite{George2023,Paczkowski2022,Houba2022-Estimation}) and calibration maneuvers as an alternative option to estimate the coefficients (\cite{Houba2022-maneuver}) prior to subtraction. Finally, the TTL noise in the \gls{TDI} Michelson combinations~\cite{Houba2022-Estimation} and also in \gls{TDI} infinity were modeled and compared in~\cite{Houba2023-TDI-inf}.

This paper is focused on noise modeling prior to the fit and subtraction step. For this purpose, we derive a linear TTL model that has already been introduced in~\cite{Houba2022-Estimation} and which was partly used for the noise generation in~\cite{Paczkowski2022}. Already in the derivation of the model, but also when the model is reduced to the dominating contributions, a number of important assumptions are made, which have not been addressed in previous publications. We discuss these assumptions and show that they are valid only for modeling the noise in \gls{TDI} observables, while we consider them invalid for individual interferometers. The reason for this is that there exist strong TTL effects, that is TTL effects involving optical bench longitudinal motion, which are common in the individual interferometers of a single link. These cancel when the individual interferometer signals are added to form the \gls{TDI} single-link signals. Therefore, we discuss in detail the difference between TTL in individual interferometers versus in a single link or in \gls{TDI} observables.

Additionally, we use the reduced linear model to estimate the expected noise levels per single link (i.e.\ interferometric connection between two test masses). 
For this, we need both estimates for the jitters, as well as for the coupling factors. Yet, deriving these factors is a substantial task in itself (see e.g.~\cite{Hartig2022-G,Hartig2023-NG,Weaver2022,Sasso2018far-field}) and is, therefore, beyond the scope of this paper. For this reason, we only very shortly argue the expected magnitude of the total coupling factors per degree of freedom, without going much into detail.\\
Depending on whether the noise is previously suppressed by realignment or not, we find median noise levels in a single link in the order of \SI{13}{pm/\sqrt{Hz}} or \SI{58}{pm/\sqrt{Hz}} prior to fit and subtraction. We confirmed these findings with two non-statistical simulation results in previous publications \cite{Paczkowski2022,Houba2023-TDI-inf}.  
The worst-case estimates for our two cases are approximately  \SI{40}{pm/\sqrt{Hz}} or \SI{172}{pm/\sqrt{Hz}} per single link prior to fit and subtraction. The noise levels we find would mostly exceed the entire mission noise budget and are the reason why the third mitigation step of fitting and subtracting the TTL noise from the data is indispensable and planned for. 

Finally, we analyze the derived TTL coupling models and show that noise contributions from the \acrshortpl{LAI} and \glspl{TMI} become indistinguishable when studied in data analysis or fitted to a model. Likewise, we highlight a correlation between the TTL coupling from transmitter and receiver jitters in the individual \gls{TDI}-$X$, -$Y$, and -$Z$ observables, but this correlation can be broken if all three observables are used for fitting the noise contributions. 

The outline of this paper is as follows: \\
We describe TTL coupling as a generic interferometric noise source in \Cref{Sec:TTL-basics}. 
We then describe in \Cref{Sec:generic-TTL-in-LISA} which of the LISA interferometers are subject to TTL-coupling. We add generic noise terms into the interferometric phase signals and derive the resulting contributions to the \gls{TDI} Michelson observables.
In \Cref{Sec:From-Frames-to-Explicit-TTL-Model}, we then derive an explicit first-order TTL model applicable for noise estimates in single-links or \gls{TDI}. This substitutes the generic form used until this point. \\
In \Cref{Sec:TTL-noise-estimate}, we reduce the explicit model to its most significant contributions, specify current assumptions for jitters and coupling factors, and compute the resulting noise in the second-generation \gls{TDI} Michelson observables. In order to validate the analytic model, we additionally tested one set of data numerically with the simulation software tools LISA Instrument \cite{LISAInstrument} and PyTDI \cite{PyTDI} and show a perfect agreement between the resulting analytic and numeric noise estimate. After this, we show the TTL noise estimates for the second-generation \gls{TDI}-$X$ variable, derived from two Monte-Carlo simulations.\\
In \Cref{Sec:Implicit-Assumptions+Implications}, we show why the assumptions made in the previous section do not hold for the individual interferometers and define another linear TTL model that should be used instead when describing the noise in individual interferometers.
In \Cref{Sec:DA}, we discuss a delineation of the shown model for noise estimation from the models used for data analysis, and show the mixing of different signals in \gls{TDI}, which causes correlations.
Finally, we conclude in \Cref{Sec:Conclusions} with the key findings. 
%%%%%%%%%%%%%%%%%%%%%%%%%%%%%%%%%%%%%%%%%%%%%%%%%%%%%%%%%%%%%
%%%%%%%%%%%%%%%%%%%%%%%%%%%%%%%%%%%%%%%%%%%%%%%%%%%%%%%%%%%%%
%
%
\section{Basic principles of TTL coupling} \label{Sec:TTL-basics}
%%%%%%%%%%%%%%%%%%%%%%%%%%%%%%%%%%%%%%%%%%%%%%%%%%%%%%%%%%%%%
For any type of interferometer, we can define the axis along which the interferometer measures distance variations as its `longitudinal' or $x$-direction. The interferometric longitudinal readout signal can then be written as
\begin{equation}
	S = c_x x + N
	\label{Eq:intro_N}
\end{equation} 
which means the interferometer has a readout signal $S$ which is proportional to displacement $x$ along its longitudinal direction, and $c_x$ is the factor of proportionality, while $N$ denotes noise disturbing the measurement. In the context of this paper, $N$ denotes TTL noise originating from any type of component jitter within the system along any orthogonal degree of freedom.
\begin{figure}
	\subfloat[Angular jitter TTL  \label{Fig:angTTL-TMI}]{\includegraphics*[page=1,width=0.49\columnwidth]{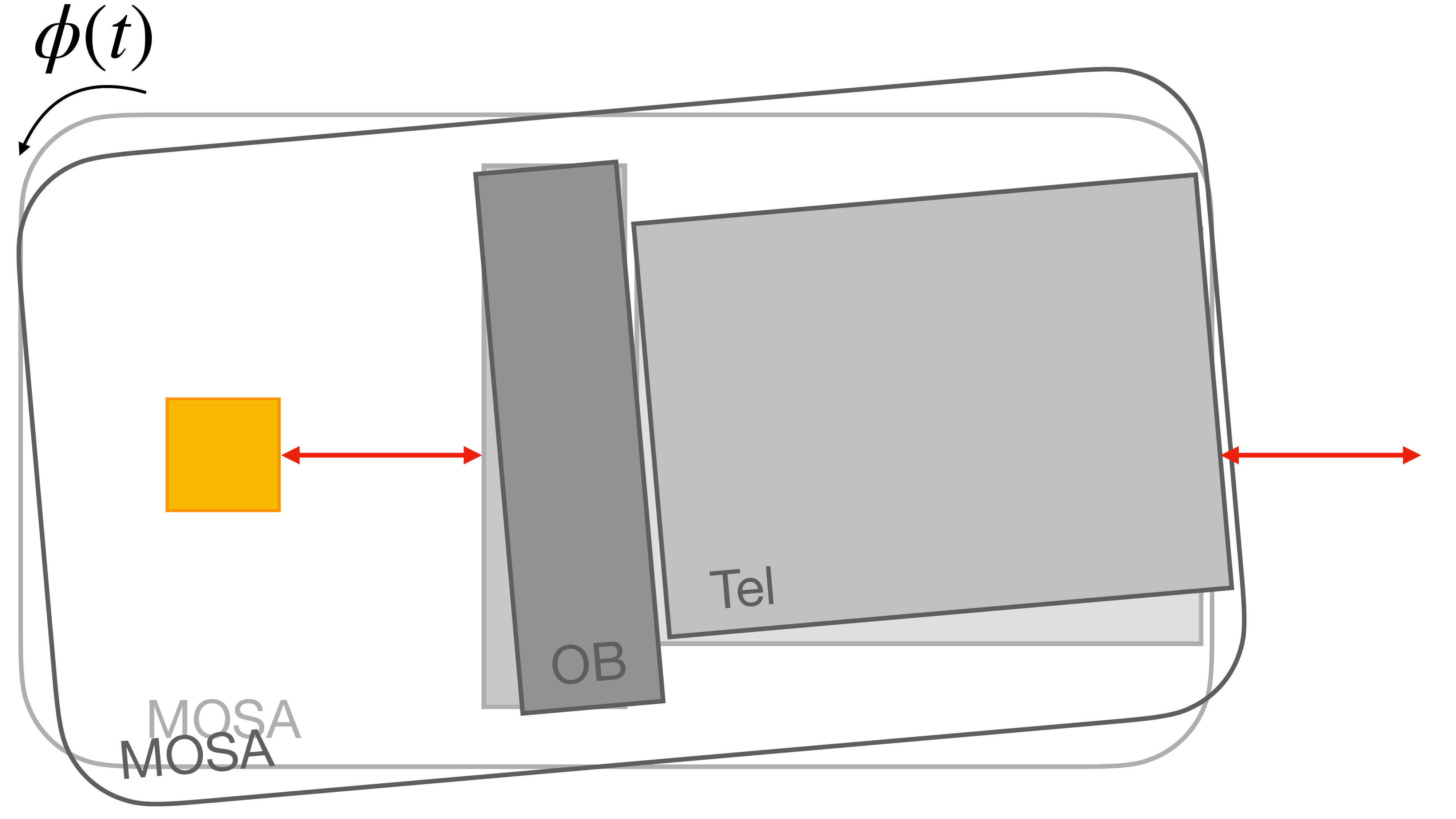}}
	\hspace{0.01 \columnwidth}
	\subfloat[Lateral jitter TTL \label{Fig:lateralTTL}]{\includegraphics*[page=2,width=0.48\columnwidth,  trim = 0 0 200 30]{figures/TTL-effects.pdf}}\\
	\subfloat[Angular transmitter jitter TTL \label{Fig:FF-TTL}]{\includegraphics*[page=3,width=\columnwidth,  trim = 0 0 0 630]{figures/TTL-effects.pdf}}\\
	\caption{Examples for the different types of \gls{TTL} coupling. Subfigure (a): Angular jitter $\phi(t)$ affects the interferometric phase because the jitter alters the distances between the components.  Subfigure (b): \Acrfull{SC} jitter along $y$ affects the phase in the four indicated interferometers because the $x$-axes (red double-arrows) of all interferometers are not orthogonal to the jitter direction.
Subfigure (c): angular transmitter jitter coupling due to wavefront errors (left-hand side) and absence of this coupling (right-hand side image) if the beam is rotating around the center of curvature of a perfectly spherical wavefront. }
	\label{Fig:TTLcoupling}
\end{figure}

There are numerous mechanisms for how angular jitter can cause TTL coupling noise, one of which is depicted in \Cref{Fig:angTTL-TMI}. Here, an angular jitter $\phi(t)$ causes distance variations between the indicated components, particularly between the test mass (orange square), and the \gls{OB}. This distance variation is measured along the beam axes indicated by the red double arrows. \\
A second type of angular jitter coupling is depicted in \Cref{Fig:FF-TTL}. Here, angular jitter causes the spread-out wavefront to be scanned over the receiver due to the rotation of the transmitter. In the case of a perfectly spherical wavefront and a center of rotation that coincides with the wavefront's center of curvature, no effect will be seen (right-hand side image). However, deviations from the sphericity (or if the centers of curvature and rotation do not coincide), a coupling will occur (left-hand side image).  
For a general 3D case, angular TTL coupling can principally occur in all three angular degrees of freedom $\eta, \phi, \theta$ and can be written in a linear form as:
\begin{equation}
    N=  c_\eta \eta + c_\phi \phi + c_\theta \theta\;.
    \label{Eq:ang_gen_ttl}
\end{equation}
We then define rotations around the $x$-axis as `roll' $\theta$, around the $y$-axis as `pitch' $\eta$, and around $z$-axis as `yaw' $\phi$. The terms $c_\eta, c_\phi, c_\theta$ are the corresponding coupling coefficients. 

As described for instance in \cite{Hartig2022-G} and applied in the LISA Pathfinder case \cite{Hartig2023-LPF-DA,Hartig2023-LPF-Ana,Wanner2017}, we also include lateral jitter coupling into TTL, because it usually relates to a static tilt. This is illustrated in \Cref{Fig:lateralTTL} for the case of a laterally jittering \gls{SC}. Here, the \gls{SC} jitters along its $y$-direction, which is not orthogonal to the interferometric longitudinal $x$-directions, indicated by the red double-arrows. Here, every red double-arrow indicates the nominal axis of one interferometer. Consequently, the components move into- or out-of the beam paths, resulting in phase changes in all four interferometers. This is caused by the SC-$y$-direction being statically tilted against all four interferometric $y$-directions. 

A complete first-order generic TTL model, therefore, includes also lateral jitter coupling and reads
\begin{subequations}
\begin{align}
    N &= c_y y + c_z z + c_\eta \eta + c_\phi \phi + c_\theta \theta
    \label{Eq:gen_ttl} \\
   \nopagebreak
    & =: \sum_\alpha c_\alpha \alpha \;,%
    \label{Eq:gen_ttl_alpha}%
\end{align}
\label{Eq:gen_ttl_complete}%
\end{subequations}
whereby we introduce a short notation by a sum over all degrees of freedom $\alpha  \in \{y, z, \eta, \phi, \theta\}$. \\
Here, $y,z$ are referred to as lateral displacement degrees of freedom, which are orthogonal to $x$. 

Please note: it was argued in \cite{Paczkowski2022} that the TTL coupling from lateral jitters ($y,z$) can be neglected, and e.g.\ also \cite{Houba2022-Estimation,Houba2022-maneuver} focus entirely on angular jitter coupling. Yet, we define this general TTL model and also the detailed models in the later sections for all five degrees of freedom, which are orthogonal to the longitudinal direction $x$. We do so not only for completeness but also to discuss assumptions, such as the mentioned statement that lateral jitter TTL is negligible. 

We then assume that the variables $y, z, \eta, \phi, \theta$ represent the angular and lateral jitters of an individual component, such as the \gls{TM}, the \gls{SC} or others. 
Such a jitter needs to be defined relative to another component or with respect to a reference frame.  There are several possible reference frames one could choose, not all of which are inertial due to the motion of the LISA satellites on a heliocentric, and thereby accelerated orbit. We consider here a coordinate frame co-moving with the satellite along a hypothetical noise-free orbit within a short time frame of about 30 minutes. Within this time, the orbit can be linearized and assumed to be non-accelerated and therefore inertial. All mentioned components are then jittering with respect to this inertial reference system, which we refer to as \gls{FS}.

Every coupling coefficient $c_\alpha$ represents, in general, the sum of many individual geometric and non-geometric TTL effects \cite{Hartig2022-G, Hartig2023-NG}. However, it is not the aim of this paper to derive the magnitude of these coefficients from the multitude of underlying coupling mechanisms. Instead, we mostly assume here the total magnitude of all coefficients to be known from other studies and only roughly argue their magnitude when needed. 

In this paper, we will model TTL coupling only up to the first order. That means we assume for this study that all higher-order effects (i.e.\ $c_{y^2} y^2, c_{z^2} z^2, c_{yz} yz, c_{\phi^2} \phi^2, ... $) are negligible. By this, we mean the following:\\
We know from fundamental theoretical TTL studies and laboratory experiments (e.g.~\cite{Hartig2022-G,Hartig2023-NG,Chwalla2020,Troebs2018,Schuster2016}) that particularly angular jitter TTL coupling is often non-linear. These observed higher-order curves are usually plotted over large angular ranges of hundreds of microradians. However, the angular and lateral jitters in LISA will be very small: in the order of a few \SI{}{nrad/\sqrt{Hz}} or \SI{}{nm/\sqrt{Hz}}. The higher-order TTL coupling curves can therefore be linearized around a certain operation point, for instance, an offset angle of a few tens of micro radians. This set point or offset angle is usually defined by the system's alignment. Therefore, higher-order TTL noise terms are not as such irrelevant but instead contribute to the linear TTL-coupling noise.\\ 
An exception might be TTL calibration manoeuvres in which larger motion is intentionally applied in order to calibrate the TTL coupling coefficients (cf. \cite{Wanner2017,Hartig2023-LPF-DA}). In such a case, second-order TTL coupling might become observable, as was the case in LISA Pathfinder. However, experience from LISA Pathfinder shows that even in that case, a linear TTL model would likely be sufficient for the coefficient fit and subsequent noise subtraction \cite{Hartig2023-LPF-DA}.

%%%%%%%%%%%%%%%%%%%%%%%%%%%%%%%%%%%%%%%%%%%%%%%%%%%%%%%%%%%%%
%%%%%%%%%%%%%%%%%%%%%%%%%%%%%%%%%%%%%%%%%%%%%%%%%%%%%%%%%%%%%
%%%%%%%%%%%%%%%%%%%%%%%%%%%%%%%%%%%%%%%%%%%%%%%%%%%%%%%%%%%%%
\section{Generic TTL $N$-terms in LISA} 
\label{Sec:generic-TTL-in-LISA}
Within this section, we define for every of LISA's interferometers a generic TTL model of the type of \Cref{Eq:gen_ttl_complete}, which we refer to as ``$N$-term'', and derive how these generic $N$-terms contribute noise to the \gls{TDI} Michelson combinations. For this, we introduce in \Cref{Sec:rough TTL model} the subsystems of LISA relevant to this paper, the laser interferometers, and the corresponding generic TTL $N$-terms.
In \Cref{Sec:PMeqs}, we show how the generic $N$-terms are added into the model of interferometric readout signals. In \Cref{Sec: TDI}, finally, we propagate the signals through \gls{TDI} and derive how the generic $N$-terms contribute noise to the second-generation Michelson observables.

%%%%%%%%%%%%%%%%%%%%%%%%%%%%%%%%%%%%%%%%%%%%%%%%%%%%%%%%%%%%%
%%%%%%%%%%%%%%%%%%%%%%%%%%%%%%%%%%%%%%%%%%%%%%%%%%%%%%%%%%%%%
%
%
\subsection{LISA's subsystems, interferometers, and generic TTL $N$-terms} 
\label{Sec:rough TTL model}
\begin{figure}
\includegraphics*[width=\columnwidth, angle = 0, trim = 0 80 0 50]{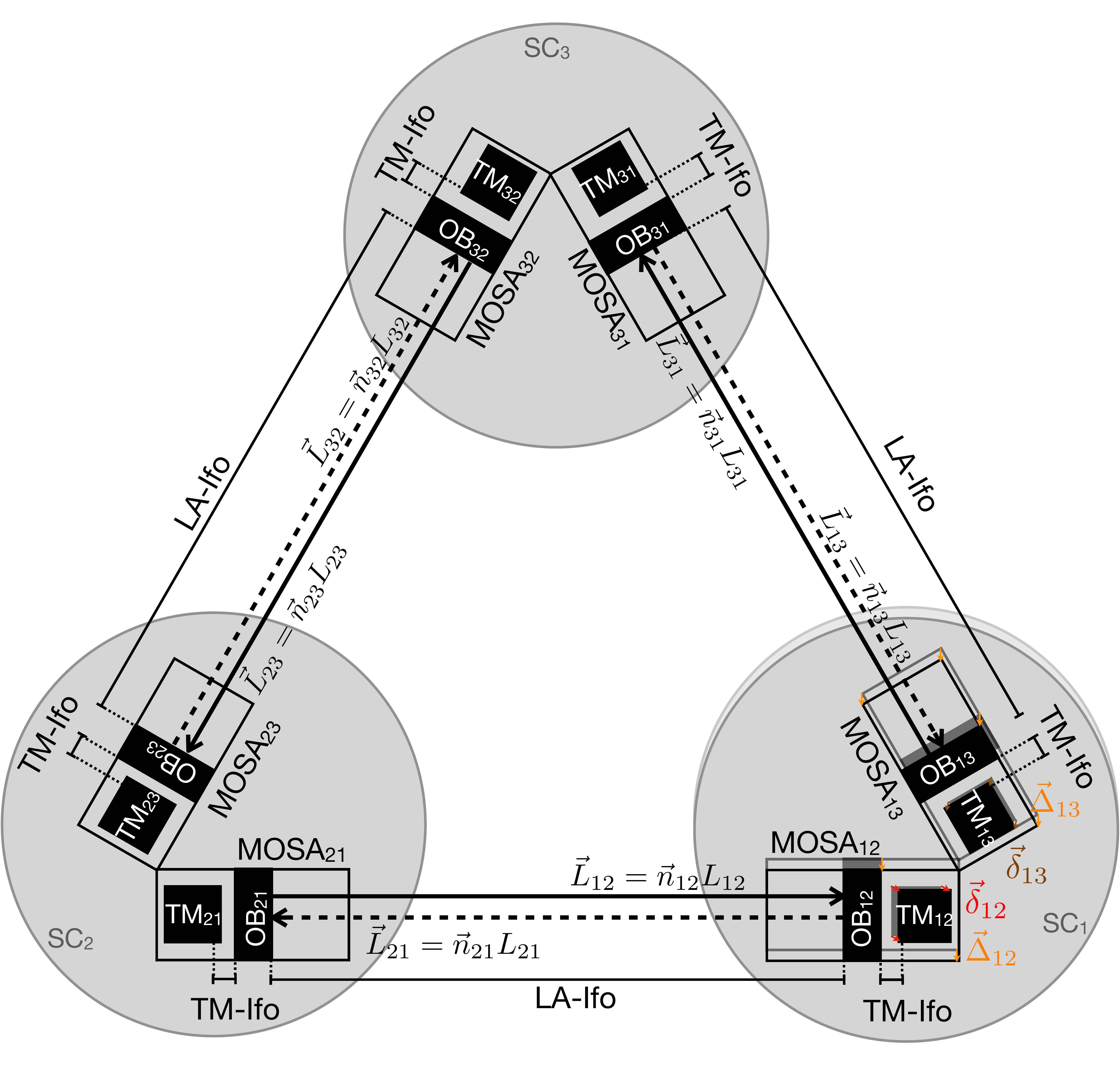}
\caption{Schematic of LISA, showing the various \acrfullpl{MOSA}, \acrfullpl{OB}, \acrfullpl{TM}, laser links and \acrfullpl{Ifo}. Reference interferometers are not shown since their TTL contributions are expected to be negligible.}
\label{Fig:lisa}
\end{figure}

A schematic of LISA is depicted in \Cref{Fig:lisa}, which shows the three spacecraft (SC) housing two \glspl{MOSA} each. 
%
%%%%%%%%%%%%%%%%%%%
%%%%%%%%%%%%%%%%%%%
%
Each \gls{MOSA} consists of a telescope (not shown in the figure), an \glsfirst{OB}, and a \gls{GRS} (again not shown in the image), inside which a freely floating \gls{TM} is located. 

We are using here the indexing convention from \cite{Hartwig2021, LISAConventionTN2021, Bayle2021} established by the LISA Consortium. The three SC, where each of the SC is labelled as $i \in \{1,2,3\}$ form a clockwise triangular constellation as seen down into the solar panels. The \gls{MOSA} (and other subsystems fixed to it) are labelled with two indices, $ij$ with $i$ representing the SC hosting the MOSA and $j$ representing the SC with which the laser beam is exchanged. We then speak of left-hand side \glspl{MOSA} if ${ij} \in \{12,23,31\}$ and right-hand side ones if ${ij} \in \{13,21,32\}$. %

LISA comprises three main interferometer types, whose signals are combined on ground to form a synthetic equal arm length interferometer. The three interferometer types are:
\begin{itemize}
    \item The \acrfullpl{TMI}, which measure distance variations between a free-floating test mass and its adjacent optical bench, with the corresponding signals $\varepsilon_{ij}$ on each SC$i$;
    \item The \glspl{LAI}, which measure distance variations between two optical benches over the large separation of about 2.5\,million\,km; with signals: $s_{ij}$ measuring changes along the arm $\vec L_{ij}$ (see \Cref{Fig:lisa});
    \item The \glspl{Ref-Ifo}, in which two local laser beams interfere aboard one spacecraft in order to measure common noise contributions; their main signals are $\tau_{ij}$.
\end{itemize}

There are 6 instances of each type of interferometer (cf. the 6 MOSAs in \Cref{Fig:lisa}).
We now assume that there are no jittering components in the reference interferometers, such that we will not consider any TTL noise in $\tau_{ij}$.
We therefore model TTL only in the 6 \glspl{TMI} $\varepsilon_{ij}$ and the 6 \glspl{LAI} $s_{ij}$.

For the \glspl{LAI}, we distinguish here two types of \gls{TTL} effects: firstly, noise originating from local effects, i.e.\ the jitter is occurring on the same SC where the resulting TTL effect is then measured. Secondly, noise originating from remote effects, i.e.\ the interferometer that is affected by the TTL noise is not part of the jittering MOSA or aboard the jittering SC. We therefore model 6 $N$-terms for the TTL noise contributions in the TMI, but 12 for the \gls{LAI}. Each of these is of the form of \Cref{Eq:gen_ttl_complete}, and we use the following syntax to distinguish the individual \gls{TTL} $N$-terms:
\begin{itemize} [leftmargin=21mm, labelsep=2mm]
	\item [$N^{\varepsilon_{ij}}_{ij}$:] \gls{TMI} \gls{TTL} effects on \gls{OB}$_{ij}$. 
	\item [$N^{s_{ij}}_{ij}$:] \gls{LAI} receiver jitter coupling, also referred to as local jitter coupling or RX-TTL coupling. Here, the upper index denotes the interferometer affected by the TTL contribution, while the lower index describes the source of the jitter. The lower index also indicates where the jitter itself is being measured, for instance via \gls{DWS} readout signals. In short: ``the lower index causes TTL, the upper index measures TTL''. Here, both indices are identical, which implies that the jitter is originating from the same \gls{SC} on which the TTL is also being measured.
	\item [$N^{s_{ij}}_{ji:ij}$:] \gls{LAI} transmitter jitter coupling, also known as remote jitter coupling or TX-TTL coupling. Again, the upper index $s_{ij}$ denotes the affected interferometer, while the lower index $ji$ describes the source of the jitter, this time on the corresponding remote MOSA, such that consequently a delay of :$ij$ along arm $L_{ij}$ needs to be considered (see \Cref{Fig:lisa}).
\end{itemize}
Here, we use a colon and two indices as short notation for a time-delay operator $\delay{ij}$. This means, we define for any arbitrary variable $A=A(t)$ the following time-delay notations:
\begin{subequations}
\begin{align}
	A_{:ij} &:=  \delay{ij} A := A(t- L_{ij} / c) \;.
	\label{Eq:Delay_Definitions}
\end{align}
\end{subequations}

Please note that we define the signals $s_{ij}$ of the \glspl{LAI} to measure distance variations along the \gls{RX-beam} direction $\vec L_{ij} = L_{ij} \vec n_{ij}$. Additionally, please note the slightly different meaning of the indices for $\vec L_{ij},  \vec n_{ij}$ and also for delays $\delay{ij}$ compared to the \glspl{MOSA}, \glspl{OB} and \glspl{TM}. For the \glspl{MOSA}, \glspl{OB} and \glspl{TM}, the first index defines aboard which SC they are located, while the second index denotes towards which SC they are pointing. However, for $\vec L_{ij}, \vec n_{ij}$ and also for delays $\delay{ij}$ that a beam experiences when propagating along $\vec L_{ij}$, the first index denotes the receiving spacecraft, and $j$ the transmitting one. This can be seen as an index inversion but has the advantage of making the indices in \gls{TDI} more harmonic, which makes indexing errors more obvious and the equations less prone to error.\\

We can now revisit the examples depicted in \Cref{Fig:TTLcoupling}. The angular jitter TTL depicted in \Cref{Fig:angTTL-TMI} (caused by the jitter of either the \gls{TM}, the local \gls{MOSA}, or the \gls{SC}) affects the TMI because the distance between the free-floating test mass and the angularly jittering MOSA varies along the axis sensed by the TMI. Additionally, the LAI is affected by the angular jitter because the \gls{MOSA}, together with the \gls{OB} and telescope (Tel), are pushed into the received beam direction. This motion of the OB simultaneously pushes the \gls{TX-beam} towards the remote spacecraft. If we now assume that \Cref{Fig:angTTL-TMI} illustrates jitter of MOSA$ij$, we can say that it illustrates angular jitter TTL contributions to $N^{\varepsilon_{ij}}_{ij}, N^{s_{ij}}_{ij}$, and $N^{s_{ji}}_{ij:ji}$.

The same argument also holds for \Cref{Fig:lateralTTL}, where \gls{SC} lateral jitter causes MOSA motion in the direction (or opposing the direction) of the indicated laser beams. Assuming that this image shows jitter of \gls{SC}$i$, we would therefore say that it illustrates lateral jitter noise contributions to $N^{\varepsilon_{ij}}_{ij},N^{s_{ij}}_{ij}$, and $N^{s_{ji}}_{ij:ji}$ for the left-hand side MOSA, and $N^{\varepsilon_{ik}}_{ik},N^{s_{ik}}_{ik}$, and $N^{s_{ki}}_{ik:ki}$ for the right-hand side MOSA.

Finally, \Cref{Fig:FF-TTL} shows how the angular jitter of the transmitting MOSA or SC affects the receiving laser interferometer. The angular jitter causes the imperfect wavefront to be scanned over the receiving spacecraft. If we again assume angular jitter of \gls{SC}$i$, then the image illustrates TTL noise contribution to either $N^{s_{ji}}_{ij:ji}$, or $N^{s_{ki}}_{ik:ki}$ .

These examples, however, are only contributions to the TTL $N$-terms, and we will not go into further detail to model the mechanisms forming the TTL coupling noise in the individual interferometers. Yet, we can have a look at the number of TTL terms we have defined by now. In total, we consider 18 general \gls{TTL} $N$-terms in the constellation-wide \glspl{TMI} and \glspl{LAI}:
\begin{subequations}\label{Eq:ttlterm} 
\begin{eqnarray}
N^{\varepsilon_{ij}}_{ij} &\mkern-12mu=&\mkern-12mu \left[ N^{\varepsilon_{12}}_{12}, N^{\varepsilon_{23}}_{23}, N^{\varepsilon_{31}}_{31}, N^{\varepsilon_{13}}_{13},N^{\varepsilon_{21}}_{21},N^{\varepsilon_{32}}_{32}  \right] \\
N^{s_{ij}}_{ij} &\mkern-12mu=&\mkern-12mu \left[ N^{s_{12}}_{12},N^{s_{23}}_{23},N^{s_{31}}_{31}, N^{s_{13}}_{13},N^{s_{21}}_{21},N^{s_{32}}_{32} \right] \\
N^{s_{ij}}_{ji:ij}&\mkern-12mu=&\mkern-12mu \left[ N^{s_{12}}_{21:12}, N^{s_{13}}_{31:13}, N^{s_{23}}_{32:23}, N^{s_{21}}_{12:21}, N^{s_{31}}_{{13}:31}, N^{s_{32}}_{23:32} \right] \;\;\;
\end{eqnarray}
\end{subequations}
Each of these comprises 5 coupling coefficients (\Cref{Eq:gen_ttl_complete}) per jittering component. These jittering components are primarily the \gls{SC} and \gls{MOSA}, but additional components like the telescopes, \glspl{TM}, and the \glspl{PAAM} could be considered as well. This results in a significant number of TTL coupling coefficients if all of these are considered, even though we model TTL here only to first order. However, such a complete model is usually not needed, because either the jitters or the coupling coefficients are considered to be small. The model can then be reduced to the most significant contributions. 

Before we reduce the model here, we show in \Cref{Sec:PMeqs} how the generic TTL $N$-terms are added into the phasemeter equations and propagated through \gls{TDI} in \Cref{Sec: TDI}. After that, we replace the generic model with an explicit one in \Cref{Sec:Derivation-of-1st-order-model}, which we then reduce to the most significant contributions in \Cref{Sec:TTL-noise-estimate}.

%%%%%%%%%%%%%%%%%%%%%%%%%%%%%%%%%%%%%%%%%%%%%%%%%%%%%%%%%%%%%
%%%%%%%%%%%%%%%%%%%%%%%%%%%%%%%%%%%%%%%%%%%%%%%%%%%%%%%%%%%%%
\subsection{Phasemeter equations with generic TTL}
\label{Sec:PMeqs}
We can now add the generic TTL $N$-terms into the models of the interferometric phase signals usually used in \gls{TDI}. These models are often referred to as phasemeter equations and are more detailed than our initial model \Cref{Eq:intro_N}. 
For easy comparison with previous publications, we use the notation of \cite{Otto2015} except that we do not distinguish between delays for constant or varying arm lengths and use ``:'' in either case. Additionally, we adapted to the more recent double-index notation. Consequently, we define the phasemeter equations in units of phase radian. Therefore, the TTL $N$-terms need to be converted from their units of meters to phase radian by multiplying with the appropriate wavenumber $k$, when adding these terms in. We find:
\begin{widetext}
\begin{subequations} \label{Eq:PM-equations}
\begin{align}
\check s_{12}(t) 	=&  H_{12} + p_{21:12} - p_{12}  + k_{21:12} \left( \vec n_{21} \cdot \vec \Delta_{21:12} + \vec n_{12} \cdot \vec \Delta_{12}  + N^{s_{12}}_{12}  + N^{s_{12}}_{21:12} \right) 
 \label{Eq:PM-LAI12} 
 \\
\check \varepsilon_{12}(t)  	=&  p_{12}  - p_{13} - \mu_{13}  + k_{12} \left( - 2 \vec n_{12} \cdot \vec \Delta_{12} + 2 \vec n_{12} \cdot \vec \delta_{12}  + N^{\varepsilon_{12}}_{12} \right)
\label{Eq:PM-TMI12}
\\
\check \tau_{12}(t)			=& p_{12} - p_{13}  - \mu_{13}
\\
\check s_{13}(t) 			= & H_{13} + p_{31:13} - p_{13}  + k_{31:13} \left( \vec n_{31} \cdot \vec \Delta_{31:13} +\vec n_{13} \cdot \vec \Delta_{13} + N^{s_{13}}_{13} +  N^{s_{13}}_{31:13}  \right)
\\ 
\check \varepsilon_{13}(t)	=&p_{13} - p_{12} - \mu_{12} + k_{13} \left(- 2 \vec n_{13}  \cdot \vec \Delta_{13} + 2 \vec n_{13} \cdot \vec \delta_{13}  + N^{\varepsilon_{13}}_{13} \right) \label{Eq:PM-TMI13} 
\\
\check \tau_{13}(t) 	=&p_{13} - p_{12} - \mu_{12}  
\;.
\end{align}
\end{subequations}
\end{widetext}
Here, we denote gravitational wave signals by $H$, laser frequency noise by $p$, fiber backlink noise by $\mu$, all defined in phase radian. Contrary to this, \gls{MOSA} (or equivalently \gls{OB}) displacement $\vec \Delta$, and test mass motion $\vec \delta$ both defined relative to free space and mapped along the beam direction $\vec n$, are defined in units of meters. These terms ($\vec n \vec \Delta$, and $\vec n \vec \delta$) are explicit substitutions of the longitudinal displacement $x$ in \Cref{Eq:intro_N}. \\
Please note that the projection directions $\vec n$ might not be intuitive. They are chosen such that the received and transmit beam directions may be different, and yet the cancellation of OB-motion between \gls{LAI} and \gls{TMI} are assured. Further information, and a more precise description allowing deviations of the beam directions in the \gls{LAI} and \gls{TMI} are given in \Cref{Sec:Initial_Phaseter_eqns}. 

With \Cref{Eq:PM-equations}, we have defined calibrated signals, i.e.\ we have divided each signal by a specific signum function to achieve a fixed sign convention that is independent of the sign of the heterodyne beatnote frequency. This calibration and suppression of the signum function is indicated by the check symbol (i.e.\ $\check s, \check \varepsilon$, rather than $s, \varepsilon$).
A detailed derivation of \Cref{Eq:PM-equations} including the suppression of signum functions and all defined signs is given in \Cref{Sec:Initial_Phaseter_eqns}. Please note that whether the $N$-terms are added into the phasemeter equations, or placed with an explicit minus sign, is a choice. We choose all TTL effects to be added in and care for intrinsic signs only once the generic $N$-terms are replaced by explicit models (see particularly \Cref{Sec:path_ttl_model,Sec:Extended TTL model}).

The phasemeter equations of all other MOSAs can be obtained by the usual cyclic permutation of the individual indices, i.e.\ $1\rightarrow 2; 2\rightarrow 3; 3\rightarrow 1$.

%%%%%%%%%%%%%%%%%%%%%%%%%%%%%%%%%%%%%%%%%%%%%%%%%%%%%%%%%%%%%%%%
\subsection{Propagating the generic TTL terms through TDI} 
\label{Sec: TDI}
To estimate the magnitude of TTL noise in the \gls{TDI} observables, we need to propagate the various TTL contributions through \gls{TDI}. We do this here on the example of the second-generation Michelson combinations $X_2$, from which the $Y_2$ and $Z_2$ combinations can be derived as usual via cyclic permutation of the indices \cite{Otto2015,Bayle2021}. 
When deriving the TTL contributions in TDI, we neglect clock noise as well as the clock noise suppression step. This is based on the assumption that the clock noise suppression step does not affect the TTL contributions to the TDI observables.

Following the process described in \cite{Otto2015}, but adapted for the signal calibration defined in \Cref{Eq:Cal_displacement_readout}, we first construct the intermediate variable $\check \xi_{ij}^\text{rad}$ in units of phase radian:
\begin{align}
	\check \xi_{ij}^\text{[rad]} = \check s_{ij} +  \frac{k_{ji:ij}}{k_{ij}}  \frac{\check \varepsilon_{ij} - \check \tau_{ij}}{2} + \delay{ij}
	\frac{\check \varepsilon_{ji} - \check \tau_{ji}}{2} 
	\label{Eq:xi} \;,
\end{align}
which is free of noise that is caused by the \gls{OB} longitudinal jitter $\vec \Delta$. This intermediate variable describes the various single-link readouts and their noise. 
Denoting only the TTL terms in $\xi_{ij}$ either in units of phase radian or meters, we find:
\begin{subequations}
\begin{align}
   \check \xi_{ij}^\text{TTL,[rad]} &=  k_{ji:ij} \left( N^{s_{ij}}_{ij}  + N^{s_{ij}}_{ji:ij}  
   +\frac{1}{2} N^{\varepsilon_{ij}}_{ij}  + \frac{1}{2} N^{\varepsilon_{ji}}_{ji:ij} 
\right)  \\
   \check \xi_{ij}^\text{TTL} &= N^{s_{ij}}_{ij}  + N^{s_{ij}}_{ji:ij}  
   +\frac{1}{2} N^{\varepsilon_{ij}}_{ij}  + \frac{1}{2} N^{\varepsilon_{ji}}_{ji:ij} \;.
\end{align}
\label{Eq:xi_TTL}% 
\end{subequations}

In the next step, all laser frequency noises $p$ of lasers belonging to right-hand side \glspl{MOSA} are removed by forming the $\check\eta$ variables (not to be confused with pitch angles $\eta_{ij}, \eta_{ik}$). This makes the definitions for the variable $\check\eta_{ij}$ from left-hand side \glspl{MOSA} different from corresponding variable $\check \eta_{ik}$ from right-hand side \glspl{MOSA}:
\begin{subequations}
\begin{align}
	 \check \eta_{ij}^\text{[rad]}& = \check \xi_{ij}^\text{[rad]} + \delay{ij} \frac{\check \tau_{jk}-\check \tau_{ji}}{2}  \\
	 \check \eta_{ik}^\text{[rad]} &= \check \xi_{ik}^\text{[rad]} + \frac{\check \tau_{ik}-\check \tau_{ij}}{2}  \;.
\end{align}
\label{Eq:hat_eta}%
\end{subequations}%
Since we assume there is no \gls{TTL} coupling in the reference interferometers, we recover the exact same \gls{TTL} terms in the $\check\eta$ variables as in the $\check \xi$ variables:
\begin{equation}
    \check\eta_{ij}^\ttl = \check \xi_{ij}^\text{TTL}  \;.
    \label{Eq:eta_ttl} 
\end{equation}
Please note that the signs in \Cref{Eq:hat_eta} differ from \cite{Otto2015} as a consequence of the notation introduced in \Cref{Sec:Initial_Phaseter_eqns} and the chosen calibration which suppresses the signum functions. However, the equation here and in \cite{Otto2015} agree, if the calibration is considered. Furthermore, this discrepancy exists only in the calibrated intermediate variables $\check \eta$, while the signs in the final step of generating the second-generation Michelson $X$-combination, $X_2$, are unaffected \cite{Otto2015}. Therefore, $X_2$, in which the residual laser frequency noise contributions are suppressed, has the same form as usual  (e.g.~\cite{Otto2015, Bayle2021,Tinto2020})
\begin{align}
    X_2^\text{[rad]} = \hspace{-8mm} &\nonumber \\ 
   & (1 - \delay{121} - \delay{12131} + \delay{1312121}) (\check\eta^\text{[rad]}_{13} + \delay{13} \check\eta^\text{[rad]}_{31}) \nonumber\\
      -&(1 - \delay{131} - \delay{13121} + \delay{1213131}) (\check\eta^\text{[rad]}_{12} + \delay{12} \check \eta^\text{[rad]}_{21})  \;.
\end{align}
The TTL noise contributions are simply found by substituting $\check \eta \rightarrow \check \eta^\text{TTL} = \check \xi_{ij}^\text{TTL}$, and are in units of meters given by:
\begin{align}
    X_2^\text{TTL} = \hspace{-8mm}  \nonumber\\
	& (1 - \delay{121} - \delay{12131} + \delay{1312121}) (\check\xi_{13}^\text{TTL} + \delay{13} \check \xi_{31}^\text{TTL}) \nonumber\\
      -&(1 - \delay{131} - \delay{13121} + \delay{1213131}) (\check\xi_{12}^\text{TTL} + \delay{12} \check\xi_{21}^\text{TTL})
          \label{Eq:X2} \;.
\end{align}
%
%%%%%%%%%%%%%%%%%%%%%%%%%%%%%%%%%%%%%%%%%%%%%%%%%%%%%%%%%%%%%
We use here another short notation contracting a sequence of delay operators defined by:
\begin{subequations}
\begin{align}
	\delay{ijk} &:= \delay{ij}\delay{jk} \\
	\delay{ijkl} &:= \delay{ij}\delay{jk}\delay{kl} \;,
\end{align}
\end{subequations}
with $i,j,k,l \in\{1,2,3\}$ as previously introduced in \cite{Bayle2021}. This contraction definition can be extended to define an arbitrary number of lower indices, as long as the shown pairing of indices occurs. In the \gls{TDI} Michelson combinations, this is always the case if the index syntax used in this paper is applied: for all variables, the first index describes the transmitting SC, while the second index names the receiving SC, except for arm lengths  $L_{ij}$ and delays  $\delay{ij}$, where this is chosen vice versa (cf. \Cref{Fig:lisa}). This ensures the pairing of indices in the Michelson combinations.

As shown in \cite{Nam2023}, the \gls{PSD} of $X_2^\text{TTL}$ defined by \Cref{Eq:X2} can be directly denoted as
\begin{align}
	\text{PSD}(X_2)(f) & =: S_{X_2}(f) 	\nonumber\\
 		&= C_{XX}^{123}  \cdot \text{PSD}\left( \check\xi_{13}^\text{TTL} + \delay{13} \check \xi_{31}^\text{TTL}\right)
		\nonumber\\
      & + C_{XX}^{132} \cdot \text{PSD}\left(\check\xi_{12}^\text{TTL} + \delay{12} \check\xi_{21}^\text{TTL}\right)
\label{Eq:PSD-X2_generell}
\end{align}
using the transfer function
\begin{align}
	C_{XX}^{ijk} = 16 \sin^2\left(\frac{2\pi f \bar{L}_{ij}}{c}\right) \sin^2\left(\frac{4\pi f \bar{L}_{ijk}}{c}\right)
		\label{Eq:CXXijk}	
\end{align}
and mean arm lengths
\begin{subequations}
\begin{align}
	\bar L_{ij} &= \frac{L_{ij}+ L_{ji} }{2} \\
	\bar L_{ijk} &= \frac{L_{ij}+ L_{ji} + L_{ik}+ L_{ki}}{4} \;.
	\label{Eq:mean-L}	
\end{align}
\end{subequations}

Please note, the prefactors of $1/2$ for the \gls{TMI} TTL terms in \Cref{Eq:xi_TTL} are an artifact from the partial calibration we have defined for these interferometers in order to allow easy comparison with previous publications such as \cite{Otto2015,Tinto2020,Houba2022-Estimation}. By partial calibration, we mean the following: while we suppressed the signum functions by calibration, we have not calibrated the TMI readout signal $\varepsilon$ to read out the longitudinal displacement of the optical bench relative to the test mass with a prefactor of 1, despite that this measurement is the primary goal of each TMI. In other words, $\vec n  (\vec\Delta - \vec \delta) $ corresponds to $x$ in \Cref{Eq:intro_N}. If we assume $S$ in \Cref{Eq:intro_N} to be in units of meters, we see a prefactor of $c_x = -2$ in the phasemeter equations \Cref{Eq:PM-TMI12,Eq:PM-TMI13}. The minus sign is of little or no relevance here. It depends on whether we wish to measure OB motion relative to the test mass, or test mass motion relative to the \gls{OB} (or spacecraft), and it defines whether the TMI and LAI signals need to be added or subtracted in \gls{TDI} to cancel the \gls{OB} motion $\vec \Delta$. But the factor of 2, which is of course present in the raw data, could be suppressed by calibration, making $c_x = \pm 1$ thereafter, which would remove the factors of $1/2$ from \Cref{Eq:xi,Eq:xi_TTL,Eq:hat_eta} and all equations derived from these throughout this paper.   

%%%%%%%%%%%%%%%%%%%%%%%%%%%%%%%%%%%%%%%%%%%%%%%%%%%%%%%%%%%%%
%%%%%%%%%%%%%%%%%%%%%%%%%%%%%%%%%%%%%%%%%%%%%%%%%%%%%%%%%%%%%
%%%%%%%%%%%%%%%%%%%%%%%%%%%%%%%%%%%%%%%%%%%%%%%%%%%%%%%%%%%%%

\section{Derivation of an explicit first-order TTL model}\label{Sec:From-Frames-to-Explicit-TTL-Model}
\label{Sec:Derivation-of-1st-order-model}
So far, we have only given a very generic description of how each noise term is modeled (\Cref{Eq:gen_ttl_complete}) and named the various TTL noise contributions (\Cref{Eq:ttlterm}). Each of these generic $N$ terms can be written explicitly as a function of the individual involved jitters.
In order to do so, we first discuss in \Cref{Sec: Jitters+coord} what jittering components we consider throughout this paper to cause TTL coupling and define the relevant coordinate frames. We then describe the necessary mapping for \gls{SC} jitter into the \gls{MOSA} frame in \Cref{Sec:Mapping}. This enables the explicit modeling which is given in \Cref{Sec:path_ttl_model}.

%%%%%%%%%%%%%%%%%%%%%%%%%%%%%%%%%%%%%%%%%%%%%%%%%%%%%%%%%%%%%
%%%%%%%%%%%%%%%%%%%%%%%%%%%%%%%%%%%%%%%%%%%%%%%%%%%%%%%%%%%%%
\subsection{Jitters and coordinate frames} \label{Sec: Jitters+coord}

Within this paper, we assume only three types of jittering objects: the \gls{SC}, \glspl{MOSA}, and \glspl{TM}. There are also other jitters of relevance, such as the jitter of the \glspl{PAAM} and, particularly, the jitter of the telescopes relative to their OB. However, these are not considered here and can be the subject of future publications.

If we want to describe how the jitter of a test mass or MOSA affects the interferometry, it is best to do so in a coordinate system aligned with the primary beam axis. We call such a reference frame a \gls{MF}. 
There are two MOSAs aboard each SC, and hence two MOSA coordinate frames. However, there is only one SC, and when we want to describe its motion, it is easiest to describe the motion in a dedicated \gls{SF}, before mapping it into either of the MOSA frames.

\begin{figure}
\includegraphics*[width=\columnwidth]{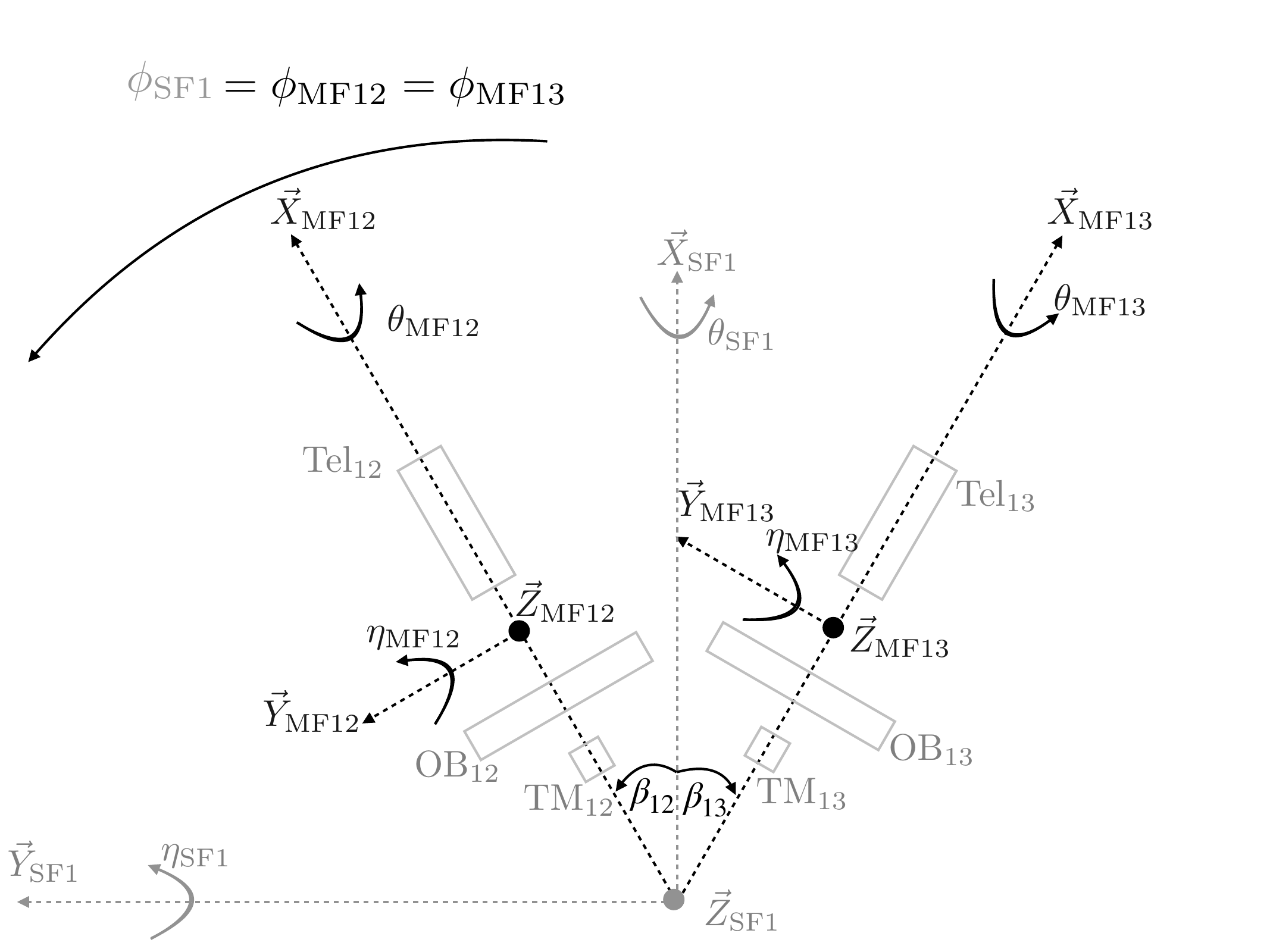}
\caption{Illustration of LISA reference frames on the example of SC1. The \glsentrylong{SF} is indicated in grey dashed lines, while the \glsentrylong{MF}s are shown in dashed black lines. Each \glspl{MOSA} is displayed via its elements: a \acrfull{TM}, \acrfull{OB} and a telescope (Tel). The opening angles $\beta_{12} \approx \SI{30}{\degree}, \beta_{13}\approx \SI{-30}{\degree}$ are defined as angles between the SF and MF $x$-axes.}
\label{Fig:CooFrames}
\end{figure}

We have depicted the two types of coordinate frames in \Cref{Fig:CooFrames} for the example of \gls{SC}1 and define them in detail in the following.
%
%%%%%%%%
%%%%%%%%

The origins of both coordinate frames, the \gls{SF} and \gls{MF} move along a hypothetical perfectly noise-free orbit and thus in synchrony with our definition of free space. They are, therefore, inertial reference frames on short time scales and not body-fixed, which allows observing SC jitter in either of the two frames while the nominal orbital motion of the SC is not accessible. 

The $x$-axis $\vec X_{\text{SF}i}$ of \gls{SC}$i$ frame is the bisector between the local MOSA $x$-axes in their nominal state, where nominal is considered to be a full opening angle of \SI{60}{\degree}. The \gls{SF} $xy$-plane is the plane in which both local \glspl{MOSA} are located, the \gls{SF} $z$-axis $\vec Z_{\text{SF}i}$ is pointing from the \gls{CoM} through the solar panel. 
We use this reference frame to describe \gls{SC} motion relative to free space. 

The axes of the \gls{MOSA} frame are rotated in $\phi$ by approximately \SI{\pm 30}{\degree} against the axes of the SF: 
The $x$-axes $\vec X_{\text{MF}ij}, \vec X_{\text{MF}ik}$ of the \glspl{MF} are defined via the telescope axes, which are pointing towards the incident wavefront from the far \gls{SC}. The  \gls{MF} $x$-axis can, therefore, be nominally aligned to the \glsentryshort{RX-beam} axis (however, with opposing direction). Contrary to this, the \gls{TX-beam} is tilted against $\vec X_{\text{MF}ij}$ with an angle defined by the \gls{PAAM}. The $z$-axis $\vec Z_{\text{MF}ij}$ is chosen to be identical to the \gls{SC} frame $z$-axis $\vec Z_{\text{SF}i}$. 

In principle, we should indicate for every motion the used coordinate frame. In order to reduce the notation, we mostly suppress the reference frame information by defining that any jitter described in its natural coordinate frame, carries no indicator, i.e.\ test mass and \gls{MOSA} motion in \gls{MF} and spacecraft jitter in SF do not carry an indicator. Since we use \gls{MF} as our primary reference frame, only spacecraft jitter expressed in the MOSA frame carries an explicit upper index MF to indicate the mapping.

Now, we can specify in detail the jitters affecting the interferometric signals:
For the LAI, the jitter of relevance is MOSA jitter with respect to the \gls{RX-beam}: $\alpha_\text{MO/RX}$ (given in MF).
We consider the RX wavefront to be effectively planar due to the long propagation distance. Angular transmitter jitter causes no considerable beam tilt in the far field and so the \gls{RX-beam} axis is considered to be static at the receiving SC.
Consequently, the jitter affecting the LAI is equivalent to the local MOSA jitter wrt.\ free space: 
$\alpha_\text{MO/RX} =  \alpha_\text{MO/FS}$.
Here, we have defined $\alpha_\text{MO/RX}$ to denote the relative motion between two geometric components: the \gls{MOSA}, and the line describing the RX beam axis, described in MF. With the notation $\alpha_\text{MO/FS}$, we treat FS as if it was a geometric body. We do so because it is not important which point in free space is chosen as a reference for the MOSA motion, provided the point rests in our definition of FS. 

Since each MOSA is mounted to its \gls{SC}, it jitters mostly with the spacecraft. Yet, these mounts allow residual individual jitter of the MOSAs relative to their spacecraft. Therefore, for every degree of freedom, the total MOSA jitter wrt.\ free space $\alpha_\text{MO/FS}$ can be split into the motion of the \gls{SC} wrt.\ \gls{FS} mapped along the MF axes ($\alpha_\text{SC}^\text{MF}$) plus the motion $\alpha_\text{MO}$ of the MOSA wrt.\ the SC. The jitter affecting the LAI is, therefore, given by:
\begin{equation}
	\alpha_\text{MO/RX} = \alpha_\text{MO/FS} =  \alpha_\text{MO/SC} + \alpha_\text{SC/FS}^\text{MF}\;.
	\label{Eq:alpha-MO/RX-long}
\end{equation}
Splitting up the motion in this way has the advantage that correlated and uncorrelated contributions to TTL noise are separated. SC motion affects both \glspl{LAI} aboard the SC and causes, therefore, correlated TTL noise. Contrary, MOSA jitter relative to the SC affects only the MOSA's own LAI.

The \glspl{TMI} are affected by the jitter of the \glspl{MOSA} relative to their test masses. This motion is equivalent to the difference in the motion of the MOSA wrt.\ FS and the test mass motion wrt.\ FS. We can split this jitter up like before:
\begin{subequations}
\begin{align}
	\alpha_\text{MO/TM} &= \alpha_\text{MO/FS}- \alpha_\text{TM/FS}\\
	&= \alpha_\text{MO/SC} + \alpha_\text{SC/FS}^\text{MF} - \alpha_\text{TM/FS}\;. 
\end{align}
	\label{Eq:alpha-MO/TM-long}
\end{subequations}
%

%\Structure{Shorten notation}
We now define default reference bodies. Within this paper, the reference for MOSA jitter is by default the SC, while the reference for TM and SC jitter is by default FS. If no reference is given, the default references apply. This allows us to shorten phrasings and reduce the indices by suppressing the reference information for default cases, i.e.\ MO/SC $\rightarrow$ MO, SC/FS $\rightarrow$ SC, TM/FS $\rightarrow$ TM and can shorten the notation of \Cref{Eq:alpha-MO/RX-long,Eq:alpha-MO/TM-long}:
\begin{subequations}
\begin{align}
	\alpha_\text{MO/RX} &= \alpha_\text{MO/FS} = \alpha_\text{MO} +  \alpha_\text{SC}^\text{MF} \\
	\alpha_\text{MO/TM} &= \alpha_\text{MO/FS}- \alpha_\text{TM}\\
	&= \alpha_\text{MO} + \alpha_\text{SC}^\text{MF} - \alpha_\text{TM}\;. 
\end{align}
	\label{Eq:alpha-MO/TM+alpha-MO/RX-short}
\end{subequations}

In the next step, we need to define the mapping of SC-jitter into \gls{MF}, i.e.\ $\alpha_\text{SC}^\text{MF}$ as a function of all degrees of freedom $\alpha_\text{SC}$ in SF.

%%%%%%%%%%%%%%%%%%%%%%%%%%%%%%%%%%%%%%%%%%%%%%%%%%%%%%%%%%%%%
\subsection{Mapping between SC and MOSA coordinate frames}
\label{Sec:Mapping}
%%%%%%%%%%%%%%%%%%%%%%%%%%%%%%%%%%%%%%%%%%%%%%%%%%%%%%%%%%%%%
The MFs are our main coordinate frames. Therefore, we describe their axes by:
\begin{subequations}
\begin{align}
	\vec X_\text{MF} = (1,0,0)^T \\
	\vec Y_\text{MF} = (0,1,0)^T \\
	\vec Z_\text{MF} = (0,0,1)^T
\end{align}
\end{subequations}
where the upper index $T$ indicates a transpose. Please note: as always in this paper, we use the lower index to describe the origin or cause of the variable. So one can ask for the lower index of a variable by `of what?'. In this case, we describe the axes *of* the MOSA frame. Contrary to that, the upper index describes in which coordinate frame we have expressed the vector or variable. As usual, we suppress the upper index MF because it is the default reference frame.

The directions of the axes of the \gls{SF} (expressed in \gls{MF}) are then found by rotation (cf. \Cref{Fig:CooFrames}):
\begin{subequations}
\label{Eq:SF_expressed_in_MF}
\begin{align}
	\vec X_\text{SF} &= R(-\beta, \vec Z_\text{MF}) \vec X_\text{MF} = (\cos(\beta),- \sin(\beta),0)^T \\
	\vec Y_\text{SF} &= R(-\beta, \vec Z_\text{MF}) \vec Y_\text{MF} = (\sin(\beta),\cos(\beta),0)^T \\
	\vec Z_\text{SF} &= R(-\beta, \vec Z_\text{MF}) \vec Z_\text{MF}  = \vec Z_\text{MF} \;,
\end{align}
\end{subequations}
where $R(\vec Z_\text{MF},-\beta) $ denotes a $3\times3$ rotation matrix describing a tilt around the axis $\vec Z_\text{MF}$ by an angle $-\beta$. 
SC translational jitter is then described by $
	 x_\text{SC} \vec X_\text{SF} +	 y_\text{SC} \vec Y_\text{SF} + 	 z_\text{SC} \vec Z_\text{SF} 
$
which will show in MF by:
\begin{subequations}
\begin{align}
	x_\text{SC}^\text{MF} &=  \cos(\beta) x_\text{SC} + \sin(\beta) y_\text{SC}   \label{Eq:x_SC_mapping_simple} \\
	y_\text{SC}^\text{MF} &= - \sin(\beta) x_\text{SC} + \cos(\beta) y_\text{SC}  \label{Eq:y_SC_mapping_simple}  \\
	z_\text{SC}^\text{MF} &= z_\text{SC}  \label{Eq:z_SC_mapping_simple}\;.
\end{align}
\label{Eq:SC_displ_jitter_mapping_simple}%
\end{subequations}

As shown in \Cref{Fig:CooFrames}, this tilt angle $\beta$ is about $\ang{30}$ if a left-hand side MOSA is considered (case $ij$), and approximately  $\ang{-30}$ if a right-hand side MOSA is considered (case $ik$). The total opening angle of $\ang{60}$ may vary up to $\pm\ang{1}$ during the mission \cite{Martens2021}. 

In the next step, we now derive the mapping of the angular SC-jitter into the MFs to see how SC jitter will appear in the MOSAs. 
For this, we assume that the SC will jitter angularly in all three angles: roll ($\theta_\text{SC}$), pitch ($\eta_\text{SC}$), and yaw ($\phi_\text{SC}$). 
All jitters will have a specific but currently unknown spectral density with a magnitude in the \SI{}{nrad/\sqrt{Hz}} regime. This amplitude is sufficiently small to allow linearization, which is an essential assumption for our mathematical description. We would generally need to consider that rotations do not commute. However, since all considered angular jitters are small, we can successively apply all three transformations and then linearize. Due to the linearization, the matrices do commute, and the sequence of rotation transformations becomes irrelevant. Consequently, the effect of SC jitter in all three angular degrees of freedom will affect an arbitrary vector $\vec x$ in MF by
\begin{align}
	R(\theta_\text{SC}, \vec X_\text{SF}) \, R(\eta_\text{SC}, \vec Y_\text{SF}) \, R(\phi_\text{SC}, \vec Z_\text{SF})  \, \vec x  
\label{Eq:complete_Rotation}
\end{align}
with 
\begin{align}
R&(\theta_\text{SC}, \vec X_\text{SF}) \, R(\eta_\text{SC}, \vec Y_\text{SF}) \, R(\phi_\text{SC}, \vec Z_\text{SF}) = \nonumber \\
	& 	\begin{pmatrix} 
	 1 	& -\phi_\text{SC} & c_\beta \eta_\text{SC}  - s_\beta \theta_\text{SC}  \\
	 \phi_\text{SC} & 1 & -(  s_\beta \eta_\text{SC} + c_\beta \theta_\text{SC}) \\
	-(c_\beta \eta_\text{SC}  - s_\beta \theta_\text{SC} )& c_\beta \theta_\text{SC} + s_\beta \eta_\text{SC} & 1 
	\end{pmatrix} 
	\label{Eq:SCRotMatrWOdispl}
\end{align}
using the short notation $c_\beta,\, s_\beta$ for $\cos(\beta),\,\sin(\beta)$, respectively.
This complete rotation matrix can be directly compared with the analogon for MOSA angular jitter around the origin of the MF:
\begin{align}
	R(\theta_\text{MO}, \vec X_\text{MF}) \, R(\eta_\text{MO}, \vec Y_\text{MF}) \,  R(\phi_\text{MO}, \vec Z_\text{MF}) = \nonumber\\
	\begin{pmatrix} 
	 1 	& -\phi_\text{MO} & \eta_\text{MO} \\
	 \phi_\text{MO} & 1 & -\theta_\text{MO} \\
	 -\eta_\text{MO} & \theta_\text{MO} & 1 
	\end{pmatrix}  \;.
\end{align}
This shows that SC angular jitter will be observable as angular jitter in the MOSA frames with the following mapping:
\begin{subequations}
\begin{align}
	\phi^\text{MF}_\text{SC}&= \phi_\text{SC}  	\\
	\eta^\text{MF}_\text{SC} &= \cos(\beta) \eta_\text{SC}  -  \sin(\beta) \theta_\text{SC}    \label{Eq:angular_mapping_eta-MF}  \\
	\theta^\text{MF}_\text{SC}  &= \sin(\beta) \eta_\text{SC} +  \cos(\beta) \theta_\text{SC}  \;.
\end{align}
\label{Eq:angular_mapping}%
\end{subequations}
The total angular jitter of the MOSA relative to \gls{FS} is therefore given by:
\begin{subequations}
\begin{align}
	\phi_\text{MO/FS} &= \phi_\text{MO} + \phi_\text{SC}  	\\
	\eta_\text{MO/FS} &= \eta_\text{MO} + \cos(\beta) \eta_\text{SC}  -  \sin(\beta) \theta_\text{SC}    \label{Eq:angular_mapping_eta}  \\
	\theta_\text{MO/FS} &= \theta_\text{MO} +\sin(\beta) \eta_\text{SC} +  \cos(\beta) \theta_\text{SC}  \;.
\end{align}
\label{Eq:angular-MOSA/FS-jitter}%
\end{subequations}
%

%%%%%%%%%%%%%%%%%%%%%%%%%%%%%%%%%%%%%%%%%%%%%%%%%%%%%%%%%%%%%
%%%%%%%%%%%%%%%%%%%%%%%%%%%%%%%%%%%%%%%%%%%%%%%%%%%%%%%%%%%%%
\subsection{TTL model as functions of component jitter}
\label{Sec:path_ttl_model}
With the information from the previous sections, we can now replace the generic TTL model from \Cref{Eq:gen_ttl_complete} with an explicit one:
\begin{subequations}
\begin{alignat}{2}
N^{\varepsilon_{ij}}_{ij} &= &&\sum_\alpha 
 c^{\varepsilon_{ij}}_{\alpha_{ij}}  \left(\alpha_{\text{MO}ij} + \alpha_{\text{SC}i}^{\text{MF}ij} - \alpha_{\text{TM}ij} \right)
 \label{Eq:N_{ij}^{varepsilon_{ij}}_simple}  \\
 N_{ij}^{s_{ij}} &=    &&\sum_\alpha   
 c^{s_{ij}}_{\alpha_{ij}}  \left(\alpha_{\text{MO}ij} + \alpha_{\text{SC}i}^{\text{MF}ij} \right)  
\label{Eq:N_{ij}^{s_{ij}}_simple} \\
N_{ji:ij}^{s_{ij}}  &=   &&\sum_\alpha 
c^{s_{ij}}_{\alpha_{ji}} \left( \alpha_{\text{MO}ji:ij} + \alpha^{\text{MF}_{ji}} _{\text{SC}j:ij}   \right)\;.
\label{Eq:N_{ji:ij}^{s_{ij}}_simple} 
\end{alignat}
\label{Eq:Complete_model_not_subst_simple}%
\end{subequations}
This model was previously published in \cite{Houba2022-Estimation,Paczkowski2022} for $\alpha \in \eta,\theta,\phi$ and the \gls{LAI}, and is extended here to include the \gls{TMI} TTL and TM jitter.

This model is strictly valid only if applied to estimate TTL in a single link or in \gls{TDI}, or if used to fit TTL noise in \gls{TDI} variables. 
It is based on several assumptions, out of which the first two are not valid for the individual interferometers. These assumptions are described in the following.
\paragraph*{1. One coefficient per total jitter \gls{DOF}:}
Each total jitter degree of freedom (i.e.\ $\alpha_\text{MO/RX}, \alpha_\text{MO/TM}$ for the LAI and TMI, respectively), is scaled with one coupling coefficient, rather than an individual coefficient per contributing jitter degree of freedom ($\alpha_{\text{MO}ij}, \alpha_{\text{SC}i}^{\text{MF}ij}, \alpha_{\text{TM}ij}$). This is a major assumption, which should only be made if the noise in a single link or in \gls{TDI} is being modeled. This will be discussed further in \Cref{Sec:Implicit-Assumptions+Implications}.

\paragraph*{2. Lateral jitter coupling is not modeled:}
The model is usually applied only to angular jitter coupling, with the assumption that lateral jitter coupling is negligible. We follow this assumption for the moment and discuss it further in \Cref{Sec:Implicit-Assumptions+Implications}, where we extend the model to include lateral jitter coupling explicitly.
Please note that the model above (\Cref{Eq:Complete_model_not_subst_simple}), could of course be interpreted for lateral jitter coupling by summing $\alpha$ additionally over $y,z$. However, to evaluate the resulting model, the mapping of SC lateral jitter into MF is needed, which we have not defined yet and postpone to  \Cref{Sec:Implicit-Assumptions+Implications}.

\paragraph*{3. Coefficients are not delayed:}
In \Cref{Eq:N_{ji:ij}^{s_{ij}}_simple} one could indicate a delay in the coupling coefficient $c^{s_{ij}}_{\alpha_{ji}}$. However, we have not done so and we generally do not delay coupling coefficients throughout this paper. This means we assume that the coefficients are constant over the time period in which the \gls{TDI} observable is formed. It can be seen in \Cref{Eq:X2} that a maximum of 7 delays are applied to an individual term. Given that each delay represents a time period of about \SI{8.3}{s}, this means that we assume that the coefficients are constant over a period of about one minute. 

\paragraph*{4. Different interferometers have different coupling factors:}
Additionally, the coupling of one specific degree of freedom of one component but in different interferometers is considered with different coefficients here. For instance, assume the test mass to be in perfect rest, i.e.\ $\alpha_{\text{TM}ij}=0$. In that case, the TMI and LAI are subject to the very same jitter $\alpha_{\text{MO}ij} + \alpha_{\text{SC}i}^{\text{MF}ij}$, and yet the interferometers have different coupling coefficients: $ c^{\varepsilon_{ij}}_{\alpha_{ij}}, c^{s_{ij}}_{\alpha_{ij}}$. 
We assume this because coupling coefficients originate from the precise alignment in an interferometer and from properties of the interfering wavefronts. By assuming different coupling coefficients, we account for different alignments as well as different wavefront properties in the different interferometers. 
Likewise, the component jitter coupling $c^{s_{ji}}_{\phi_{ji}}$ into the LAI of the local transmitting SC originates from different mechanisms than the coupling of the very same jitter into the receiving LAI $c^{s_{ij}}_{\phi_{ji:ij}}$ and is therefore considered with different coefficients.

Please note that there is an implicit swap for the $\eta_{\text{SC}}$ jitter contributions in  \Cref{Eq:N_{ij}^{s_{ij}}_simple} in comparison to \Cref{Eq:N_{ji:ij}^{s_{ij}}_simple}. 
This is caused by a different mapping sign. Both use \Cref{Eq:angular_mapping_eta} for the mapping of SC jitter, but with inversed order of the indices: there is the mapping into MF$_{ij}$ with $\beta_{ij}$ in \Cref{Eq:N_{ij}^{s_{ij}}_simple}, but MF$_{ji}$ with $\beta_{ji}$ in \Cref{Eq:N_{ji:ij}^{s_{ij}}_simple}. This states that in a left-hand side MOSA (case $ij \in {12,23,31}$), the local SC jitter coupling (\Cref{Eq:N_{ij}^{s_{ij}}_simple}) naturally uses the mapping into a left-hand side MOSA ($\beta_{ij}\approx \ang{30}$). The remote SC jitter coupling, however, originates from a right-hand side MOSA, (case $ji \in {21, 32, 13}$) and has therefore  ($\beta_{ji}\approx \ang{-30}$).

%%%%%%%%%%%%%%%%%%%%%%%%%%%%%%%%%%%%%%%%%%%%%%%%%%%%%%%%%%%%%
%%%%%%%%%%%%%%%%%%%%%%%%%%%%%%%%%%%%%%%%%%%%%%%%%%%%%%%%%%%%%
%%%%%%%%%%%%%%%%%%%%%%%%%%%%%%%%%%%%%%%%%%%%%%%%%%%%%%%%%%%%%
\section{TTL noise in TDI-$X$ for the most significant noise contributions}
\label{Sec:TTL-noise-estimate}
%%%%%%%%%%%%%%%%%%%%%%%%%%%%%%%%%%%%%%%%%%%%%%%%%%%%%%%%%%%%%
We derive in this section estimates of the TTL noise levels expected in LISA. For this, suitable estimates for all jitter noise levels and all coupling coefficients are needed. Jitter estimates for LISA have been published before. However, most coupling coefficients have not been published yet and cannot be derived here in the scope of this paper due to the complexity of the coupling (compare, for instance, \cite{Hartig2022-G, Hartig2023-NG}). 
We, therefore, reduce in \Cref{Sec:TTL-reduced} the equations to the most significant contributions and argue only shortly why we consider these most relevant. In \Cref{Sec:Value-collection}, we roughly estimate the remaining coupling coefficients of the reduced model.
In \Cref{Sec:Analytic-TDI-X-PSD-model}, we then show the resulting analytic \gls{PSD} model of the \gls{TDI}-$X_2$ variable and validate this model in \Cref{Sec:Noise-Model Validation} by comparing with a numeric simulation. Finally, we perform a Monte Carlo simulation and compute the expected TTL noise magnitude in \Cref{Sec:Noise-Estimates}. We show that the noise is expected to violate the mission displacement noise requirement, resulting in the known and planned-for need to fit and subtract the noise in data post-processing.

%%%%%%%%%%%%%%%%%%%%%%%%%%%%%%%%%%%%%%%%%%%%%%%%%%%%%%%%%%%%%
%%%%%%%%%%%%%%%%%%%%%%%%%%%%%%%%%%%%%%%%%%%%%%%%%%%%%%%%%%%%%
\subsection{Model reduction}
\label{Sec:TTL-reduced}

It is known and expected that the coupling factors listed in \Cref{Eq:Complete_model_not_subst_simple} have considerably different magnitudes. In order to estimate the expected noise levels, we can, therefore, significantly reduce the given model and consider only dominant contributors. We do this in three steps below. First we argue why \gls{TMI} angular jitter coupling can be neglected. In the next step, we argue why we neglect all coupling of roll in \gls{MF}. Finally, we argue shortly why we neglect lateral jitter coupling. Finally, we show the resulting reduced model consisting of only yaw and pitch jitter coupling in \gls{MF}.

%%%
\subsubsection{Neglecting angular jitter coupling in the \glspl{TMI}}
Let us compare the magnitude of the \gls{TMI} TTL contributions to those of the \gls{LAI} contributions due to receiver jitter.
For this, we need to express the coupling in terms of beam jitters rather than component jitters and then consider the imaging performed in LISA.
We, therefore, assume that an angular component jitter in an arbitrary degree of freedom $\gamma$ (i.e.\ $\gamma \in \{ \phi,\eta,\theta\}$), causes a beam jitter in a certain degree of freedom $\alpha$. The two degrees of freedom $\gamma$ and $\alpha$ can be identical (this is the case for all MOSA jitters and for SC $\phi$ jitters, but they could also be different: e.g.\ SC jitter in $\theta$ partially maps into a beam jitter in $\eta$). The magnitude of the resulting beam jitter depends on where it is being measured. 

For example, let us consider the \gls{LAI} and jitter of the MOSA relative to free space. For the case of receiver jitter (in our notation the case where the upper and lower indices agree), we can assume a static incoming large wavefront, relative to which the MOSA is jittering. As observer located in the telescope's large pupil, we experience instead a jitter of the received beam relative to the MOSA. 
Thereby, any jitter $\alpha^\text{FS}_\text{MO/FS}$ of the MOSA relative to \acrfull{FS}, results in a beam jitter  $\alpha_\text{beam}^\text{FS}$ relative to the telescope measured in \gls{FS} at the telescope's large pupil, but with an inverse sign.

 The telescope images the beam to its small pupil on \gls{OB} level, thereby decreasing the spot size and likewise all lateral beam jitters $y^\text{OB} =y^\text{FS}/m_\text{tel}, z^\text{OB} =z^\text{FS}/m_\text{tel}$ while magnifying the angular jitters in pitch and yaw $\eta^\text{OB} = m_\text{tel} \eta^\text{FS}, \phi^\text{OB} = m_\text{tel} \phi^\text{FS}$ and leaving the roll $\theta$ unaffected.
Here, $m_\text{tel}$ is the telescope's angular magnification.
We then assume that additional imaging optics are used to image the telescope's small pupil onto the photodiodes. Thereby, the beam jitters are scaled further with the angular magnification $m_\text{IO}$ of the imaging optics. Consequently, the beam jitters on the photodiode are given by: 
\begin{subequations}
\begin{align}
	\phi_\text{beam}^{s,\text{PD}} &= m_\text{tel} m_\text{IO} \phi_\text{beam}^\text{FS} 
		= - m_\text{tel} m_\text{IO} \phi_\text{MO/FS}\\
	\eta_\text{beam}^{s,\text{PD}} &= m_\text{tel} m_\text{IO} \eta_\text{beam}^\text{FS}
		= - m_\text{tel} m_\text{IO} \eta_\text{MO/FS}\\
	\theta_\text{beam}^{s,\text{PD}} &= \theta_\text{beam}^\text{FS} \;.%\\
\end{align}
\label{Eq:LAI-PD-vs-FS-level}
\end{subequations}
The very same \gls{MOSA} jitter relative to free space will result in a jitter of the measurement beam in the test mass interferometer, when the measurement beam leaves the optical bench and reflects from the test mass. Thereby, the relative jitter of the test mass with respect to the MOSA is imprinted onto the measurement beam with an inverse sign and an additional factor of 2 due to the reflection. We expect that also the test mass interferometer comprises imaging optics, which we assume to have nominally the same magnification factor as in the \gls{LAI}. Consequently, we would expect the following beam angles in the test mass interferometer on photodiode level:
\begin{subequations}
\begin{align}
	\phi_\text{beam}^{\varepsilon,\text{PD}} &= - 2  m_\text{IO} \phi_\text{MO/FS}\\
	\eta_\text{beam}^{\varepsilon,\text{PD}} &= - 2  m_\text{IO} \eta_\text{MO/FS}\\
	\theta_\text{beam}^{\varepsilon,\text{PD}} &= -\theta_\text{MO/FS}^\text{FS} \;.
\end{align}
\label{Eq:TMI-PD-vs-FS-level}
\end{subequations}

Finally, we assume that the TTL coupling factors are caused by unavoidable small misalignments on either \gls{OB} or \gls{PD} level.
The alignment tolerances achievable during manufacturing on these levels are identical for both types of interferometers. It is, therefore, to be expected that the same level of coupling factors could occur:
\begin{subequations}
\begin{align}
	c_{\phi_\text{beam}}^{\varepsilon,\text{PD}} &\approx c_{\phi_\text{beam}}^{s,\text{PD}} \\
	c_{\eta_\text{beam}}^{\varepsilon,\text{PD}} &\approx c_{\eta_\text{beam}}^{s,\text{PD}} \;,
	\end{align}
	\label{Eq:Beam-Angle-Couplings-Coeffs}%
\end{subequations}%
which scale the beam angles, resulting in a TTL coupling of 
\begin{subequations}
\begin{align}
	N_{\phi_\text{MO/FS}}^{\varepsilon} &= c_{\phi_\text{beam}}^{\varepsilon,\text{PD}} \phi_\text{beam}^{\varepsilon,\text{PD}} \\
	N_{\phi_\text{MO/FS}}^s &= c_{\phi_\text{beam}}^{s,\text{PD}} \phi_\text{beam}^{s,\text{PD}} \\
	N_{\eta_\text{MO/FS}}^{\varepsilon} &= c_{\eta_\text{beam}}^{\varepsilon,\text{PD}} \eta_\text{beam}^{\varepsilon,\text{PD}} \\
	N_{\eta_\text{MO/FS}}^s &= c_{\eta_\text{beam}}^{s,\text{PD}} \eta_\text{beam}^{s,\text{PD}} \;.
\end{align}
\label{Eq:TMI-vs-LAI-angular-coupling}%
\end{subequations}

We can now conclude that the angular jitter TTL coupling in the \glspl{LAI} is expected to be significantly larger than in the \glspl{TMI}. This originates from beam jitter coupling coefficients on PD-level of the same magnitude (\Cref{Eq:Beam-Angle-Couplings-Coeffs}), but significantly different levels of beam jitters on PD-level in the different types of interferometers. The jitters differ by the telescope angular magnification factor $m_\text{tel}$. 
If we assume that the telescope and imaging optics jointly reduce the beam size from \SI{30}{cm} diameter to fit onto a \SI{1}{mm} diameter photodiode, and if we assume a factor of approximately $|m_\text{IO}| \approx 2$, we find $|m_\text{tel}|\approx 150$. 
Combining the equations above, and considering that the TMI-readout is divided by 2 when added to the LAI-signal in \gls{TDI} (see \Cref{Eq:xi,Eq:xi_TTL}), we find that MOSA pitch and yaw jitter relative to FS are expected to couple in the order of 150 times stronger to \gls{TDI}-$X$ via the \glspl{LAI} than via the \glspl{TMI}. Consequently, we can neglect the angular jitter coupling in the \gls{TMI} in the simplified model below.\\

%%%
\subsubsection{Neglecting roll in \gls{MF}:}
For rotationally symmetric beams and a roll around the beam axis, no TTL effect would occur at all. Only if either the beam is not rotationally symmetric, e.g.\ due to wavefront errors, or the center of rotation is not on the beam axis, a small effect could occur. Roll is therefore neglected here, by setting $ 0 = c^{\varepsilon_{ij}}_{\theta_{ij}} =  c^{s_{ij}}_{\theta_{ij}} =   c^{s_{ij}}_{\theta_{ji:ij}} $ in  \Cref{Eq:Complete_model_not_subst_simple}.\\

%%%
\subsubsection{Neglecting lateral jitter coupling}
\label{Par:Neglection-of-lateral-jitter-coupling}
Only jitter in yaw and pitch ($\phi,\eta$) are magnified by the telescope and imaging optics, while lateral beam jitter in \gls{MF} is demagnified by the magnification factors. Therefore, even misalignments on free-space level are expected to contribute less than the magnified effects for angular jitter coupling.  We, therefore, assume here that lateral jitter coupling contributes less strongly to \gls{TDI} than angular jitter coupling in the \gls{LAI}. So, we neglect lateral jitter coupling both in the \gls{LAI} and \gls{TMI} in the simplified model below. This assumption is further discussed in \Cref{Sec:Implicit-Assumptions+Implications}.\\

%%%
\subsubsection{Relevance of transmitter angular jitter coupling}
Angular jitter of a transmitting SC or MOSA contributes significant TTL coupling (e.g.~\cite{Sasso2018misalignment,Sasso2018far-field,Sasso2019}). The exceptionally long lever arm of \SI{2.5}{Gm} translates any nanoradian of angular jitter into a lateral (horizontal or vertical) displacement of the beam axis of $\SI{2.5}{Gm} \cdot \SI{1}{nrad} = \SI{2.5}{m}$ at the receiving end. Thereby, all angular transmitter jitter strongly shifts the RX wavefront over the receiving telescope.  This results in phase changes in the interferometer due to the \gls{RX-beam}'s wavefront errors, which need to be considered.  \\

%%%
\subsubsection{Reduced $N$-terms}
In summary, we consider angular jitter coupling in the \gls{LAI} to be the dominating TTL coupling terms. We, therefore, neglect all contributions from lateral jitter in the \glspl{LAI}, and all TTL noise contributions from angular and lateral jitter from the \glspl{TMI}.
Consequently, the TTL-model reduces to 
\begin{widetext}
\begin{subequations}
\begin{align}
	N^{\varepsilon_{ij}}_{ij} &\approx 0 \\
    N^{s_{ij}}_{ij} &= \left[c^{s_{ij}}_{\phi_{ij}} \phi_{ij} + c^{s_{ij}}_{\eta_{ij}} \eta_{ij}\right] 
    = \left[c^{s_{ij}}_{\phi_{ij}} \left(\phi_i^\mathrm{SC} + \phi_{ij}^\mathrm{MO}\right) 
    + c^{s_{ij}}_{\eta_{ij}}\left (\eta_{ij}^\mathrm{MO} + \cos(\beta_{ij})\eta_i^\mathrm{SC} + \sin(\beta_{ij})\theta_i^\mathrm{SC} \right)\right] \\
    N^{s_{ij}}_{ji:ij} &= \delay{ij}\left[c^{s_{ij}}_{\phi_{ji}} \phi_{ji} + c^{s_{ij}}_{\eta_{ji}} \eta_{ji}\right] %\nonumber \\
    = \delay{ij}\left[c^{s_{ij}}_{\phi_{ji}} (\phi_j^\mathrm{SC} + \phi_{ji}^\mathrm{MO}) + c^{s_{ij}}_{\eta_{ji}} (\eta_{ji}^\mathrm{MO} + \cos(\beta_{ji})\eta_j^\mathrm{SC} + \sin(\beta_{ji})\theta_j^\mathrm{SC})\right] \;.
\end{align}
\label{Eq:reduced_TTL_model}
\end{subequations}
\end{widetext}

%%%%%%%%%%%%%%%%%%%%%%%%%%%%%%%%%%%%%%%%%%%%%%%%%%%%%%%%%%%%%
%%%%%%%%%%%%%%%%%%%%%%%%%%%%%%%%%%%%%%%%%%%%%%%%%%%%%%%%%%%%%
\subsection{Magnitudes of coefficients}
\label{Sec:Value-collection}
%%%%%%%%%%%%%%%%%%%%%%%%%%%%%%%%%%%%%%%%%%%%%%%%%%%%%%%%%%%%%
%
Without going into much detail, we want to give now estimates of the magnitude of the coupling coefficients that we use below for estimating the resulting noise levels. 

For the receiver jitter coupling, we stated in the previous section that we assume alignment tolerances on either \gls{OB}- or \gls{PD}-level to be the primary cause of the coupling coefficients. Let us assume here that there is a misalignment on \gls{OB}-level causing a lateral piston effect (see \cite{Hartig2022-G} for more information on this geometric TTL-effect). This means we assume there is a lateral misalignment, causing the OB to push into, or out of, the beam path during the jitter. A typical estimate for such an alignment tolerance would be \SI{30}{\upmu m}. The resulting beam jitter coupling coefficient on OB-level is then $ \SI{30}{\upmu m/rad}$ \cite{Hartig2022-G}. On OB-level, the beam jitter is $m_\text{tel}$ times larger than MOSA-jitter relative to FS. Using $m_\text{tel} = 134$ \cite{Gehler2019,Sasso2019}, we find that the coupling of MOSA yaw or pitch jitter is given by $ 134 \cdot \SI{30}{\upmu m/rad} \approx \SI{4}{mm/rad}$. Considering that this would be only one of the TTL coupling mechanisms out of many (cf. \cite{Hartig2022-G, Hartig2023-NG} for a general list of possible coupling mechanisms), we estimate that the magnitude of the total coupling factor could likewise be in the order of \SI{10}{mm/rad}. 

For the magnitude of the coefficients describing the transmitter jitter coupling, we can make the very same argument as for the receiver jitter coupling. Additionally, we know from simulations \cite{Sasso2018misalignment,Sasso2018far-field,Sasso2019,UKOB-INST-TN1,UKOB-STOP-Model} that wavefront errors cause mm/rad-level coupling coefficient contributions. In total, we, therefore, estimate also for transmitter jitter coupling coefficient levels in the order of \SI{10}{mm/rad}.

These coupling coefficients refer to the case after mitigation step 1: mitigation by design (cf.~\Cref{Sec:intro}). The coupling coefficients after a realignment optimization (mitigation step 2) were published previously in \cite{Paczkowski2022} and stated to be in the order of \SI{2.3}{mm/rad} for both the receiver and transmitter jitter couplings coefficients.

We, therefore, use two different levels of coupling coefficients for the noise estimates in the subsections below:  \SI{10}{mm/rad} assuming noise levels prior to a realignment for noise minimization, and \SI{2.3}{mm/rad} assuming a system realignment was already performed. 

In either case, a TTL model will be fitted to the mission data and afterward subtracted from it. It is only after this that the resulting noise levels have to meet the corresponding requirements. For more information on this subtraction in LISA, see \cite{Paczkowski2022}, and \cite{Armano2016} for the successful subtraction in LISA Pathfinder.

%%%%%%%%%%%%%%%%%%%%%%%%%%%%%%%%%%%%%%%%%%%%%%%%%%%%%%%%%%%%%
%%%%%%%%%%%%%%%%%%%%%%%%%%%%%%%%%%%%%%%%%%%%%%%%%%%%%%%%%%%%%
\subsection{Analytic TDI-$X$ noise model for the reduced TTL model}
\label{Sec:Analytic-TDI-X-PSD-model}
%%%%%%%%%%%%%%%%%%%%%%%%%%%%%%%%%%%%%%%%%%%%%%%%%%%%%%%%%%%%%

To estimate the TTL noise contribution in $X_2$ we evaluate \Cref{Eq:X2} using \Cref{Eq:xi_TTL}. The power spectral density (PSD) of the resulting term $X_2^\text{TTL}$ was evaluated using \Cref{Eq:reduced_TTL_model} and the simplifying assumption that the jitters in $\phi_\text{MO/FS}$ are uncorrelated to those in $\eta_\text{MO/FS}$ and that $\phi_{\text{MO/FS}_{21}},\eta_{\text{MO/FS}_{21}}$ are uncorrelated to $\phi_{\text{MO/FS}_{31}},\eta_{\text{MO/FS}_{31}}$.
Additionally, we set all arm lengths to be equal ($L_{ij} = L_{ik} = L$ for all $ijk$), and, consequently, the constellation opening angles are equal ($\beta = \beta_{ij}= - \beta_{ik}$ with $ \beta>0$ where $ij$ denotes a left-hand side MOSA and $ik$ a right-hand side MOSA). The resulting \gls{PSD} $S_{X_2}(f) $ of the TTL noise in the \gls{TDI}-$X_2$ observable is then given by:
\begin{subequations}
\begin{align}
	S_{X_2}^\text{TTL}(f)  &=   
	S_{X_2}^{\phi_\mathrm{SC}}(f) +  S_{X_2}^{\eta_\mathrm{SC}}(f) + S_{X_2}^{\theta_\mathrm{SC}}(f)  \nonumber \\
	& + S_{X_2}^{\phi_\mathrm{MO}}(f) + S_{X_2}^{\eta_\mathrm{MO}}(f)
\end{align}%
	\label{Eq:PSDX2}%
\end{subequations}%
using the spectral densities of the individual jitter contributions
\begin{widetext}
\begin{subequations}
\begin{align}
    S_{X_2}^{\phi_\mathrm{SC}}(f) &= C_{XX}(f) \left[\left((c_{\phi_{13}}^{s_{13}} - c_{\phi_{12}}^{s_{12}})^2 + (c_{\phi_{13}}^{s_{31}} - c_{\phi_{12}}^{s_{21}})^2 + 2(c_{\phi_{13}}^{s_{13}} - c_{\phi_{12}}^{s_{12}})(c_{\phi_{13}}^{s_{31}} - c_{\phi_{12}}^{s_{21}})\cos(4\pi f L/c)\right)S_{\phi_{\mathrm{SC}1}}(f)\right. \notag\\
    &\hspace{20ex}\left.
    + (c_{\phi_{21}}^{s_{12}} + c_{\phi_{21}}^{s_{21}})^2 S_{\phi_{\mathrm{SC}2}}(f)
    + (c_{\phi_{31}}^{s_{13}} + c_{\phi_{31}}^{s_{31}})^2 S_{\phi_{\mathrm{SC}3}}(f) 
    \right] \label{Eq:X2_sc_phi}\\
    S_{X_2}^{\eta_\mathrm{SC}}(f) &= C_{XX}(f) \left[\left((c_{\eta_{13}}^{s_{13}} - c_{\eta_{12}}^{s_{12}})^2 + (c_{\eta_{13}}^{s_{31}} - c_{\eta_{12}}^{s_{21}})^2 + 2(c_{\eta_{13}}^{s_{13}} - c_{\eta_{12}}^{s_{12}})(c_{\eta_{13}}^{s_{31}} - c_{\eta_{12}}^{s_{21}})\cos(4\pi f L/c)\right)S_{\eta_{\mathrm{SC}1}}(f)\right. \notag\\
    &\hspace{20ex}\left.
	+ (c_{\eta_{21}}^{s_{12}} + c_{\eta_{21}}^{s_{21}})^2 S_{\eta_{\mathrm{SC}2}}(f) 
	+ (c_{\eta_{31}}^{s_{13}} + c_{\eta_{31}}^{s_{31}})^2 S_{\eta_{\mathrm{SC}3}}(f) 
    \right] \cos^2(\beta) \label{Eq:X2_sc_eta}\\
    S_{X_2}^{\theta_\mathrm{SC}}(f) &= C_{XX}(f) \left[\left((c_{\eta_{13}}^{s_{13}} + c_{\eta_{12}}^{s_{12}})^2 + (c_{\eta_{13}}^{s_{31}} + c_{\eta_{12}}^{s_{21}})^2 + 2(c_{\eta_{13}}^{s_{13}} + c_{\eta_{12}}^{s_{12}})(c_{\eta_{13}}^{s_{31}} + c_{\eta_{12}}^{s_{21}})\cos(4\pi f L/c)\right)S_{\theta_{\mathrm{SC}1}}(f)\right. \notag\\
    &\hspace{20ex}\left.
    + (c_{\eta_{21}}^{s_{12}} + c_{\eta_{21}}^{s_{21}})^2 S_{\theta_{\mathrm{SC}2}}(f) 
    + (c_{\eta_{31}}^{s_{13}} + c_{\eta_{31}}^{s_{31}})^2 S_{\theta_{\mathrm{SC}3}}(f) 
    \right] \sin^2(\beta) \label{Eq:X2_sc_theta}\\
    S_{X_2}^{\phi_\mathrm{MO}}(f) &= C_{XX}(f) \left[\left({c_{\phi_{13}}^{s_{13}}}^2 + {c_{\phi_{13}}^{s_{31}}}^2 + 2c_{\phi_{13}}^{s_{13}}c_{\phi_{13}}^{s_{31}}\cos(4\pi f L/c)\right)S_{\phi_{\mathrm{MO}13}}(f)
        + (c_{\phi_{31}}^{s_{13}} + c_{\phi_{31}}^{s_{31}})^2 S_{\phi_{\mathrm{MO}31}}(f) 
    \right. \notag\\
    &\hspace{20ex}\left.
    +\left({c_{\phi_{12}}^{s_{12}}}^2 + {c_{\phi_{12}}^{s_{21}}}^2 + 2c_{\phi_{12}}^{s_{12}}c_{\phi_{12}}^{s_{21}}\cos(4\pi f L/c)\right)S_{\phi_{\mathrm{MO}12}}(f) 
    + (c_{\phi_{21}}^{s_{12}} + c_{\phi_{21}}^{s_{21}})^2 S_{\phi_{\mathrm{MO}21}}(f) 
    \right] \label{Eq:X2_mosa_phi} \\
    S_{X_2}^{\eta_\mathrm{MO}}(f) &= C_{XX}(f) \left[\left({c_{\eta_{13}}^{s_{13}}}^2 + {c_{\eta_{13}}^{s_{31}}}^2 + 2c_{\eta_{13}}^{s_{13}}c_{\eta_{13}}^{s_{31}}\cos(4\pi f L/c)\right)S_{\eta_{\mathrm{MO}13}}(f)
    + (c_{\eta_{31}}^{s_{13}} + c_{\eta_{31}}^{s_{31}})^2 S_{\eta_{\mathrm{MO}31}}(f) 
    \right. \notag\\
    &\hspace{20ex}\left.+\left({c_{\eta_{12}}^{s_{12}}}^2 + {c_{\eta_{12}}^{s_{21}}}^2 + 2c_{\eta_{12}}^{s_{12}}c_{\eta_{12}}^{s_{21}}\cos(4\pi f L/c)\right)S_{\eta_{\mathrm{MO}12}}(f)  
       + (c_{\eta_{21}}^{s_{12}} + c_{\eta_{21}}^{s_{21}})^2 S_{\eta_{\mathrm{MO}21}}(f)
     \right] \label{Eq:X2_mosa_eta}
\end{align}
\label{Eq:PSDs-per-DOF}
\end{subequations}
\end{widetext}
and the equal arm length transfer function $C_{XX}(f)$ 
\begin{equation} 
	 C_{XX}(f) = 16 \sin^2\left(\frac{2\pi f L}{c}\right) \sin^2\left(\frac{4\pi f L}{c}\right) \;.
\end{equation}
As usual, the corresponding equations for the Michelson $Y_2$ and $Z_2$ \glspl{PSD} can be found by cyclic permutation of the indices.
It is important to note that the magnitude of \gls{TTL} noise does not only depend on the absolute value of the coupling coefficients but also on their signs, since they mostly appear as sums or differences. 

%%%%%%%%%%%%%%%%%%%%%%%%%%%%%%%%%%%%%%%%%%%%%%%%%%%%%%%%%%%%%
%%%%%%%%%%%%%%%%%%%%%%%%%%%%%%%%%%%%%%%%%%%%%%%%%%%%%%%%%%%%%
\subsection{Noise model validation}
\label{Sec:Noise-Model Validation}
To validate the TTL noise model \Cref{Eq:PSDX2,Eq:PSDs-per-DOF}, we directly compare the so-defined \glspl{PSD} for one test case with the results of a numerical simulation performed with LISA Instrument \cite{LISAInstrument} and PyTDI \cite{PyTDI}. In these simulations, LISA Instrument computed a time series of the interferometric displacement readout signals. This corresponds to a simulation of $ N^{s_{ij}}_{ij}$ and $ N^{s_{ij}}_{ji:ij} $ since no other noise was assumed than TTL originating from angular jitters of the \glspl{SC} and the \glspl{MOSA}. PyTDI was used to numerically propagate these signals through \gls{TDI}. The \gls{ASD} of the resulting time series was then computed in Python. This resulted in a numerical analogon of \Cref{Eq:PSDX2}. 

The simulation was run 6 times: once with all noise contributions active (numerical equivalent of \Cref{Eq:PSDX2}), and five simulations with each assuming only angular jitter of one degree of freedom (numerical equivalent of \Cref{Eq:PSDs-per-DOF}). All simulations cover $\SI{50000}{\s}\approx\SI{14}{\hour}$ of data.

For both types of simulations, i.e.\ the evaluation of the analytical noise model \Cref{Eq:PSDX2,Eq:PSDs-per-DOF} and the numerical simulation with LISA Instrument and PyTDI, we made the following assumptions:\\
For the jitters, we assume here values from the LISA noise budget listed in \Cref{Tab:Noises-for-Fig6}, which have been previously published in \cite{Paczkowski2022}. We simplify the noise spectral shapes to white noise because the pole-zero model defined in  \cite{Paczkowski2022} is effectively white in the frequency band of \SIrange{3}{300}{mHz}, and TTL usually dominates at frequencies above approximately \SI{3}{mHz}. Naturally, we do not assume the SC or the MOSA to have a motion that is well-described by a white noise spectral density, particularly not up to \SI{1}{Hz}. Instead, this assumption should be understood as an upper limit representing preliminary requirements.
\begin{table}  
     \centering
    \caption{Properties of all angular jitters, used for the computation of \Cref{Fig:ttl_models} and \Cref{Fig:mc}. Assumed are white noise shapes and \glspl{ASD} with the listed amplitudes.}
       \label{Tab:Noises-for-Fig6}
    \begin{tabular}{rc }
        \toprule
	 $\sqrt{S_{\phi_{i}}^\mathrm{SC}(f)}$,  $\sqrt{S_{\eta_{i}}^\mathrm{SC}(f)}$, $\sqrt{S_{\theta_{i}}^\mathrm{SC}(f)}$:  & \SI{5}{nrad\per\sqrt\Hz}\\
	 $\sqrt{S_{\phi_{i}}^\mathrm{MO}(f)}$: & \SI{2}{nrad\per\sqrt\Hz}\\
	 $\sqrt{S_{\eta_{i}}^\mathrm{MO}(f)}$: & \SI{1}{nrad\per\sqrt\Hz}\\
	    \bottomrule
    \end{tabular}
\end{table}
\begin{figure}
    \includegraphics{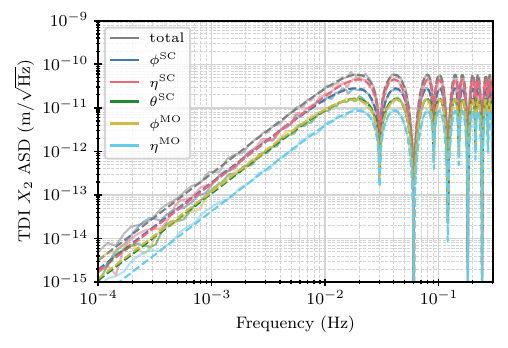}
    \caption{\Gls{TTL} noise in the second-generation Michelson $X_2$ combination for the randomly drawn coefficients listed in \Cref{Tab:CoeffsFig6} and using white angular jitter noises with the amplitudes listed in \Cref{Tab:Noises-for-Fig6}. Shown is the resulting total TTL noise (grey), as well as the individual noise contributions coming from jitters in \gls{SC} yaw ($\phi^\mathrm{SC}$, blue), \gls{SC} pitch ($\eta^\mathrm{SC}$, red), \gls{SC} roll ($\theta^\mathrm{SC}$, green), \gls{MOSA} yaw ($\phi^\mathrm{MO}$, yellow) and \gls{MOSA} pitch ($\eta^\mathrm{MO}$, cyan). For each of the curves, we show the  \acrfull{ASD} of the numerically computed time series (solid) and the analytical results computed from \Cref{Eq:PSDX2,Eq:PSDs-per-DOF}  (dashed). The results of both methods agree.}
    \label{Fig:ttl_models}
\end{figure}
We chose to draw the TTL coupling coefficients randomly from a uniform distribution with limits \SI{\pm2.3}{\milli\m\per\radian}.
The resulting coefficients are listed in \Cref{Tab:CoeffsFig6}.  %
\begin{table}  
     \centering
    \caption{Coefficients in \SI{}{mm/rad} rounded to the fourth digit used for generating \Cref{Fig:ttl_models}. }
       \label{Tab:CoeffsFig6}
    \begin{tabular}{rc || r c}
        \toprule
		$c^{s_{12}}_{\phi_{12}}$ = & +0.3799		 		& $c^{s_{12}}_{\eta_{12}}$ =& +0.6261, \\
		$c^{s_{23}}_{\phi_{23}}$ =&  -1.2744		 		& $c^{s_{23}}_{\eta_{23}}$ =& +0.2130\\
		$c^{s_{31}}_{\phi_{31}}$ =&  +0.8191				& $c^{s_{31}}_{\eta_{31}}$ =& -0.6629 \\
		$c^{s_{13}}_{\phi_{13}}$ =&  +0.0504		 		& $c^{s_{13}}_{\eta_{13}}$ =& -0.5469\\
		$c^{s_{32}}_{\phi_{32}}$ =&  -0.8243				& $c^{s_{32}}_{\eta_{32}}$ =& +0.950 \\	
		$c^{s_{21}}_{\phi_{21}}$ =&  -1.8077		 		& $c^{s_{21}}_{\eta_{21}}$ =& -0.2721\\			
        \midrule
        		$c^{s_{21}}_{\phi_{12}}$ = &+1.2481				& $c^{s_{21}}_{\eta_{12}}$ =&- 1.8766  \\
        		$c^{s_{32}}_{\phi_{23}}$ = &-0.7490				& $c^{s_{32}}_{\eta_{23}}$ =& +1.2681\\
		$c^{s_{13}}_{\phi_{31}}$ = & +0.0651				& $c^{s_{13}}_{\eta_{31}}$ = &- 0.8641\\
		$c^{s_{31}}_{\phi_{13}}$ =&+1.3336		 		& $c^{s_{31}}_{\eta_{13}}$ =& +0.4544\\
		$c^{s_{23}}_{\phi_{32}}$ =&  -0.6130				& $c^{s_{23}}_{\eta_{32}}$ =& +1.4640 \\	
		$c^{s_{12}}_{\phi_{21}}$ =& +0.2540 				& $c^{s_{12}}_{\eta_{21}}$ = & +0.2069\\			
    \bottomrule
    \end{tabular}
\end{table}
The numerical simulations considered realistic \gls{SC} orbits based on an orbit file provided by ESA. For the evaluation of the analytical noise model, we assumed constant and equal arm lengths, and hence $\beta = \SI{30}{\degree}$. For the value of the constant arm length, we averaged all three arm lengths derived from the orbit file over the duration of the simulation (i.e.\ \SI{50000}{s}), which evaluated to $L/c = \SI{8.28}{s}$.

The resulting \glspl{ASD} of the second-generation Michelson variable $X_2$ are shown in \Cref{Fig:ttl_models} in a direct comparison of the analytical and numerical method. 
\begin{figure}
    \includegraphics*[trim = 5 0 5 0]{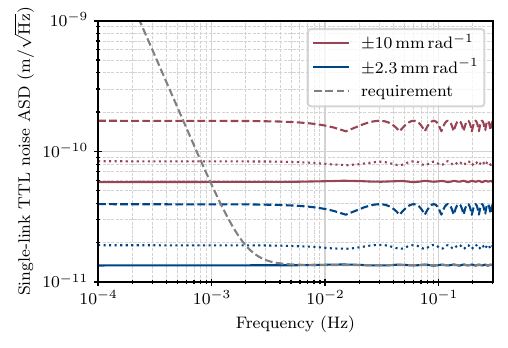}
    \caption{Range of \gls{TTL} noise coupling in \gls{TDI}-$X_2$ converted to single-link equivalents using \Cref{Eq:TDI-toSL-conversion}. The solid and dotted lines show the median and 95th percentile of a simulation where 10000 coupling coefficients were drawn from a uniform distribution with limits of \SI{\pm2.3}{\milli\m\per\radian} (blue lines) or \SI{\pm10}{\milli\m\per\radian} (red lines). The dashed line represents the worst case computed from \Cref{Eq:SX2-WorstCase} (signs chosen individually for every frequency, see text for more information).} %
	\label{Fig:mc}%
\end{figure}
Please note that we calculated the \glspl{ASD} by taking the square root of the logarithmically scaled \gls{PSD}, which was computed using the method described in \cite{lpsd_paper}. The numerical and analytical curves match perfectly, which validates the analytical models in \Cref{Eq:PSDX2,Eq:PSDs-per-DOF}. 
Only at the low-frequency tail around \SI{0.1}{mHz} small deviations are visible, which we attribute to higher variances of the logarithmically scaled \gls{PSD} estimate. These higher variances originate from fewer possible averages than at high frequencies due to longer data stretches required to resolve the \gls{ASD} at low frequencies.

Analyzing the individual contributions in \Cref{Fig:ttl_models} is of little meaning since the result of only one set of random TTL coupling coefficients is shown. The figure is, therefore, a proof of principle: the shown analytical models can be used to estimate the TTL noise in LISA and the contributions from individual degrees of freedom. Likewise, the shown TTL $N$-terms and method for analytically computing a \gls{TDI} variable can be used to derive any of the other \gls{TDI} variables (cf. \cite{Muratore2020} for a list of \gls{TDI} variables).

\subsection{Noise estimates}
\label{Sec:Noise-Estimates}

In order to estimate the noise level expected for \gls{TDI}-$X_2$ we ran a Monte Carlo simulation by drawing 10000 random sets of coupling coefficients from a uniform distribution with limits of \SI{\pm2.3}{\milli\m\per\radian}. These limits correspond to the current estimate of the magnitude of the in-flight coefficients, as described in \Cref{Sec:TTL-noise-estimate} and \cite{Paczkowski2022} under the assumption that a previous coefficient reduction by realignment was performed. We assume it to hold for all coupling coefficients, i.e.\ for angular jitter in $\eta$ and $\phi$, and for contributions of receiver jitter coupling to the \glspl{RX-beam} (i.e.\ $c^{s_{ij}}_{\eta/\phi_{ij}}$), as well as the transmitter jitter coupling via the \glspl{TX-beam}  (i.e.\ $c^{s_{ij}}_{\eta/\phi_{ji}}$). We then repeated the process and drew a second set of 10000 random coefficients from a uniform distribution with limits of \SI{\pm10}{\milli\m\per\radian}, roughly representing the coefficient levels if no prior TTL-noise minimization by realignment was performed.

The coefficients were then used to evaluate the analytic noise model. We then convert the noise estimates to equivalent single-link contributions by dividing the \gls{PSD} by the \gls{TDI} transfer function $C_{XX}$ and the number of involved links, which is 4 for \gls{TDI}-$X_2$ (cf. \Cref{Eq:PSD-X2_generell}), and finally taking the square root:
\begin{equation}
	S^{1/2,\text{TTL}}_\text{SL} =  \sqrt{S_{X_2}^\text{TTL}/(4C_{XX})} 	\;.
	\label{Eq:TDI-toSL-conversion}
\end{equation}
The resulting \glspl{ASD} are nearly white since we assume white noise angular jitter, and can be directly compared with requirements like the mission displacement noise requirement 
\begin{equation}
	S^{1/2}_\text{req LISA} \leq \SI{13.5}{\frac{pm}{\sqrt{Hz}}} \, \sqrt{1+\left( \frac{\SI{2}{mHz}}{f}\right)^4}
	\label{Eq:MissionRequirement}
\end{equation}
taken from \cite{PerfModel} and previously published in \cite{Paczkowski2022}. This requirement is defined for single links and the sum of all types of displacement noises in LISA.
For every frequency, we show in \Cref{Fig:mc} the median and 95th percentile noise level of the resulting \gls{ASD} as solid and dotted blue lines for coefficients within \SI{\pm2.3}{\milli\m\per\radian}, and as solid and dotted red lines for coefficients within \SI{\pm10}{\milli\m\per\radian}, in direct comparison with the mission requirement.

Additionally, we searched for the worst-case outcome by setting all absolute values of the coefficients to be equal to  $c_\text{TTL}$ and then searching for the sign combinations that maximized \Cref{Eq:PSDX2}. This maximal PSD, however, is frequency dependent and, likewise, depends on the contributing jitter spectra and their correlations. Assuming all spectra to be uncorrelated and the spectra of the individual MOSAs and the individual SC to be identical, i.e.\ $S_{\alpha_{\mathrm{SC}i}}= S_{\alpha_{\mathrm{SC}}}, S_{\alpha_{\mathrm{MO}ij}} = S_{\alpha_{\mathrm{MO}}}$ and additionally $S_{\theta_{\mathrm{SC}}} = S_{\eta_{\mathrm{SC}}}$ we found a maximal PSD of:
\begin{widetext}
\begin{align}
S_{X_2}^\text{max}(f) = 
\begin{cases}
\!\begin{aligned}
% phi part
 2c^2 C_{XX}(f) \big[& (6 - 2 \cos(4\pi f L/c))S_{\phi_{\mathrm{MO}}} + (8 -4 \cos(4\pi f L/c))S_{\phi_{\mathrm{SC}}} \\
% eta part
& + 4\cos(2\beta)\sin^2(2\pi f L/c)S_{\eta_{\mathrm{SC}}} + (6 -2 \cos(4\pi f L/c))(S_{\eta_{\mathrm{MO}}}  + S_{\eta_{\mathrm{SC}}} )
 \big]
\end{aligned}&, \text{if } \cos(4\pi f L/c) \leq 0 \\
% else part
\!\begin{aligned} 2c^2 C_{XX}(f)\big[
& (6 + 2\cos(4\pi f L/c))S_{\phi_{\mathrm{MO}}}+ (8 + 4\cos(4\pi f L/c))S_{\phi_{\mathrm{SC}}} \\
&  + 4\cos(2\beta)\cos^2(2\pi f L/c)S_{\eta_{\mathrm{SC}}} + (6 +2 \cos(4\pi f L/c))(S_{\eta_{\mathrm{MO}}}  + S_{\eta_{\mathrm{SC}}} )\big]
\end{aligned} &, \text{else}\;.
\end{cases}
\label{Eq:SX2-WorstCase}
\end{align}
\end{widetext}
The sign choices resulting in this maximal PSD are listed in \Cref{Tab:worst_case_ccs}. 
\begin{table}[tbh]
    \centering
  \caption{Sign combinations resulting in the worst case TTL coupling noise in the second-generation Michelson variable $X_2$, provided $S_{\theta_{\mathrm{SC}}}=S_{\eta_{\mathrm{SC}}}$. Each column of the table shows a pair of signs where the first one yields for frequencies where $\cos(4\pi f L / c) \le 0$ and the second sign holds else (cf. the cases in \Cref{Eq:SX2-WorstCase}). Please note that the table holds individually for $\alpha = \eta$ and $\alpha = \phi$. Since for each, there are four sign combinations, there is a total of 16 sign combinations resulting in the worst case coupling defined in \Cref{Eq:SX2-WorstCase}.}
    \begin{tabular}{l  | C{1.2cm} C{1.2cm}  C{1.2cm}  >{\hspace{3mm}}r}
        \toprule
        $c_{\alpha_{12}}^{s_{12}}$ & $+/+$ & $+/+$ & $+/+$ & $+/+$ \\
        $c_{\alpha_{21}}^{s_{21}}$ & $+/+$ & $+/+$ & $-/-$ & $-/-$ \\
        $c_{\alpha_{13}}^{s_{13}}$ & $-/-$ & $-/-$ & $-/-$ & $-/-$ \\
        $c_{\alpha_{31}}^{s_{31}}$ & $+/+$ & $-/-$ & $+/+$ & $-/-$ \\
        $c_{\alpha_{21}}^{s_{12}}$ & $+/+$ & $+/+$ & $-/-$ & $-/-$ \\
        $c_{\alpha_{12}}^{s_{21}}$ & $-/+$ & $-/+$ & $-/+$ & $-/+$ \\
        $c_{\alpha_{31}}^{s_{13}}$ & $+/+$ & $-/-$ & $+/+$ & $-/-$ \\
        $c_{\alpha_{13}}^{s_{31}}$ & $+/-$ & $+/-$ & $+/-$ & $+/-$ \\
        \bottomrule
    \end{tabular}
    \label{Tab:worst_case_ccs}
\end{table}
Please note that this table contains only the 16 coefficients contributing to \gls{TDI}-$X_2$. However, the worst cases for \gls{TDI}-$Y_2$ and $Z_2$, as well as the sign combinations for their coupling coefficients can be found as usual via cyclic permutation of the indices in the given equation and table.

We then choose at each frequency the worst case sign option and computed the resulting single-link \gls{ASD}, i.e.\ $\sqrt{S_{X_2}^\text{max}/(4 C_{XX})}$. The results for $c_\text{TTL} = {\SI{2.3}{\milli\m\per\radian}}$ and $\SI{10}{\milli\m\per\radian}$ are shown as dashed blue and red lines in \Cref{Fig:mc}, representing the worst case at every individual frequency.

To quantify the comparison with the mission requirement, we list in \Cref{Tab:mc} the maximum value of each curve in \Cref{Fig:mc}, which can be compared with the requirement for frequencies above the corner frequency of \SI{2}{mHz} defined in \Cref{Eq:MissionRequirement}. 
\begin{table}
    \centering
    \caption{Summary of single-link displacement noise levels in \Cref{Fig:mc}. Shown are the maximal values 
    of each curve (i.e.\ of the median and 95th percentile of the Monte Carlo simulation, and of the worst case) within the frequency range shown in \Cref{Fig:mc}.}
    \label{Tab:mc}
    \begin{tabular}{l c c c}
        \toprule
        range & median & 95th perc. & worst \\
        & [$\si{\pico\m\per\sqrt{\Hz}}$] & $[\si{\pico\m\per\sqrt{\Hz}}]$ & $[\si{\pico\m\per\sqrt{\Hz}}]$ \\
        \midrule
        \SI{\pm2.3}{\milli\m\per\radian} & 13.43 & 19.34 &  39.50 \\
        \SI{\pm10}{\milli\m\per\radian} &  58.22 & 83.78 & 171.8 \\
        \bottomrule
    \end{tabular}
\end{table}
This shows, under the given assumptions the median TTL noise level for coefficients within \SI{\pm 2.3}{mm/rad} just marginally fits into the mission displacement noise budget, leaving no room for any other displacement noise sources. All remaining results would violate the entire mission displacement noise budget, with factors of up to 2.9  or 12.7 for the lower and higher coefficient levels, respectively, at frequencies higher than \SI{2}{mHz}. 

These high noise levels have been known already since 2018 \cite{Shah2018Presentation} and they are not a show-stopper for LISA. They hold prior to the final noise mitigation step of fitting and subtracting. This step will reduce the noise levels to within the requirements, as shown by \cite{Paczkowski2022, LCST-INST-RP-002, LCST-INST-TN-017}.

Fitting and subtracting is considered a reliable mitigation strategy. However, it has limitations if the initial noise level and the involved coupling coefficients are too high. A typical rule of thumb is that noise with factors of 10 to 20 above a requirement can be fitted and subtracted. This factor, however, depends significantly on the noise in the signals used for subtraction, so on DWS readout noise provided that only angular jitter coupling needs to be subtracted, as described here. The given rule of thumb phrases, therefore, only in simple terms the known problems that the method can fail to achieve the wanted suppression if the coupling coefficients (and thereby the noise level prior to subtraction) are too high.

In that case, sensing noise is added to the displacement readout during the subtraction process. This additional sensing noise scales with the coupling coefficients, such that the higher the coupling coefficients, the higher the additional sensing noise added to the \gls{TDI} variables (cf. Eq.\,(25) in \cite{Paczkowski2022}). This will eventually limit the noise subtraction quality. 
This was already observed in the early times of the LISA Pathfinder mission (until March 2016), where the coefficient of the horizontal lateral $y$-jitter was particularly high and GRS-sensing noise was added in during the subtraction process (\cite[yellow vs. red curve in Fig.~13]{Hartig2023-LPF-DA}). In LISA Pathfinder, this was resolved by the in-flight test mass realignment performed in March 2016.

Judging whether or not the presented noise levels could be fully subtracted is beyond the scope of this paper. However, the high levels presented for the coefficient ranges of \SI{\pm 10}{mm/rad} indicate a risk. They are the reason why realignment for noise mitigation is currently planned, and in-flight coupling coefficients within \SI{\pm 2.3}{mm/rad} are considered to be achievable. We know from \cite{Paczkowski2022, LCST-INST-RP-002, LCST-INST-TN-017} that the noise originating from these coefficient levels can be suppressed to the required noise levels by fitting and subtracting in post-processing.

\subsection{Worst case vs. equal coefficients and comparison with other publications}
It is not trivially visible from \Cref{Eq:PSDs-per-DOF}, for which assumptions the noise is maximal. In the past, it was therefore a simple work-around to assume all coefficients to be equal to a maximal value. So rather than assigning random values with limits of e.g.\ \SI{\pm 2.3}{mm/rad}, one set all to have the very same value of  \SI{2.3}{mm/rad}. This was e.g.\ done in \cite{Paczkowski2022}, and we compare now, how this case relates to the worst case and the statistics we have presented here.

\begin{figure}%
\includegraphics*[width=\columnwidth, trim =5 0 5 5]{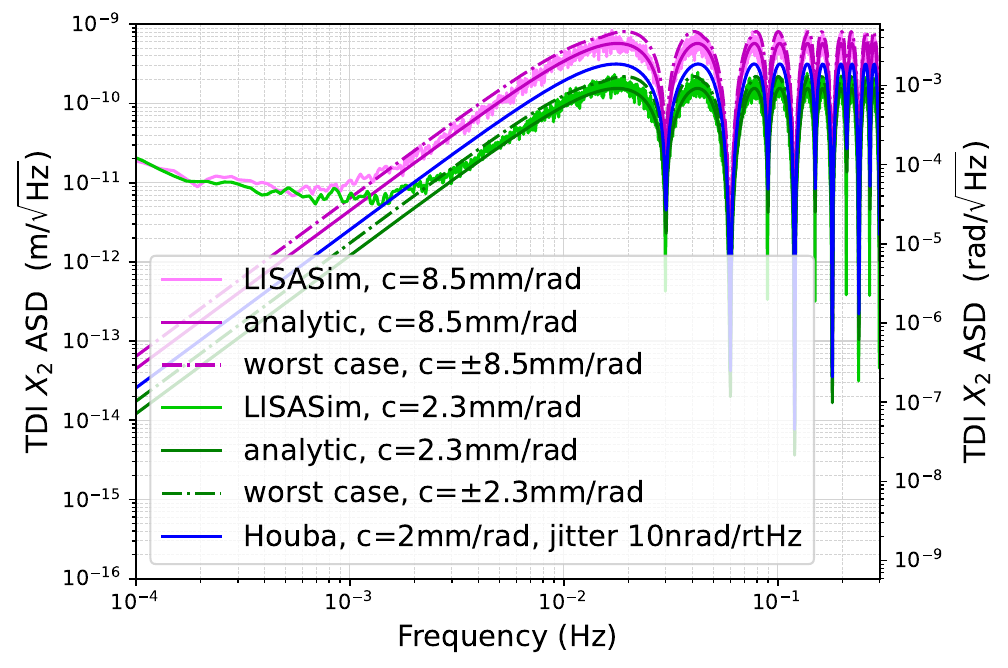}
\caption{Comparison between the analytic results for the worst case (\Cref{Eq:SX2-WorstCase},  $c^{s_{ij}}_{\alpha_{ij}} = c^{s_{ij}}_{\alpha_{ji:ij}} = \pm c$) and the case of equal coefficients ( $c^{s_{ij}}_{\alpha_{ij}} = c^{s_{ij}}_{\alpha_{ji:ij}} = c$) generated either analytically (\Cref{Eq:SX2-EqualCoeffs}), or by the simulator LISASim (data reused from \cite[Fig.~1]{Paczkowski2022}). The analytic model fits the data, while the worst case is approximately a factor of $\sqrt{2}$ times larger than the case of equal coefficients. The blue curve shows our analytic equation applied to the settings of \cite[Fig. 8]{Houba2023-TDI-inf}.}
\label{Fig:WorstCase-vs-equal-signs}
\end{figure}

For this, we evaluate \Cref{Eq:PSDs-per-DOF}. This time, we assume that all coefficients are equal to $c_\text{TTL}$, unlike in the previous section where only the absolute value was assumed to be equal to $c_\text{TTL}$. For this case of equal coefficients (EC) with equal signs, we find a PSD of
\begin{align}
S_{X_2}^\text{EC}(f) =& 
   C_{XX}(f) 2 c^2_\text{TTL}\bigg[  4 S_{\phi_{\mathrm{SC}}}(f) + 4 \cos^2(\beta)S_{\eta_{\mathrm{SC}}}(f)  \nonumber \\
&+\left(8 + 4\cos(4\pi f L/c)\right) \sin^2(\beta) S_{\theta_{\mathrm{SC}}}(f) \nonumber \\
&+\left(   5 +    2\cos(4\pi f L/c)\right)S_{\phi_{\mathrm{MO}}}(f)   \nonumber \\
&+\left(6 + 2\cos(4\pi f L/c)\right)S_{\eta_{\mathrm{MO}}}(f)      
\bigg]
	\label{Eq:SX2-EqualCoeffs}
\end{align}
assuming again the constellation opening angles to be equal ($\beta = \beta_{ij}= - \beta_{ik}$ with $ \beta>0$ ).
\Cref{Fig:WorstCase-vs-equal-signs} shows the \glspl{PSD} from \cite[Fig.~1]{Paczkowski2022}, together with the result of the corresponding analytic description \Cref{Eq:SX2-EqualCoeffs}, and compare with the worst case estimate according to \Cref{Eq:SX2-WorstCase}.
%%%%%%%%
%%%%%%%% 
The data from \cite[Fig.~1]{Paczkowski2022} was generated using the open loop simulator LISASim, under the assumptions of unequal arm lengths and nearly white angular jitter noise in the frequency range from \SIrange{2}{200}{mHz}. The LISASim data likewise included a number of other secondary noise sources. The included test mass force noise dominated at low frequencies. This explains the deviation between the analytic models and the simulated data at frequencies below approximately \SI{2}{mHz}.

We find that our results presented here for equal coefficients match the data from \cite{Paczkowski2022} well at frequencies where TTL is dominant (i.e.\ $f\gtrsim$ \SI{2}{mHz}). 
The match of our model with the data is possible because the jitter spectra used in \cite{Paczkowski2022} are nearly white at $f\gtrsim$ \SI{2}{mHz}, and therefore roughly match our simplified assumption of white jitter spectra.
The data and model where all coefficients were set to $\SI{8.5}{mm/rad}$ matches particularly well, while the match is slightly less perfect for $\SI{2.3}{mm/rad}$. This small deviation originates from the other secondary noises included in the LISASim data.  
Additionally, we see that the worst-case noise estimates are approximately a factor of $\sqrt{2}$ larger than the corresponding curves for equal coefficients with equal signs.   
 
Additionally, we compared our analytic models with the results presented in \cite[Fig. 8]{Houba2023-TDI-inf} assuming all coefficients to be \SI{+2.0}{mm/rad} and found a qualitative agreement (the blue curve shows maxima at approximately \SI{2}{mrad/\sqrt{Hz}}, which matches the maxima in \cite[Fig. 8]{Houba2023-TDI-inf}). Please note that for this comparison, white jitter of not \SI{1.6}{nrad/\sqrt{Hz}} (as stated in the paper) but \SI{10}{nrad/\sqrt{Hz}} need to be assumed \cite{HoubaPrivateCom}, except for the MOSA jitter in $\eta$, which is neglected.

%%%%%%%%%%%%%%%%%%%%%%%%%%%%%%%%%%%%%%%%%%%%%%%%%%%%%%%%%%%%%
%%%%%%%%%%%%%%%%%%%%%%%%%%%%%%%%%%%%%%%%%%%%%%%%%%%%%%%%%%%%%
%%%%%%%%%%%%%%%%%%%%%%%%%%%%%%%%%%%%%%%%%%%%%%%%%%%%%%%%%%%%%
\section{Discussion of the assumptions and their implications}
\label{Sec:Implicit-Assumptions+Implications}
Within \Cref{Sec:Derivation-of-1st-order-model,Sec:TTL-noise-estimate}, we have made strong assumptions that are commonly made in the LISA community. We highlight in \Cref{Sec:Contradicting-assumptions} that these quickly result in contradictions. We resolve these contradictions shortly in \Cref{Sec:Resolving-assumption} by highlighting that they hold only for TTL modeled for \gls{TDI}, but not for the TTL in the individual interferometers. We then derive a more complete model for the TTL coupling in the individual interferometers in \Cref{Sec:Extended TTL model}. With this extended model and a new delineation between OB motion $\vec \Delta$ and TTL $N$, we conclude in \Cref {Sec:Delineation-N-vecDelta} the mathematical description for TTL in the individual interferometers vs.\ in \gls{TDI}. Finally, we shortly discuss in \Cref{Sec:Completeness-of-Delta-cancellation} how the phasemeter equations can be adapted to account for imperfect cancellation of OB displacement.

%%%%%%%%%%%%%%%%%%%%%%%%%%%%%%%%%%%%%%%%%%%%%%%%%%%%%%%%%%%%%
%%%%%%%%%%%%%%%%%%%%%%%%%%%%%%%%%%%%%%%%%%%%%%%%%%%%%%%%%%%%%
\subsection{Three contradicting assumptions}
\label{Sec:Contradicting-assumptions}
\begin{figure}%
\includegraphics*[width=\columnwidth, trim = 10 370 0 10]{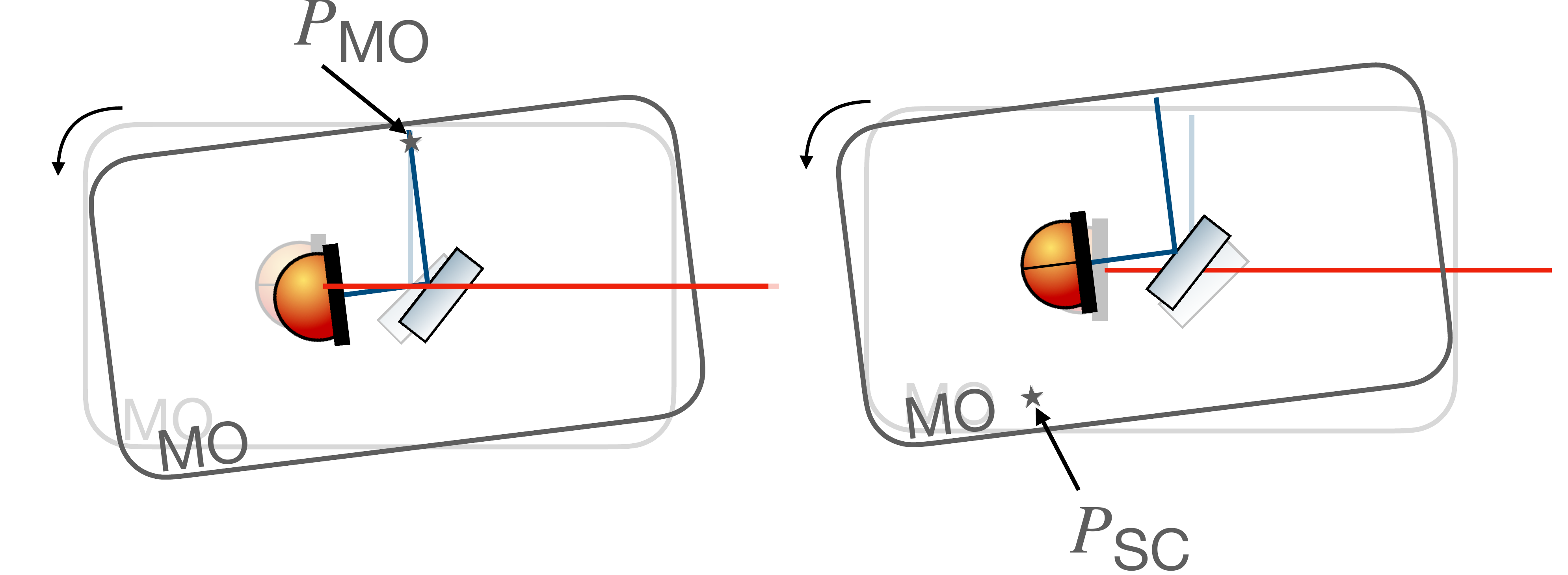}
\caption{Relevance of different pivot points for the MOSA and SC. This sketch illustrates how a rotation by the very same angle around either the pivot $P_\text{MO}$ of the MOSA (left-hand side) or the pivot $P_\text{SC}$ of the \gls{SC} (right-hand side) causes different magnitudes of TTL-coupling. Despite that the MOSA rotates by the same amount in both images, it moves into the received beam indicated by the red trace on the left-hand side, but out of the received beam on the right-hand side. The received beam would, therefore, have to propagate a shorter distance to the diode in the rotation around the MOSA pivot, but a longer distance for the rotation around the SC pivot point. Consequently, MOSA and SC angular jitter need to be modeled with individual coupling coefficients. See \Cref{Sec:Resolving-assumption} for a description of why the models in \Cref{Sec:Derivation-of-1st-order-model,Sec:TTL-noise-estimate} are still valid. Please note: The shown pivot points are placed fairly arbitrarily with the sole purpose of illustrating the described principle.}
\label{Fig:Implication-of-CoRs}
\end{figure}

Let us revisit three assumptions we have made in \Cref{Sec:Derivation-of-1st-order-model,Sec:TTL-noise-estimate}.
%%%
\paragraph{Neglection of the pivot location}
First of all, we have neglected in \Cref{Sec:Derivation-of-1st-order-model} that all SC angular jitter will be occurring with a pivot located in the SC's \gls{CoM}. This pivot location should be considered when mapping SC-jitter into the MOSA-frame.  
Likewise, all MOSA angular jitter will have a certain pivot point. This pivot will be defined by the MOSA hinges, and will not be located in the SC's \gls{CoM}. The location of the pivot is a key parameter affecting the magnitude of particularly the geometric TTL coupling \cite{Hartig2022-G}, such that one can directly deduce: Different locations of the pivot points imply that the coupling of MOSA and SC-jitter will have different coupling factors. This property is illustrated in \Cref{Fig:Implication-of-CoRs},  agrees with the findings of  \cite{Hartig2022-G}, and will likewise be mathematically shown in \Cref{Sec:Extended TTL model}.
% 
 %%%%%%%%
%%%%%%%%
%
%%%
\paragraph{Neglection of TMI TTL}
The angular jitter around a remote pivot point described in the previous paragraph affects the \gls{TMI} in the same way as it affects the \gls{LAI}. Due to the shift of the pivot location against the origin of the coordinate frame, the MOSA is displaced both against the incoming beam in the \gls{LAI}, as well as against the test mass. \\
We know from \cite{Hartig2022-G} that a lateral displacement of $d$\,mm of the pivot against the beam axis causes a TTL coupling of magnitude $d$\,mm/rad (and we will show this again in \Cref{Eq:SC_displ_jitter_mapping,Eq:MO_displ_jitter_mapping} below). Consequently, the TTL contributions originating from lateral displacements of the pivot points could be considerable, and the angular jitter coupling in the \gls{TMI} should not be neglected.

%%%
\paragraph{Neglection of lateral jitter coupling}
We have stated in \Cref{Par:Neglection-of-lateral-jitter-coupling} that the TTL-coupling originating from lateral jitter can be neglected. In fact, this is directly a contradiction to \Cref{Eq:x_SC_mapping_simple}, which states that SC $y$-jitter couples with $\sin(\beta) \approx \pm 0.5$ for $\beta \approx \SI{\pm 30}{\degree}$. According to our definition, this is a type of TTL coupling, since it couples lateral jitter due to a tilt of the SC against the MOSA frame. If we assume lateral SC-jitter levels in the order of \SI{5}{nm/\sqrt{Hz}}, the resulting TTL noise contributions would be about \SI{2.5}{nm/\sqrt{Hz}}, and thereby not at all negligible. On the contrary: this TTL contribution is very high both in the \glspl{LAI} and \glspl{TMI}.

All three shown contradictions originate from the same assumption and are resolved in the next subsection.

%%%%%%%%%%%%%%%%%%%%%%%%%%%%%%%%%%%%%%%%%%%%%%%%%%%%%%%%%%%%%
%%%%%%%%%%%%%%%%%%%%%%%%%%%%%%%%%%%%%%%%%%%%%%%%%%%%%%%%%%%%%
\subsection{Resolving the contradiction: we neglected contributions that cancel in TDI }
\label{Sec:Resolving-assumption}
The contradictions described above are resolved in the following way:
In each of the described cases, a longitudinal motion of the OB is induced. Either by lateral SC jitter in the spacecraft frame, or by angular jitter of the MOSA or SC around remote pivot points. However, OB motion in beam direction is suppressed in the single link readout, and hence also in \gls{TDI} (cf. \Cref{Eq:PM-equations} and the removal of $\vec n \vec \Delta$ in $\check \xi$ \Cref{Eq:xi}). Rather than modeling these elements and canceling them again in \gls{TDI}, they were suppressed already in the original model.

This means, for the case of SC-jitter in $y$-direction, we find indeed considerable coupling to motion in $x$-direction, existent in the individual interferometers. Yet, the resulting phase contribution is suppressed in \gls{TDI} when the signals of the \gls{LAI} and \gls{TMI} are added (see \Cref{Eq:xi,Eq:xi_TTL}). 

Likewise, angular SC- and MOSA-jitters cause considerable OB displacements, resulting in considerable levels of TTL in the individual interferometers. Furthermore, the different pivot points for MOSA and SC angular jitters cause different magnitudes of longitudinal OB displacements. This causes the coupling factors of MOSA and SC angular jitter to be different in the individual interferometers. \\
Yet, both the considerable TTL coupling magnitude, as well as the individual coupling factors for MOSA and SC angular jitters originate from OB motion induced by the jitters. Hence, when the phase changes originating from longitudinal OB motion are suppressed in \gls{TDI}, TTL in the \gls{TMI} becomes comparably small which allowed us to neglect it in \Cref{Sec:TTL-noise-estimate}.
Likewise, the effect of the different pivot points ideally cancels. This allowed us to assume only one factor for the residual coupling of MOSA and SC angular jitters in \Cref{Sec:Derivation-of-1st-order-model,Sec:TTL-noise-estimate}. 

In summary, the TTL in the individual \glspl{LAI} and  \glspl{TMI} is considerably different than in a single link $\check \xi$ or in \gls{TDI}. When modeling the TTL in the individual interferometers, the OB longitudinal motion originating from lateral and angular jitters contributes significantly. The TTL in the TMI is then non-negligible, and lateral SC-jitter coupling cannot be neglected. Also, the couplings of the angular MOSA and SC jitters need to be modeled with different coupling factors. 
This is different for modeling the TTL in \gls{TDI}. For this, we can logically invert the previous sentences: \gls{TMI}-TTL and lateral SC-jitter coupling is negligible, and MOSA and SC angular jitters may be modeled to couple with the same coefficient.

In the next subsections, we shortly derive a mathematical description for the statements made here.

%%%%%%%%%%%%%%%%%%%%%%%%%%%%%%%%%%%%%%%%%%%%%%%%%%%%%%%%%%%%%
%%%%%%%%%%%%%%%%%%%%%%%%%%%%%%%%%%%%%%%%%%%%%%%%%%%%%%%%%%%%%
\subsection{An extended linear model for TTL in individual interferometers}
\label{Sec:Extended TTL model}
Extending the model defined in \Cref{Eq:Complete_model_not_subst_simple} to consider individual coupling factors for MOSA and SC-jitter results in
\begin{subequations}
\begin{alignat}{2}
N^{\varepsilon_{ij}}_{ij} &= &&\sum_\alpha
  \left( 
    c^{\varepsilon_{ij}}_{\alpha_{\text{MO}ij}}\alpha_{\text{MO}ij} 
  + c^{\varepsilon_{ij}}_{\alpha_{\text{SC}i}}\alpha_{\text{SC}i}^{\text{MF}ij} 
 - c^{\varepsilon_{ij}}_{\alpha_{\text{TM}ij}} \alpha_{\text{TM}ij} 
  \right)
 \label{Eq:N_{ij}^{varepsilon_{ij}}}  \\
 N_{ij}^{s_{ij}} &=    &&\sum_\alpha   
  \left(c^{s_{ij}}_{\alpha_{\text{MO}ij}} \alpha_{\text{MO}ij} + c^{s_{ij}}_{\alpha_{\text{SC}i}} \alpha_{\text{SC}i}^{\text{MF}ij} \right)  
\label{Eq:N_{ij}^{s_{ij}}} \\
N_{ji:ij}^{s_{ij}}  &=   &&\sum_\alpha 
 \left( c^{s_{ij}}_{\alpha_{\text{MO}ji}}\alpha_{\text{MO}ji:ij} + c^{s_{ij}}_{\alpha_{\text{SC}j}} \alpha^{\text{MF}_{ji}} _{\text{SC}j:ij}   \right)
\label{Eq:N_{ji:ij}^{s_{ij}}} 
\end{alignat}
\label{Eq:Complete_model_not_subst}
\end{subequations}

As a next step, we now repeat the computation for the mapping from SC-jitter into MF but consider this time that the SC jitters about its \gls{CoM}. 
We start with an explicit definition of the location of the origins of the \gls{MF} and \gls{SF}, which we omitted in \Cref{Sec: Jitters+coord} like it is usually done. For the complete model, we now define the origin of the \gls{SF} to be located in the time-averaged location of the \gls{SC}'s center of mass. All \gls{SC} angular jitter is then described by rotation matrices around the coordinate system's origin.\\
The origin of the \gls{MF} is the point around which a MOSA rotation would not cause a geometric TTL response in the \gls{LAI}. Assuming ideal imaging, the origin would be located in the telescope's large pupil. 
In \Cref{Sec: Jitters+coord} we defined the MF to be inertial on short time scales by placing its origin on a hypothetical perfectly noise-free orbit. In order to achieve this behavior of the MF origin, we assume its location to be an average over a short time period of interest.

The origin of the \gls{SF}, i.e.\ the spacecraft's \gls{CoM}, is now shifted by a vector $\vec P_\text{SCC}$ from the MF origin. 
SC angular jitter is then described by rotation transformations $\mathbf R$:
\begin{subequations}
\begin{align}
	\mathbf{R}( \theta_\text{SC}, \vec X_\text{SF}, \vec P_\text{SCC}) =:\mathbf{R}_\theta \\
	\mathbf{R}( \eta_\text{SC}, \vec Y_\text{SF}, \vec P_\text{SCC}) =:\mathbf{R}_\eta \\
	\mathbf{R}( \phi_\text{SC}, \vec Z_\text{SF}, \vec P_\text{SCC}) =:\mathbf{R}_\phi
\end{align}
\end{subequations}
defined by
\begin{subequations}
\begin{align}
	\mathbf{R}( \alpha, \vec X, \vec P_\text{SCC}) [\vec x]  \hspace{-1.5cm} &\nonumber\\
	&:=	R(\alpha, \vec X) \, \vec x  - (R( \alpha, \vec X) - \mathbf E)  \vec P_\text{SCC} \\
	&=	R(\alpha, \vec X) \, (\vec x-\vec P_\text{SCC}) + \vec P_\text{SCC} \;,
\end{align}
\end{subequations}
where $\vec X$ is the rotation axis, $\alpha$ the angle through which it is rotated, and $\mathbf E$ is the $3 \times 3$ identity matrix. 

We assume again that the rotation matrices can be linearized since the magnitude of the angular jitter is small. Consequently, the effect of SC jitter in all three angular degrees of freedom will affect an arbitrary vector $\vec x$ in MF by
\begin{align}
	\mathbf{R}_\theta[	\mathbf{R}_\eta[   \mathbf{R}_\phi[  \vec x]]]  =\hspace{-2.4cm}& \nonumber\\
	&R(\theta_\text{SC}, \vec X_\text{SF}) \, R(\eta_\text{SC}, \vec Y_\text{SF}) \, R(\phi_\text{SC}, \vec Z_\text{SF})  \, \vec x  - \nonumber\\
	&(R(\theta_\text{SC}, \vec X_\text{SF}) \, R(\eta_\text{SC}, \vec Y_\text{SF}) \, R(\phi_\text{SC}, \vec Z_\text{SF}) - \mathbf E)  \vec P_\text{SCC} \;.
\label{Eq:complete_RotTrafo}
\end{align}
The first summand has already been evaluated in \Cref{Eq:SCRotMatrWOdispl} to \Cref{Eq:angular_mapping}.
The second term describes the displacement noise in the MOSAs and originates from the center of rotation for the SC angular jitters to not coincide with the origins of the two MFs. This displacement evaluates to:
\begin{widetext}
\begin{align}
(R(\theta_\text{SC}, \vec X_\text{SF}) \, R(\eta_\text{SC}, \vec Y_\text{SF}) \, R(\phi_\text{SC}, \vec Z_\text{SF}) - \mathbf E)  \vec P_\text{SCC}=
	 	\begin{pmatrix} 
	 0 	& -\phi_\text{SC} & c_\beta \eta_\text{SC}  - s_\beta \theta_\text{SC}  \\
	 \phi_\text{SC} & 0 & -(  s_\beta \eta_\text{SC} + c_\beta \theta_\text{SC}) \\
	-(c_\beta \eta_\text{SC}  - s_\beta \theta_\text{SC} )&  s_\beta \eta_\text{SC} + c_\beta \theta_\text{SC}& 0 
	\end{pmatrix}  \vec P_\text{SCC} \;.
	\label{Eq:displacement_due_to_SC_rot}
\end{align}
\end{widetext}
If we define $\vec P_\text{SCC} =: (P_{\text{SCC}x},P_{\text{SCC}y},P_{\text{SCC}z})^T$, SC-jitter will cause displacements in the MF given by:
\begin{subequations}
\begin{align}
	x_\text{SC}^\text{MF} &=  \cos(\beta) x_\text{SC} + \sin(\beta) y_\text{SC}  \nonumber\\ 
	& - P_{\text{SCC}y} \phi_\text{SC} +  P_{\text{SCC}z} (c_\beta \eta_\text{SC}  - s_\beta \theta_\text{SC} ) \label{Eq:x_SC_mapping} \\
	y_\text{SC}^\text{MF} &= - \sin(\beta) x_\text{SC} + \cos(\beta) y_\text{SC} \nonumber\\ 
	&+ P_{\text{SCC}x} \phi_\text{SC} - P_{\text{SCC}z} (s_\beta \eta_\text{SC}  + c_\beta \theta_\text{SC} )   \label{Eq:y_SC_mapping}  \\
	z_\text{SC}^\text{MF} &= z_\text{SC} - P_{\text{SCC}x}(c_\beta \eta_\text{SC}  - s_\beta \theta_\text{SC} ) \nonumber \\
	& + P_{\text{SCC}y} (s_\beta \eta_\text{SC}  + c_\beta \theta_\text{SC} ) \label{Eq:z_SC_mapping}
\end{align}
\label{Eq:SC_displ_jitter_mapping}%
\end{subequations}
which naturally depends on the pivot point $\vec P_\text{SCC}$. 
Please note that even though there is, of course, only one SC CoM, the $\vec P_\text{SCC}$ pointing to it will depend on the coordinate frame, such that $\vec P_\text{SCC}$ will be different for left- and right-hand side MOSAs.

We can now consider that also MOSA jitter occurs with a center of rotation shifted by $\vec P_\text{MOC} = (P_{\text{MOC}x},P_{\text{MOC}y},P_{\text{MOC}z})^T$ against the origin of the \gls{MF}. The mathematical description is the very same as for SC-jitter, except that there is no mapping with $\beta$ into the frame. The coupling of MOSA jitter is, therefore, easily derived from \Cref{Eq:SC_displ_jitter_mapping} by adjusting the indices for MOSA jitter and setting $\beta=0$:
\begin{subequations}
\begin{align}
	x_\text{MO} &=  x_\text{MO}^{ \text{PT}}
	- P_{\text{MOC}y} \phi_\text{MO} +  P_{\text{MOC}z}  \eta_\text{MO}
	\label{Eq:x_MO_mapping} \\
	y_\text{MO} &=  y_\text{MO}^{\text{PT}}
	+ P_{\text{MOC}x}\phi_\text{MO} - P_{\text{MOC}z}  \theta_\text{MO} 
	\label{Eq:y_MO_mapping}  \\
	z_\text{MO}&= z_\text{MO}^{ \text{PT}}
	- P_{\text{MOC}x} \eta_\text{MO}   + P_{\text{MOC}y}  \theta_\text{MO}  \;.
	 \label{Eq:z_MO_mapping}
\end{align}
\label{Eq:MO_displ_jitter_mapping}%
\end{subequations}
Here, the upper index PT stands for purely translational and indicates the part of $(x_\text{MO},y_\text{MO},z_\text{MO})^T$ that is not tilt induced.

Comparing \Cref{Eq:x_SC_mapping,Eq:x_MO_mapping} with \Cref{Eq:gen_ttl_complete,Eq:Complete_model_not_subst}, we have modeled explicitly several contributions to coupling factors. Considering that MF $x$-direction is defined to be facing towards the received beam, we find that a MOSA displacement in $x-$direction decreases the optical pathlength in the \gls{LAI} but increases it in the TMI. We then use the sign conventions used for deriving the phasemeter equations (cf. \Cref{Sec:Initial_Phaseter_eqns}), i.e.\ the phase increases if the optical pathlength of the interferometer's measurement beam increases. This gives the following contributions to the \gls{LAI} and \gls{TMI} coupling factors:
\begin{subequations}
\begin{align}
	 \sin(\beta)  \rightarrow & \{- c^s_{y_\text{SC}},  c^\varepsilon_{y_\text{SC}} \}    	\\
	- P_{\text{SCC}y}  \rightarrow & \{-c^s_{\phi_\text{SC}},c^\varepsilon_{\phi_\text{SC}} \}    \\
	- P_{\text{MOC}y}  \rightarrow & \{-c^s_{\phi_\text{MO}} ,c^\varepsilon_{\phi_\text{MO}} \}   \\ 
	  P_{\text{SCC}z} c_\beta \rightarrow &  \{-c^s_{\eta_\text{SC}},c^\varepsilon_{\eta_\text{SC}}   \}  \\
	  P_{\text{MOC}z}  \rightarrow &\{ -c^s_{ \eta_\text{MO}},c^\varepsilon_{ \eta_\text{MO}} \}  \\
	- P_{\text{SCC}z}  s_\beta    \rightarrow & \{ -c^s_{\theta_\text{SC}},c^\varepsilon_{\theta_\text{SC}} \} \,.   
\end{align}
\label{Eq:TTL_in_mapping}%
\end{subequations}
Here, arrows indicate that these mappings are only one particular kind of geometric contribution to the total coupling factors and not necessarily the full coupling coefficients that need to be considered. 

Since the jittering \glspl{MOSA} (or \glspl{SC}) simultaneously send out beams towards a remote \gls{SC} and receive light from the same remote \gls{SC}, the shown jitters equally affect both the transmitted and received light in the \glspl{LAI}. This means, \Cref{Eq:TTL_in_mapping} holds both for receiver jitter coefficients (i.e.\ \gls{LAI} with matching upper and lower first index, e.g.\ $c^{s_{ij}}_{\phi_{\text{SC}i}},c^{s_{ij}}_{\phi_{\text{MO}{ij}}}$), and transmitter jitter coefficients (i.e.\ \gls{LAI} with upper and lower first index not matching, e.g.\ $c^{s_{ij}}_{\phi_{\text{MO}ji:ij}}, c^{s_{ij}}_{\phi_{\text{SC}j:ij}} $).

\Cref{Eq:TTL_in_mapping} phrases mathematically several statements we have previously made:
\begin{itemize}
	\item Lateral jitter coupling is a high TTL contribution in the individual interferometers.
	\item Due to the different pivot locations, we expect different coupling factors for MOSA and SC jitters when modeling the TTL in individual interferometers.
	\item For every mm of lateral displacement of the pivot point against the beam axis, an \SI{1}{mm/rad} coupling coefficient is found, meaning that these lateral displacements of the pivot point locations cause strong TTL coupling, and might deviate considerably for the MOSA and SC jitters.
	\item The \gls{LAI} and \gls{TMI} are subject to the very same coupling but with inverse signs. This means that individually seen, the \glspl{TMI} might have high magnitudes of TTL coupling that should not be neglected. 
\end{itemize}

The result that the TTL coupling in the  \gls{LAI} and \gls{TMI} have identical magnitudes holds only under the given assumptions, particularly that the MOSA displacement is sensed identically by the ISI and TMI beams which map along the very same MOSA-frame $x$-axis. Given the naturally occurring small levels of misalignments, we expect deviations, which we will discuss further in \Cref{Sec:Completeness-of-Delta-cancellation}.

%%%%%%%%%%%%%%%%%%%%%%%%%%%%%%%%%%%%%%%%%%%%%%%%%%%%%%%%%%%%%
%%%%%%%%%%%%%%%%%%%%%%%%%%%%%%%%%%%%%%%%%%%%%%%%%%%%%%%%%%%%%
\subsection{Delineation: TTL $N$-term vs OB motion $\vec \Delta$}
\label{Sec:Delineation-N-vecDelta}

%
%\Structure{Relevance and meaning of $x_\text{SC}^\text{MF}$}
In the previous subsection, we have derived $x_\text{MO}$ and $x_\text{SC}^\text{MF}$ which in sum represent the total MOSA-$x$ displacement caused by SC and MOSA jitter. We can interpret this sum as the longitudinal displacement noise $\vec{n}_{ij}\vec\Delta_{ij}$, or for the transmitter jitter case as $\vec{n}_{ji}\vec\Delta_{ji:ij}$. 
Yet, we have shown that it contains several TTL contributions and could, therefore, likewise be partly attributed to the $N$-terms modeled here. This means it needs to be clearly defined which effects are placed into $\vec n \vec \Delta$, and which into TTL noise $N$ to avoid double-counting.

%\Structure{Delineation $\vec\Delta$ vs $N_\Delta$ for this paper}
There are several possibilities on how to avoid double counting. We resolve this here by defining 
\begin{equation}
	\vec \Delta \vec n = (x_\text{MO}^{\text{PT}}, y_\text{MO}^{\text{PT}}, z_\text{MO}^{\text{PT}}) \cdot (1,0,0)^T
\end{equation}
to hold the pure translations of the MOSA,
while placing the contributions originating from angular and lateral jitters into $N$, marking them explicitly as OB motion by an underscore $\vec \Delta$:
\begin{subequations}
\begin{align}
	N_{\vec \Delta} &= 
	\begin{pmatrix} 
	x_\text{MO}+x_\text{SC}^\text{MF}  - x_\text{MO}^{\text{PT}}\\
	 y_\text{MO}+y_\text{SC}^\text{MF} - y_\text{MO}^{\text{PT}}\\
	  z_\text{MO}+z_\text{SC}^\text{MF} -z_\text{MO}^{\text{PT}}
  	\end{pmatrix} ^T
	\cdot
	\begin{pmatrix} 
		1 \\ 0\\ 0
	\end{pmatrix} \\
	& = x_\text{MO}+x_\text{SC}^\text{MF}  - x_\text{MO}^{\text{PT}}\;.
\end{align}
\end{subequations}

Now, $N_{\vec \Delta}$ contains significant TTL contributions that exist in the individual interferometers but cancel in \gls{TDI}. This is best seen in \Cref{Eq:TTL_in_mapping}, which shows that the contributions to $N_{\vec \Delta}$ are equal but of opposite sign in the \gls{LAI} and \gls{TMI}, so that
\begin{subequations}
\begin{align}
	N_{\vec \Delta, ij}^{s_{ij}}    &= 	- N_{\vec \Delta, ij}^{\varepsilon_{ij}}  \\
	N_{\vec \Delta, ji:ij}^{s_{ij}} &= 	- N_{\vec \Delta, ji:ij}^{\varepsilon_{ji}} \;,
	\label{Eq:N_Deltas}
\end{align}
\end{subequations}
and all four terms cancel in $\check \xi_{ij}^\text{TTL}$ in \Cref{Eq:xi_TTL}. Thereby, the TTL contributions related to MOSA shifts in $x$-direction cancel. These are the terms listed in \Cref{Sec:Contradicting-assumptions} and discussed in \Cref{Sec:Resolving-assumption}:  the lateral SC-jitter contributions as well as the pivot-dependent contributions (cf. \Cref{Eq:TTL_in_mapping} which implies \Cref{Eq:N_Deltas}). 

This concludes the mathematical description of the statements made in \Cref{Sec:Resolving-assumption}.

%%%%%%%%%%%%%%%%%%%%%%%%%%%%%%%%%%%%%%%%%%%%%%%%%%%%%%%%%%%%%
%%%%%%%%%%%%%%%%%%%%%%%%%%%%%%%%%%%%%%%%%%%%%%%%%%%%%%%%%%%%%
\subsection{Completeness of the cancellation of OB-motion in TDI}
\label{Sec:Completeness-of-Delta-cancellation}

The discussed suppression of OB jitter in \gls{TDI} is based on the phasemeter equations defined in \Cref{Eq:PM-equations}. In these phasemeter equations, it is assumed that OB jitter $\vec \Delta$ couples with an identical magnitude but opposite sign in the \gls{LAI} and \gls{TMI}. This is mathematically phrased by mapping the vectorial OB motion along the very same direction $\vec n$ (e.g.\ $\vec n_{12}$ is used to map $\vec \Delta$ both in \Cref{Eq:PM-LAI12,Eq:PM-TMI12}). This means the full cancellation of OB longitudinal motion in \gls{TDI} assumes a perfect angular alignment of the \gls{TMI} and \gls{LAI} beam axes. 

In reality, the two \glspl{LAI} and \glspl{TMI} beam axes will be coaligned as well as experimentally possible, which allows a small residual angle. The mapping in the phasemeter equations should, therefore, be done with the individual beam directions $\vec n^s, \vec n^\varepsilon$. However, the small displacement noise contributions from this angular beam misaligment can easily be expressed by an additional term each in $N^\varepsilon, N^s$ of the type $\delta n^s \vec \Delta,- 2 \delta n^\varepsilon \vec \Delta$ with $\vec n^\varepsilon = \vec n + \delta n^\varepsilon$ and $\vec n^s = \vec n + \delta n^s$.
The phasemeter equations, therefore, do not need to be adapted. Instead, the corresponding term is simply one of the many contributions in modeling the TTL $N$-term.

%%%%%%%%%%%%%%%%%%%%%%%%%%%%%%%%%%%%%%%%%%%%%%%%%%%%%%%%%%%%%
%%%%%%%%%%%%%%%%%%%%%%%%%%%%%%%%%%%%%%%%%%%%%%%%%%%%%%%%%%%%%
%%%%%%%%%%%%%%%%%%%%%%%%%%%%%%%%%%%%%%%%%%%%%%%%%%%%%%%%%%%%%
\section{TTL in data analysis}
\label{Sec:DA}
Within this paper, we have derived TTL models that allow the estimation of TTL noise. Within this section, we will now discuss implications for data analysis, i.e.\ the fit and noise subtraction which was tested for instance in \cite{Paczkowski2022}. In \Cref{Sec:Model-Applicability}, we discuss the differences between the models presented here, and those used for fitting and subtracting in data post-processing. In \Cref{Sec:Mixing_of_TMI-and-LAI-TTL-in-TDI} we show that the models presented here imply that the TTL coefficients of the \gls{LAI} and \gls{TMI} are inseparable in data analysis - which, however, is not a problem. 
Finally, we show in \Cref{Sec:Mixing-local+remote-terms}  that there are pairs of coefficients for transmitter and receiver jitters which are inseparable in data analysis if only one of the \gls{TDI} Michelson combinations is used. However, the correlation resolves when all three \gls{TDI} Michelson combinations are used for coefficient estimation.

%%%%%%%%%%%%%%%%%%%%%%%%%%%%%%%%%%%%%%%%%%%%%%%%%%%%%%%%%%%%%
%%%%%%%%%%%%%%%%%%%%%%%%%%%%%%%%%%%%%%%%%%%%%%%%%%%%%%%%%%%%%
%%%%%%%%%%%%%%%%%%%%%%%%%%%%%%%%%%%%%%%%%%%%%%%%%%%%%%%%%%%%%
\subsection{Noise estimation models vs data analysis models}
\label{Sec:Model-Applicability}
%%%%%%%%%%%%%%%%%%%%%%%%%%%%%%%%%%%%%%%%%%%%%%%%%%%%%%%%%%%%%
The models presented within this paper can be understood as a description of how angular and lateral jitters couple in mother nature. 
There is a second type of TTL model that slightly differs from the presented mother nature type of model: the models for data analysis. Let us, therefore, briefly describe how the two models differ.

The mother nature model takes angular and lateral component jitters as well as coupling factors as input and describes the resulting TTL noise. It is used for modeling the magnitude of the TTL noise in an interferometer, a single link, or in \gls{TDI}.

The data analysis model uses interferometric readout signals instead of jitters. That means, all angular jitters are typically replaced by DWS-signals, and all lateral jitters by GRS signals. These signals are usually assumed to be calibrated to best estimate the jitters, but they contain readout noise.

While we can differ in a mother nature model between SC jitter relative to FS, and MOSA jitter relative to the SC, we cannot directly do such a distinction in a data analysis model. This is because the DWS signals sense the beam tilt caused by the total MOSA motion relative to free space. The same holds for the GRS signals that sense only the total motion of the MOSA relative to the free-falling test mass.

Given the just-discussed relation between a mother nature model to a data analysis model, the data analysis models corresponding to \Cref{Eq:Complete_model_not_subst_simple,Eq:reduced_TTL_model} can be easily derived. 
However, an equivalent data analysis model to our extended model for the TTL in the individual interferometers \Cref{Eq:Complete_model_not_subst} cannot be easily written down since the application of individual coupling coefficients to  MOSA and SC jitters is not directly possible in a data analysis model like the one used in \cite{Paczkowski2022}.
%%%%%%%%%%%%%%%%%%%%%%%%%%%%%%%%%%%%%%%%%%%%%%%%%%%%%%%%%%%%%
%%%%%%%%%%%%%%%%%%%%%%%%%%%%%%%%%%%%%%%%%%%%%%%%%%%%%%%%%%%%%
\subsection{Mixing of LAI and TMI TTL noise contributions in $\xi$ and TDI-$X$}
\label{Sec:Mixing_of_TMI-and-LAI-TTL-in-TDI}
%%%%%%%%%%%%%%%%%%%%%%%%%%%%%%%%%%%%%%%%%%%%%%%%%%%%%%%%%%%%%
We have stated in \Cref{Sec:TTL-reduced} that the TTL noise contributions in \gls{TDI} originating from the \gls{TMI} are expected to be minor, and we neglected it for the noise estimates in \Cref{Sec:Noise-Estimates}. However, even if the \glspl{TMI} contribute significantly less noise to the \gls{TDI} observables, they will still contribute to the total noise.
This TTL noise of the \gls{TMI} mixes in \gls{TDI} with the noise contributions from the \gls{LAI} in a way that the origin becomes inaccessible. This means it cannot be distingished in the \gls{TDI} observable what part of the noise originates from the \gls{LAI} and what from \gls{TMI}. This can be seen with either the simplified TTL model \Cref{Eq:Complete_model_not_subst_simple} or the extended one for the contributions in individual interferometers \Cref{Eq:Complete_model_not_subst}. We show this here for the extended model.

We have shown in \Cref{Eq:X2} that the TTL coupling noise in the \gls{TDI}-$X_2$-variable is a fairly simple combination of four single link TTL contributions $\check \xi^\text{TTL}$. Each of these $\check \xi^\text{TTL}$ is defined by the linear combination of four $N$-terms shown in \Cref{Eq:xi_TTL}. These noise terms can be paired into non-delayed contributions in MOSA$ij$-jitter noise and delayed MOSA$ji$-jitter contributions, and thereby into pairs with identical lower indices:
\begin{align}
    \check \xi_{ij}^\text{TTL} =  k_{ji:ij} \bigg[  \left( N^{s_{ij}}_{ij} 
    + \frac{1}{2} N^{\varepsilon_{ij}}_{ij} \right)
    +  \left(    N^{s_{ij}}_{ji:ij} 
    + \frac{1}{2}  N^{\varepsilon_{ji}}_{ji:ij} \right) \bigg] 
     \label{Eq:xi_TTL_sorted} \;.
\end{align}
We can now use \Cref{Eq:Complete_model_not_subst} to analyze the individual pairs for each degree of freedom $\alpha \in \{\phi,\eta,\theta,y,z\}$:
\begin{widetext}
\begin{subequations}
\begin{alignat}{3}
 	N^{s_{ij}}_{ij}  + \frac{1}{2} N^{\varepsilon_{ij}}_{ij}  &=  &&\sum_\alpha  \bigg\{&&
   \left( c^{s_{ij}}_{\alpha_{\text{MO}ij}} \alpha_{\text{MO}ij} 
   +c^{s_{ij}}_{\alpha_{\text{SC}i}}\alpha_{\text{SC}i}^{\text{MF}ij} \right)  +
 \frac{1}{2}\bigg[
  c^{\varepsilon_{ij}}_{\alpha_{\text{MO}ij}} \alpha_{\text{MO}ij} 
  +c^{\varepsilon_{ij}}_{\alpha_{\text{SC}i}}\alpha_{\text{SC}i}^{\text{MF}ij} 
  -	c^{\varepsilon_{ij}}_{\alpha_{\text{TM}ij}} \alpha_{\text{TM}ij} 
  \bigg] \bigg\} \\
  &=: && \sum_\alpha  \bigg\{ &&
     c^{{ij}}_{\alpha_{\text{MO}ij}} \alpha_{\text{MO}ij} 
    +c^{{ij}}_{\alpha_{\text{SC}i}}\alpha_{\text{SC}i}^{\text{MF}ij} 
    -c^{ij}_{\alpha_{\text{TM}ij}} \alpha_{\text{TM}ij}  \bigg\} \\
\label{Eq:Ns+Neps-local}
	N^{s_{ij}}_{ji:ij} + \frac{1}{2}  N^{\varepsilon_{ji}}_{ji:ij}  &= 
	 &&\sum_\alpha  \bigg\{ &&  
	   \left( c^{s_{ij}}_{\alpha_{\text{MO}{ji}}} \alpha_{\text{MO}_{ji:ij}} 
   +c^{s_{ij}}_{\alpha_{\text{SC}j}}\alpha_{\text{SC}_{j:ij}}^{\text{MF}ij} \right)  +    \frac{1}{2} \bigg[
   c^{\varepsilon_{ji}}_{\alpha_{\text{MO}{ji}}} \alpha_{\text{MO}{ji:ij}} 
  +c^{\varepsilon_{ji}}_{\alpha_{\text{SC}{j}}}\alpha_{\text{SC}{j:ij}}^{\text{MF}{ji}} 
  -c^{\varepsilon_{ji}}_{\alpha_{\text{TM}{ji}}} \alpha_{\text{TM}{ji:ij}} 
  \bigg] \bigg\} \\
   &=: &&\sum_\alpha  \bigg\{ &&
   c^{ji}_{\alpha_{\text{MO}{ji}}} \alpha_{\text{MO}{ji:ij}} 
  +c^{ji}_{\alpha_{\text{SC}{j}}}\alpha_{\text{SC}{j:ij}}^{\text{MF}{ji}} 
   - c^{ji}_{\alpha_{\text{TM}{ji}}} \alpha_{\text{TM}{ji:ij}} 
  \bigg\}\,. %
\end{alignat}
\label{Eq:single-link-mixing-TMI+LAI-c}
\end{subequations}
\end{widetext}
Here, we defined total coupling coefficients, which are the observable coefficients during the mission:
\begin{subequations}
\begin{align}
	c^{ij}_{\alpha_{\text{TM}ij}} 	&:= \frac{1}{2} c^{\varepsilon_{ij}}_{\alpha_{\text{TM}ij}} \\
	c^{{ij}}_{\alpha_{\text{MO}ij}} 	&:= c^{s_{ij}}_{\alpha_{\text{MO}ij}} +\frac{1}{2}c^{\varepsilon_{ij}}_{\alpha_{\text{MO}ij}} \\
	c^{{ij}}_{\alpha_{\text{SC}i}} 	&:= c^{s_{ij}}_{\alpha_{\text{SC}i}} +\frac{1}{2}c^{\varepsilon_{ij}}_{\alpha_{\text{SC}i}} \\
	c^{ji}_{\alpha_{\text{TM}{ji:ij}}}	&:= \frac{1}{2} c^{\varepsilon_{ji}}_{\alpha_{\text{TM}{ji:ij}}}\\
	c^{ji}_{\alpha_{\text{MO}{ji:ij}}}	&:= c^{s_{ij}}_{\alpha_{\text{MO}{ji:ij}}} + \frac{1}{2}c^{\varepsilon_{ji}}_{\alpha_{\text{MO}{ji:ij}}} \\
	c^{ji}_{\alpha_{\text{SC}{j:ij}}} &= c^{s_{ij}}_{\alpha_{\text{SC}j:ij}} + \frac{1}{2} c^{\varepsilon_{ji}}_{\alpha_{\text{SC}{j:ij}}} \;.
\end{align}
\end{subequations}
This can be further simplified if we assume again that the coefficients of \gls{MOSA} and \gls{SC} jitter match in \gls{TDI}, i.e.\ $c^{{ij}}_{\alpha_{\text{MO}ij}}  = c^{{ij}}_{\alpha_{\text{SC}i}} $ and $c^{ji}_{\alpha_{\text{MO}{ji:ij}}} = c^{ji}_{\alpha_{\text{SC}{j:ij}}}$.

A distinction between the TMI and LAI coefficients contributing to the total contributions is not expected to be possible because they are multiplying the very same jitter. This means that for every degree of freedom, the TTL coupling coefficients of the TMI and LAI add up in the \gls{TDI} observables, and only the combined effect will be measured. The only exception is the coefficient for TM jitter. However, TM-jitter is expected to be a very minor motion, such that this coefficient is not expected to be measurable and is written here rather for completeness. 

If we neglect \gls{TM} jitter, we can, therefore, simplify the TTL in a single link $\check \xi^\text{TTL}_{ij}$ to
\begin{align}
    \check \xi_{ij}^\text{TTL} =  k_{ji:ij} \bigg[   N^{{ij}}_{ij}     
    +     N^{{ij}}_{ji:ij} 
     \bigg] \;.
     \label{Eq:xi_TTL_no-LAI-TMI-distinction} 
\end{align}
Since this finding holds for every individual single link, it means that it holds likewise for every \gls{TDI} combination built from linear combinations of these single links, so in particular it holds for \gls{TDI}-$X$, $Y$, and $Z$. A clear delineation of the TTL contributions of the \gls{TMI} from the \gls{LAI} is, therefore, expected to be not possible from LISA data.
However, this is not considered a problem. \\
There is no need to separate the effects, only the need to suppress or minimize the total coupling coefficient. This can be achieved either by minimizing each individual TTL coefficient ($c^{s}=c^{\varepsilon} = 0$) or by minimizing the sum ($c^{s} = - c^{\varepsilon}$). In the latter case, it is a design choice, whether TTL mitigation strategies are implemented and applied to tune the TMI coefficients or the LAI coefficients. As described in \cite{Paczkowski2022}, the current planning favors tuning the LAI coupling coefficients.

%%%%%%%%%%%%%%%%%%%%%%%%%%%%%%%%%%%%%%%%%%%%%%%%%%%%%%%%%%%%%
%%%%%%%%%%%%%%%%%%%%%%%%%%%%%%%%%%%%%%%%%%%%%%%%%%%%%%%%%%%%%
\subsection{Mixing of local and remote jitter terms in TDI~$X$, $Y$, $Z$}
\label{Sec:Mixing-local+remote-terms}
The \gls{TDI}-$X$ observable contains two combinations of single link TTL contributions $\check \xi$:
$	(\check \xi^\text{TTL}_{13} + \delay{13} \check \xi^\text{TTL}_{31})$
and 
$(\check \xi^\text{TTL}_{12} + \delay{12} \check \xi^\text{TTL}_{21}) $.
Each of these causes indistinguishability of receiver and transmitter jitter coefficients, i.e.\ $c^{{ji}}_{\alpha_{ji}}, 
c^{{ij}}_{\alpha_{{ji}}}$, principally for $\alpha \in y,z,\phi,\eta,\theta$. Following the assumptions and description in \Cref{Sec:TTL-reduced}, however, primarily for $\alpha \in \phi,\eta$.

We can quickly find this correlation for an arbitrary sum of links $ij$ and $ji$, under the assumption of approximately equal $k_{ij},k_{ji}$ and negligible \gls{TM}-jitter relative to FS. For this, we use the combined TTL-$N$-terms of the \gls{LAI} and \gls{TMI} from \Cref{Eq:xi_TTL_no-LAI-TMI-distinction}:
\begin{subequations}
	\begin{align}
	\check \xi^\text{TTL}_{ij} + \delay{ij} \check \xi^\text{TTL}_{ji} & =
	 \bigg[   N^{{ij}}_{ij}   +     N^{{ij}}_{ji:ij}   \bigg]
	+\delay{ij}
	 \bigg[   N^{{ji}}_{ji}   +     N^{{ji}}_{ij:ji}   \bigg]
	\\
	&=    N^{{ij}}_{ij}   +     \delay{ij} \left ( N^{{ij}}_{ji}   +  N^{{ji}}_{ji}  \right) +   \delay{iji}  N^{{ji}}_{ij}    \;.
	\end{align}
\end{subequations}
Here, $ \delay{ij} \left( N^{{ij}}_{ji}     +  N^{{ji}}_{ji} \right)$ makes the involved coefficients indistinguishable for data analysis because they multiply the very same jitters with the same delays:
\begin{align}
	 N^{{ij}}_{ji}   +  N^{{ji}}_{ji}  =  &
	  \sum_\alpha  
	   \left( c^{{ij}}_{\alpha_{\text{MO}{ji}}} \alpha_{\text{MO}_{ji}} 
   +c^{{ij}}_{\alpha_{\text{SC}j}}\alpha_{\text{SC}_{j}}^{\text{MF}ji} \right)    \nonumber \\
	+& \sum_\alpha  
   \left( c^{{ji}}_{\alpha_{\text{MO}ji}} \alpha_{\text{MO}ji} 
   +c^{{ji}}_{\alpha_{\text{SC}j}}\alpha_{\text{SC}j}^{\text{MF}ji} \right)  
\end{align}
resulting for every MOSA degree of freedom $\alpha$ in $c^{{ji}}_{\alpha_{\text{MO}ji}}, c^{{ij}}_{\alpha_{\text{MO}{ji}}}$ to become indistinguishable in data analysis, and likewise the SC jitter coefficients $\alpha$ in $c^{{ji}}_{\alpha_{\text{SC}j}}, c^{{ij}}_{\alpha_{\text{SC}{j}}}$ become indistinguishable. If we assume again that the coefficients for MOSA and SC jitter are identical in the combined \gls{TTL} $N$-term, we find the initially stated indistinguishability of $c^{{ji}}_{\alpha_{ji}}, 
c^{{ij}}_{\alpha_{{ji}}}$.

This indistinguishability, however, is different from the one of \gls{LAI} and \gls{TMI} coefficients. In every Michelson combination, it occurs for jitters of the \glspl{MOSA} forming the end mirrors of the virtual Michelson interferometer. This means there is the following list of correlations and degeneracies: \\
\begin{center}
\setstretch{1.5}
\begin{tabular} {l l c r}
 TDI-$X$: &  $ c^{{31}}_{\alpha_{31}}, c^{{13}}_{\alpha_{{31}}}$  &\; and \quad \;  & $ c^{{21}}_{\alpha_{21}}, c^{{12}}_{\alpha_{{21}}}$ \\
 TDI-$Y$: &  $ c^{{32}}_{\alpha_{32}}, c^{{23}}_{\alpha_{{32}}}$  &\; and \quad \;  & $ c^{{12}}_{\alpha_{12}}, c^{{21}}_{\alpha_{{12}}}$ \\
 TDI-$Z$: &  $ c^{{23}}_{\alpha_{23}}, c^{{32}}_{\alpha_{{23}}}$  & \; and \quad \;  & $ c^{{13}}_{\alpha_{13}}, c^{{31}}_{\alpha_{{13}}}$ 
\end{tabular}
\end{center}
Consequently, a correlation of the stated pairs of coefficients will be found when estimating these from one \gls{TDI} variable, i.e.\ $X$, or $Y$, or $Z$. However, since the pairs are different in the different \gls{TDI} Michelson observables, the coefficients are distinguishable when all \gls{TDI}~$X,Y,Z$ are jointly used for fitting the coefficients in data post-processing. 
Please note that the coefficients could likewise be recovered from other sets of TDI variables, such as the set A,E,T. This surely holds for any set of TDI variables that can be expressed as a linear combination of TDI X,Y,Z from which the TDI variables X,Y,Z can be recovered by inversion.

%%%%%%%%%%%%%%%%%%%%%%%%%%%%%%%%%%%%%%%%%%%%%%%%%%%%%%%%%%%%%
%%%%%%%%%%%%%%%%%%%%%%%%%%%%%%%%%%%%%%%%%%%%%%%%%%%%%%%%%%%%%
%%%%%%%%%%%%%%%%%%%%%%%%%%%%%%%%%%%%%%%%%%%%%%%%%%%%%%%%%%%%%
\section{Summary and conclusions}
\label{Sec:Conclusions}
Within this paper, we have re-derived a TTL coupling noise model for LISA's second-generation \gls{TDI} Michelson observables, which was previously published in \cite{Houba2022-Estimation} and discussed for the first time the various important assumptions made in this derivation. We have shown that this model, as well as several assumptions made in the derivation, hold only if the model is applied to estimate the noise in single link readouts or \gls{TDI} observables. This means the assumptions and the model hold for cases where optical bench translations that are commonly sensed by the \glspl{LAI} and their corresponding \glspl{TMI} are either canceled out or explicitly neglected.

For the individual interferometers, a different model should be used, which contains the longitudinal motion of the optical bench caused by angular or lateral jitters. We have shown that the TTL coupling model for the individual interferometers deviates from the one for \gls{TDI} by having individual coupling coefficients for every jittering component in every degree of freedom. Likewise, we have shown that several assumptions made in the derivation of the \gls{TDI} model or its simplification, do not hold for the individual interferometers. In particular, the TTL noise in the \gls{TMI} itself is non-negligible, and the coupling from lateral jitter into the \glspl{LAI} cannot be neglected if one cares for the noise in the individual interferometers.
The topic of TTL in individual interferometers will likely be of no interest for the \glspl{LAI}. These interferometers will be dominated by laser frequency noise, such that TTL becomes only observable once the laser frequency noise is suppressed by TDI. Contrary, for the \glspl{TMI}, TTL might be directly visible in the interferometric readout in case of stronger motion, for instance, if TTL-calibration manoeuvres are performed (cf.~\cite{Houba2022-maneuver,Hartig2023-LPF-DA,Wanner2017} for the TTL calibration manoeuvres in LISA Pathfinder and LISA). For interpreting this TTL in the \glspl{TMI}, one would need the equations for the TTL coupling in the individual interferometers. It is expected that the observable coupling would contain strong contributions from longitudinal OB motion caused by angular or lateral jitters. Since this longitudinal OB motion is significantly suppressed in the TDI-TTL contributions, the observable coupling coefficients in the \glspl{TMI} would deviate significantly from the coupling coefficients found in TDI.

Using the model for the \gls{TDI} Michelson $X_2$ observable, we have computed the expected TTL noise levels prior to subtraction for two cases. We assumed coupling coefficients of \SI{10}{mm/rad} which roughly matches the magnitude of coefficients expected prior to performing TTL mitigation by realignment of the optics. Additionally, we used coupling coefficients \SI{2.3}{mm/rad}, resembling coefficient magnitudes after a realignment. We have shown that it is statistically expected that the noise in both cases would not fit into the LISA noise budget, such that a final step of fitting and subtracting the noise in post-processing (as shown in \cite{Paczkowski2022}) is inevitable. 

Even though we have presented an analytic model for the TTL coupling noise PSD for \gls{TDI}-$X_2$, one cannot easily see from the equation for which case the noise is maximal. Therefore, we have additionally derived an analytic equation for the worst-case TTL coupling noise in \gls{TDI}-$X_2$ and all sign combinations that result in this maximal coupling. The derived model holds under the assumption that for every degree of freedom, the jitter spectra of the 6~\glspl{MOSA} or the 3~\gls{SC} are equal, i.e., $S_{\alpha_{\mathrm{SC}i}} = S_{\eta_{\mathrm{SC}}}, S_{\theta_{\mathrm{MO}ij}} = S_{\theta_{\mathrm{MO}}}$ for $\alpha \in \eta, \theta, \phi$, and that the spectra of SC jitter in $\eta$ and $\theta$ are identical, i.e.\ $S_{\theta_{\mathrm{SC}}} =S_{\eta_{\mathrm{SC}}}$.

Furthermore, we analyzed the derived model for implications for the fit and subtraction process. We have shown that the TTL contributions from the \gls{TMI} and \gls{LAI} will be indistinguishable in the LISA data and their postprocessing. However, this is not considered a problem because a distinction of the contribution is neither needed for noise suppression by realignment nor for fitting and subtracting the noise in post-processing. So this indistinguishability is only a fact that should be considered for example for the phrasing of requirements.

Additionally, we found that there exist two sets of receiver and transmitter jitter coupling coefficients that are indistinguishable (i.e.\ fully correlated) if only one \gls{TDI} Michelson observable is used for fitting the coefficients. However, these sets are different for the three different Michelson observables, such that the coefficients can be individually resolved if all three Michelson observables are used simultaneously for fitting the coupling coefficients in post-processing.

%%%%%%%%%%%%%%%%%%%%%%%%%%%%%%%%%%%%%%%%%%%%%%%%%%%%%%%%%%%%%
%%%%%%%%%%%%%%%%%%%%%%%%%%%%%%%%%%%%%%%%%%%%%%%%%%%%%%%%%%%%%
%%%%%%%%%%%%%%%%%%%%%%%%%%%%%%%%%%%%%%%%%%%%%%%%%%%%%%%%%%%%%
\acknowledgements
We want to thank apl.~Prof. Dr. G. Heinzel, for his support and the valuable discussions on the content of this paper.
We gratefully acknowledge support by the Deutsches Zentrum für Luft- und Raumfahrt (DLR) with funding of the Bundesministerium Wirtschaft und Klimaschutz with a decision of the Deutsche Bundestag (DLR Project Reference No.\ FKZ~50OQ2301 based on funding from FKZ~50OQ1801). 
Additionally, G.W.\ acknowledges funding by Deutsche Forschungsgemeinschaft (DFG) via its Cluster of Excellence QuantumFrontiers (EXC 2123, Project ID 390837967).\\
Furthermore, we want to thank ESA for providing the LISA orbits files used in the simulations in \Cref{Sec:Noise-Estimates}.
Finally, we would like to thank Dr.\ M.-S. Hartig and Dr.\ R.\ Giusteri for interesting discussions, and Dr.\ N.\ Houba for helping us quickly resolve the seemingly contradictory results in our papers. 

%%%%%%%%%%%%%%%%%%%%%%%%%%%%%%%%%%%%%%%%%%%%%%%%%%%%%%%%%%%%%
%%%%%%%%%%%%%%%%%%%%%%%%%%%%%%%%%%%%%%%%%%%%%%%%%%%%%%%%%%%%%
%%%%%%%%%%%%%%%%%%%%%%%%%%%%%%%%%%%%%%%%%%%%%%%%%%%%%%%%%%%%%
\appendix
%%%%%%%%%%%%%%%%%%%%%%%%%%%%%%%%%%%%%%%%%%%%%%%%%%%%%%%%%%%%%
\section{Phasemeter model with primary noises}
\label{Sec:Initial_Phaseter_eqns}
%%%%%%%%%%%%%%%%%%%%%%%%%%%%%%%%%%%%%%%%%%%%%%%%%%%%%%%%%%%%%
We have specified the LISA phasemeter equations including generic TTL $N$-terms in \Cref{Eq:PM-equations}. Here, we show its derivation with a particular focus on the involved signs, the mentioned calibration, and the mapping of MOSA and test mass motions along the beam axis. \Cref{Eq:PM-equations} is based on \cite{Otto2015}, but updated to the double-index notation lately used in the LISA Consortium and within this paper. 

In order to allow better tracking of signs, we define the equations below in a two-step process, starting with the beat note $\mathcal{B}[E_j,E_k]$ of two individual laser beams with electric fields $E_j$ and $E_k$, respectively. For instance, the phase signal $s_{12}(t)$ of the \gls{LAI} in OB$_{12}$ is then given by:
\begin{equation}
s_{12}(t) = \arg\{\mathcal{B}[E^{s_{12}}_{21:12}, E^{s_{12}}_{12}] \} \;.
\end{equation}
Here, we give the electric fields an additional upper index to specify the location where they are being measured. 
We now choose $s_{12}$ and all other interferometric signals below to have units of radian, to be consistent in our notation with \cite{Otto2015}. The phase of each electric field is influenced by various effects, and we define in the following simple list notation, what parameters these are:
\begin{subequations}
\begin{align}
E^{s_{12}}_{21:12} =&  E^{s_{12}}_{21:12} (H_{12}, p_{{21}:12}, -k_{{21}:12} \vec n_{{12}} \cdot \vec \Delta_{21:12}, \nonumber\\ 
	&k_{{21}:12}\vec n_{12} \cdot \vec \Delta_{12})  \\
	E^{s_{12}}_{12}  =& E^{s_{12}}_{12} (p_{12}) \;.
\end{align}
\end{subequations}
In the \gls{LAI} phase signal $s_{12}$ we have therefore assumed that the local reference beam described by the electric field $E^{s_{12}}_{12}$ carries only laser frequency noise $p_{12}$, and no other noise. 
Contrary, the received beam from the far spacecraft $E^{s_{12}}_{21:12}$, in the role of the measurement beam, carries a number of phase changes, which we will now discuss.
\begin{itemize}[wide=\parindent]
	\item[$H_{12}$:] Phase shift caused by one or several gravitational waves. The phase shift is accumulated in the electric field during its propagation from \gls{SC}$_{2}$ along arm $L_{12}$ to \gls{SC}$_{1}$, before it is detected in $s_{12}$. We chose the index in $H_{12}$ to match the interferometer in which it is sensed. Since $H_{12}$ is a generic variable representing a phase contribution, it is likewise a choice to place it with an implicit plus sign. The actual sign needs to be modeled when $H_{12}$ is replaced by an explicit expression, which is beyond the scope of this paper.
	\item[$p_{{21}:12}$:] Laser frequency noise contribution. This, again, is a generic term, and therefore simply added in.
	\item[$- k_{{21}:12} \vec n_{{12}} \cdot \vec \Delta_{21:12}$:] Longitudinal transmitter displacement noise contribution. This term describes the phase shift caused by displacement noise of the \gls{MOSA}$_{21}$ (or its optical bench) mapped along the beam's own direction of propagation and converted to phase radian, delayed by the propagation time along arm length $L_{12}$ (cf. \Cref{Fig:lisa}).  
	Here, $ \vec \Delta_{21:12}$ is the delayed displacement noise vector in units of meters describing the motion of \gls{MOSA}$_{21}$ relative to free space.
	This displacement noise causes phase shifts that are described by mapping $ \vec \Delta_{21:12}$ along the beam's direction. Assuming that the beam direction at the time of transmission and receivel are identical (i.e.\ $\vec n_{12:12} =\vec n_{12}$), we can denote the phase shift by $ \vec n_{{12}} \vec \Delta_{21:12}$.\\
	Since the displacement noise is assumed to be given in its natural units of meters, it needs to be converted to units of radians before it can be added to other phase noise terms. This is done by multiplication with the beam's wavenumber, which is $k_{21:12}$ for $E^{s_{12}}_{21:12}$.\\
	Unlike the previous terms, $k_{{21}:12} \vec n_{12} \vec \Delta_{21:12}$ is an explicit model. Therefore, also the sign needs to be defined explicitly. For this, we define a sign convention: the phase of an electric field is stated to increase if the optical path length of the beam's axis increases. Since the optical path length decreases, if the MOSA moves into the beam direction $\vec n_{12}$, we explicitly place a minus sign.
	\item[$k_{{21}:12} \vec n_{12} \vec \Delta_{12}$:] Like the previous term, but for receiver jitter $\vec \Delta_{12}$. The motion is again projected along the beam's direction $ \vec n_{12}$ and converted from units of meters to phase radian by the beam's wavenumber $k_{{21}:12}$. 
	Since the optical path length increases for $E^{s_{12}}_{21:12}$, if the jitter direction is coaligned with the beam axis, the term is added in with an explicit plus sign.
\end{itemize}
In the next step, we can now evaluate the beat note in \gls{LAI}$_{12}$ by assuming linearity and simply subtracting the phases of $E^{s_{12}}_{12}$ and $E^{s_{12}}_{21:12}$. In this step, we assume that the beat note phase is given by the phase of the measurement beam minus the phase of the reference beam, provided the frequency of the measurement beam is higher. Else, it is the other way around:
\begin{equation}
s_{12} = 
\begin{cases}
  \arg(E^{s_{12}}_{12}) - \arg(E^{s_{12}}_{21:12})& \text{if } f_{12} > f_{21:12}\\
  \arg(E^{s_{12}}_{21:12}) - \arg(E^{s_{12}}_{12})& \text{if } f_{21:12} > f_{12} .
\end{cases}
\end{equation}
With the given syntax we denote that $f_{21:12}$ is the frequency of laser 21 which is doppler shifted when propagating along arm-length $L_{12}$ if \gls{SC}$_1$ and \gls{SC}$_2$ move relative to each other. 
Using a signum function, we can now evaluate the beat note as: 
\begin{align}
s_{12}(t) = & \sign(f_{21:12}-f_{12}) \big[ H_{12} + p_{{21}:12} \, -  \nonumber\\ 
	& k_{{21}:12} \vec n_{{12}}  \vec \Delta_{21:12} + k_{{21}:12} \vec n_{12} \vec \Delta_{12} - p_{12} \big] \nonumber \\
= & \sign(f_{21:12}-f_{12}) \bigg[ H_{12} + p_{{21}:12} - p_{12} + \nonumber\\ 
	& k_{{21}:12} \vec n_{{12}} \left( 
	 \vec \Delta_{12} -  \vec \Delta_{21:12} 
	 \right) \bigg] 
	 \;.
\end{align}
Following the same logic, we find the beat notes for \gls{TMI}$_{12}$, where $E_{12}^{\varepsilon_{12}}$ plays the role of the measurement beam, while $E_{13}^{\varepsilon_{12}}$ describes the electric field of the reference beam:
\begin{eqnarray}\label{Eq:BeatNotes}
\varepsilon_{12}(t) &=&  \arg\{ \mathcal{B}[E^{\varepsilon_{12}}_{12} (p_{12},-2  k_{12} \vec n^\varepsilon_{12} \vec \Delta_{12}, +  2 k_{12} \vec n^\varepsilon_{12} \vec \delta_{12}), 
\nonumber\\ 
	&&E^{\varepsilon_{12}}_{13} (p_{13}, \mu_{13})] \}\;.
\end{eqnarray}
The reference beam carries only phase noise $p_{13}$ and fiber backlink noise $\mu_{13}$. We assume that the beam is effectively fixed to the optical bench, which moves with the \gls{SC}, such that its phase is unaffected by \gls{SC} motion. Instead, the measurement beam picks up phase changes by both test mass displacement ($\vec \delta_{12}$) and MOSA displacement ($ \vec \Delta_{12}$) relative to free space, when it reflects from the test mass. 
These are mapped along the beam's own axis, which we denote $\vec n_{12}^\varepsilon$, and which is nominally pointing from OB$_{12}$ towards TM$_{12}$ and is ideally identical to $\vec n_{12}$.
Since a TM displacement along $\vec n^\varepsilon_{12}$ increases the distance between TM and OB we find a plus sign for the contribution of $\vec n^\varepsilon_{12} \vec \delta_{12}$. Contrary to this, \gls{MOSA} displacement along $\vec n^\varepsilon_{12}$ decreases the optical pathlength, resulting in the explicit minus sign for the term $\vec n^\varepsilon_{12} \vec \Delta_{12}$. 
Furthermore, the factor of 2 describes that the wavefront accumulates the corresponding phase change once when propagating towards the test mass, and a second time after reflection when it propagates back towards the optical bench.
Consequently, the phase signal in \gls{TMI}$_{12}$ is given by:
\begin{align}
\varepsilon_{12}(t) =& \sign(f_{12}-f_{13}) \bigg[ p_{12}  - p_{13} - \mu_{13}  \nonumber \\
	& -2 k_{12} \vec n^\varepsilon_{12} \left(\vec \Delta_{12} -  \vec \delta_{12} \right)  \bigg]  \;.
	\label{rawPMTM_app}
\end{align}
Finally, the phase readout in the reference interferometer can be described by:
\begin{subequations}
\begin{align}\label{rawPM_app}
\tau_{12}(t) 	
 		=&   \arg\{ \mathcal{B}[E^{\tau_{12}}_{12}(p_{12}), E^{\tau_{12}}_{13}(p_{13},\mu_{13})] \}  \\
		=&\sign(f_{12}-f_{13}) \left[ p_{12} - p_{13}  - \mu_{13} \right]  \label{rawPMRef_app}  \;.
\end{align}
\end{subequations}
The phasemeter equations for \gls{MOSA}$_{13}$ can be found by substituting index 2 by 3, and vice versa, i.e.:
\begin{subequations}
\begin{align}
s_{13}(t) 			=& \arg\{ \mathcal{B}[E^{s_{13}}_{31:13}, E^{s_{13}}_{13}]  \}  \\
\varepsilon_{13}(t)	=& \arg\{ \mathcal{B}[E^{\varepsilon_{13}}_{13}, E^{\varepsilon_{13}}_{12}] \} \\
\tau_{13}(t)			=& \arg\{ \mathcal{B}[E^{\tau_{13}}_{13}, E^{\tau_{13}}_{12}] \} \;,
\end{align}
\end{subequations}
such that the phasemeter equations for \gls{MOSA}$_{13}$ read:
\begin{subequations}
\begin{align}
s_{13}(t) 			= &\sign(f_{31:13}-f_{13}) \bigg[ H_{13} + p_{31:13} - p_{13} + \nonumber\\ 
	&k_{31:13} \vec n_{{13}}  \left(\vec \Delta_{13} -  \vec \Delta_{31:13} \right)\bigg] \label{rawPMLA_app} \\
\varepsilon_{13}(t)	=&\sign(f_{13}-f_{12}) \bigg[p_{13} - p_{12} - \mu_{12} -   \nonumber\\ 
	&2 k_{13} \vec n^\varepsilon_{13} \left(   \vec \Delta_{13} -  \vec \delta_{13}  \right) \bigg] \label{rawPMTMp_app} \\
\tau_{13}(t) 	=&\sign(f_{13}-f_{12}) \left[p_{13} - p_{12} - \mu_{12} \right] \label{rawPMRefp_app} \;.
\end{align}
\end{subequations}
The phasemeter equations of all remaining interferometers are found by cyclic index permutation.

Finally, we reduce the complexity of the notation by suppressing the signum functions by assuming calibrated signals:
\begin{subequations}%
\begin{align}%
	\check s_{12}(t)				:= & \sign(f_{21:12}-f_{12}) s_{12}(t)   \\
	\check \varepsilon_{12}(t)  	:=& \sign(f_{12}-f_{13}) \varepsilon_{12}(t)  \\
	\check \tau_{12}(t)			:=&\sign(f_{12}-f_{13}) \tau_{12}(t)	\\
	\check s_{13}(t) 			:= &\sign(f_{31:13}-f_{13}) s_{13}(t) \\
	\check \varepsilon_{13}(t)		:=&\sign(f_{13}-f_{12})\varepsilon_{13}(t) \\
	\check \tau_{13}(t) 			:=&\sign(f_{13}-f_{12}) \tau_{13}(t) \;.
\end{align}%
\label{Eq:Cal_displacement_readout}%
\end{subequations}%
This means we assume that the frequencies are measured during flight, the signum functions are evaluated and multiplied to the raw signals (RHS of the equation), and the resulting products are given to the users. 
An alternative interpretation, which is equally valid for the signals used in \Cref{Eq:PM-equations} and throughout the paper, is that they hold for an uncalibrated case for the frequency relations that evaluate all signum-functions in  \Cref{Eq:Cal_displacement_readout} to plus one. This means it is assumed that in every stated interferometer, the frequency of the measurement beams would be higher than the frequency of the corresponding reference beam. 

The equations presented in this appendix, still deviate from the ones in \Cref{Eq:PM-equations} by the normal vectors used for the projection of MOSA and test mass motions. The notation used in this appendix is principally more precise because it distinguishes the projection directions in the \glspl{LAI} and \glspl{TMI}. Due to these different projection directions, the OB motion terms $\vec \Delta$ would no longer fully cancel from \gls{TDI}. Again, this is more realistic, yet, we can consider the residuals as TTL coupling noise. After all, the residuals originate from angular misalignments of the \gls{TMI}$_{ij}$ and \gls{LAI}$_{ij}$ beam axes, which is a typical TTL coupling mechanism for us.

Furthermore, we state in \Cref{Sec:Delineation-N-vecDelta} that OB motion $\vec n \vec \Delta$ and TTL effects $N$ need to be clearly delineated. Imperfect OB motion cancellation due to misaligned beam axes can be considered as TTL effect. We therefore decided to assume all such effects to be modelled in N, rather than in the original phasemeter equations. 
Consequently, we can simplify the mapping and assume perfect alignment between the \gls{LAI}$_{ij}$ and the corresponding \gls{TMI}$_{ij}$ for the phasemeter equations in \Cref{Eq:PM-equations}:
\begin{align}%
	\vec n^\varepsilon_{ij} = \vec n_{ij} = - \vec n_{ji} 
	\label{Eq:projection-n-simplified}%
\end{align}%
Here, the last equality states that we now assume the received beam direction and transmit beam directions to be equal except for opposite directions. This is needed to allow perfect cancellation of the transmitter jitter in a single link (cf.~\Cref{Eq:xi}). For example, $-\vec n_{12} \vec \Delta_{21:12}$ in $\check s_{12}$ cancels only with the corresponding term $-2 \vec n_{21} \vec \Delta_{21:12}$ in $\varepsilon_{21:12}$, if we assume $-\vec n_{12} = \vec n_{21}$.
 
With this, we find the phasemeter equations defined in \Cref{Eq:PM-equations}.

\bibliography{bibl} 
\end{document}